\documentclass[floatfix,aps,pre,twocolumn,superscriptaddress,showpacs]{revtex4}
\usepackage{epsfig,amsmath,amssymb,graphicx,color,calc}

\newcommand{\beq}{\begin{equation}}
\newcommand{\eeq}{\end{equation}}
\newcommand{\ba}{\begin{align}}
\newcommand{\ea}{\end{align}}

\let\a=\alpha \let\b=\beta  \let\g=\gamma  \let\d=\delta 
         
\let\l=\lambda \let\m=\mu  \let\n=\nu       \let\p=\pi  
 \let\t=\tau   \let\ph=\varphi    \let\f=\phi 
           
\let\D=\Delta  \let\Th=\Theta        
\let\Si=\Sigma         
 \let\r=\rho \let\th=\theta

\let\io=\infty
\def\ie{{i.e. }}\def\eg{{e.g. }}

\def\CC{{\cal C}} 
\def\NN{{\cal N}} 
  
 \def\SS{{\cal S}}

\def\vx{{\vec{x}}} \def\vX{{\vec{X}}} \def\vr{{\vec{r}}}

\def\erf{\text{erf}}

\def\to{\rightarrow}
\def\la{\left\langle}
\def\ra{\right\rangle}

\def\wt{\widetilde}

\begin{document}

\title{Microscopic theory of the 
jamming transition of harmonic spheres}

\author{Ludovic Berthier}
\affiliation{Laboratoire Charles Coulomb, UMR 5221, CNRS and Universit\'e
Montpellier 2, Montpellier, France}

\author{Hugo Jacquin}
\affiliation{Laboratoire Mati\`ere et Syst\`emes Complexes, UMR 7057,
CNRS and Universit\'e Paris Diderot -- Paris 7, 10 rue Alice Domon et L\'eonie 
Duquet, 75205 Paris cedex 13, France}

\author{Francesco Zamponi}
\affiliation{Laboratoire de Physique Th\'eorique, UMR 8549, CNRS and Ecole Normale Sup\'erieure, 
24 Rue Lhomond, 75231 Paris Cedex 05, France}

\date{\today}

\begin{abstract}
We develop a microscopic theory to analyze  
the phase behaviour and compute correlation functions of dense assemblies 
of soft repulsive particles both at finite temperature, 
as in colloidal materials, and at vanishing temperature, a situation 
relevant for granular materials and emulsions. We use a mean-field statistical 
mechanical approach which combines elements of liquid state theory 
to replica calculations to obtain quantitative predictions 
for the location of phase boundaries, macroscopic 
thermodynamic properties and microstructure of the system. We focus
in particular on the derivation of scaling properties emerging in 
the vicinity of the jamming transition occurring at large density and 
zero temperature. The new predictions we obtain 
for pair correlation functions near contact are tested using 
computer simulations. Our work also clarifies the conceptual 
nature of the jamming transition, and its relation to the 
phenomenon of the glass transition observed in atomic liquids. 
\end{abstract}



\maketitle

\section{Introduction}

\subsection{From atoms to balls}

It is  common practice in physics textbooks to represent
various states of matter by replacing atomic or molecular 
arrangements by assemblies of spherical balls. The analogy goes
much deeper, since simple theoretical models of spherical, repulsive 
particles are known to display phase transitions between states 
of matter that are actually observed in atomic systems. 
About 50 years ago, Bernal~\cite{Be59,BM60} 
proposed a theory of the liquid state
based on the intuition that dense liquids have a structure similar 
to dense assemblies of macroscopic grains, and performed
with coworkers experiments to analyze the microstructure of steel
ball bearings.
Since Bernal's insightful work, a large number of detailed
experiments and numerical simulations have been made to characterize
the structure of dense assemblies of spherical particles, both 
for hard particles relevant for granular materials~\cite{rmpgrains}
and for soft particles relevant for emulsions and foams~\cite{booklarson}.

However, while the structure of dense atomic liquids is by now well 
understood using the framework of statistical mechanics~\cite{hansen}, 
the problem of sphere packing proved much more difficult to 
solve~\cite{bookpursuit,TS10}. 
Indeed, there are two major obstacles to be faced when approaching 
this problem. First, sphere packings are inherently nonequilibrium
systems where thermal fluctuations play no role. This has for instance led 
researchers to extend concepts of statistical mechanics to specifically
address athermal packings implementing ideas first put forward in particular 
by Edwards~\cite{EO89,Ed98,SWM08,HC09}. 
A second major difficulty is that one must deal with 
both fluid and solid disordered states of matter, which are poorly
understood subjects even for atomic systems driven by thermal 
fluctuations. 
In particular, while dense liquids are appropriately described 
by liquid state theory, viscous liquids and amorphous materials 
are still the focus of intense research~\cite{BB10}. 
Following Bernal's insight, 
it was suggested that the fluid to solid transition observed 
in athermal systems shares similarities with the glass transition 
of supercooled molecular fluids~\cite{LN98}. 
Since its publication, 
the ``jamming phase diagram'' has striked lively debates and a 
large literature seeking in particular to understand 
similarities and differences in parallel studies of both thermal and 
athermal amorphous media~\cite{LNSWW10}.

In this work, we are interested in describing theoretically 
fluid and solid disordered 
states of matter observed in dense assemblies of spherical 
repulsive particles both for finite temperatures relevant 
for colloidal materials, and in the athermal limit 
relevant for emulsions and foams. Unlike Bernal, we now 
have at our disposal theories for the liquid state, but this is not 
sufficient~\cite{JB10,BJZ10}.
Liquid state theory is essentially based on perturbation 
theory extrapolating from the gas state, and for this reason it 
inherently fails to describe the phase 
transitions to low temperature and high density solid phases of matter. 
However, the liquid-crystal transition is conveniently described 
by density functional theory, 
where one starts from an approximated form for the free energy as 
functional of the local density profile and looks for a transition
between uniform and periodically modulated minima of this functional.
Early studies of the glass transition~\cite{SW84,SSW85} used density 
functional theory and looked for amorphous density profiles that minimize 
the free energy. 
Theoretical developments in the field of spin glasses~\cite{KW87a,KT87a,KTW89}
suggested similarities between supercooled liquids 
and a class of mean-field spin glass models which can be 
extensively analyzed theoretically~\cite{MPV87,CC05}. 
These works gave birth to the so-called Random First Order Transition 
(RFOT) theory~\cite{KTW89}, 
which is currently one of the most consistent theories of the glass 
transition~\cite{LW07,Ca09,BB09,BB10}. 
Along with these conceptual developments, RFOT 
theory has been turned into a 
quantitative computational scheme in 
a series of pioneering works by Monasson~\cite{Mo95}, Franz, 
M\'ezard and Parisi~\cite{MP96,CFP98,MP99a,MP99b}. 
These works can schematically be seen as an extension 
of the tools developed in liquid state theory to study 
systems near the glass transition and the properties 
of amorphous solids, but they remain ``mean-field'' in nature
in the sense that the glassy phase is assumed 
to be of the same nature as the one obtained from
various mean-field perspectives~\cite{BB10}. We wish to emphasize, however, 
that ``mean-field'' has a very specific meaning
in the context of glass theories. Although the theory 
we shall develop is a mean-field theory, it remains 
fully microscopic in nature (we start from the interaction 
between the particles), and it is able to capture many-body
correlations between particles giving rise to a complex
free energy landscape, which is crucial to accurately 
predict both the phase diagram of the system and correlation 
functions. In particular, we are able to provide 
a theoretical description of the distinct nature 
of both glass and jamming transitions.  

Our goal is thus to apply concepts and tools from 
RFOT theory to analyze the glassy and jammed phases 
of systems of soft repulsive particles, thus contributing 
to better understanding amorphous, athermal 
states of matter~\cite{Vh09,LNSWW10}. In previous work,
Parisi and Zamponi (PZ) extended the scheme devised 
by M\'ezard and Parisi (MP)  to treat systems of hard 
spheres~\cite{PZ10}. 
The result is a RFOT theory of the glass transition and jamming
of hard spheres; this theory is expected to be correct in the limit
of large space dimension~\cite{PZ10} and for a modified model where
particles are randomly displaced over the box~\cite{MK11} (both limits
being of a mean-field nature), and gives
quantitatively correct results in three dimensions~\cite{PZ10}.
In a similar spirit, Mari {\it et al.}
devised a model of hard spheres with diluted interactions~\cite{MKK09} 
where ``mean-field'' replica calculations become exact, 
although the solution remains quite involved~\cite{solvable}. 
We wish to unify the approaches of MP and PZ  
to describe simultaneously the critical properties on both sides of 
the jamming transition occurring in athermal materials, and in 
its vicinity at finite temperatures.
The theory that we develop in this work is, in our opinion, 
the equivalent, for the jamming transition, 
of a Landau mean-field theory for second order phase transition.

\subsection{Glass and jamming transitions in harmonic spheres}

A model system that can interpolate between finite temperature 
glasses and hard spheres is the 
model of harmonic spheres. 
An assembly of $N$ spherical particles of diameter $\sigma$ is enclosed 
in a volume $V$ in three spatial dimensions, interacting with a soft 
harmonic repulsion of finite range:
\beq
\phi(r) = \epsilon (1-r/\sigma)^2 \, \theta(\sigma-r) \ ,
\label{pot}
\eeq 
where $r \geq 0$ is the interparticle distance, $\th(r)$ is the Heaviside step function and $\epsilon$ controls the strength of the repulsion.
The model (\ref{pot}), originally proposed to describe wet foams~\cite{Du95}, has become a paradigm in numerical studies of the $T=0$ jamming transition~\cite{OLLN02,Vh09}. It has also been studied at finite temperatures~\cite{BW09b,BW09a,ZXCYAAHLNY09}, and finds experimental realizations in emulsions, soft 
colloids and grains~\cite{bookmicrogel}. The model 
has the two needed control parameters to explore the jamming phase diagram:
the temperature $T$ (expressed in units of $\epsilon$), and the fraction of the volume occupied by the particles in the absence of overlap, the volume 
fraction $\ph = \pi N \sigma^3/ (6V)$. We now set $\sigma$ and $\epsilon$ to unity.

\begin{figure}[t]
\includegraphics[width=.95\columnwidth]{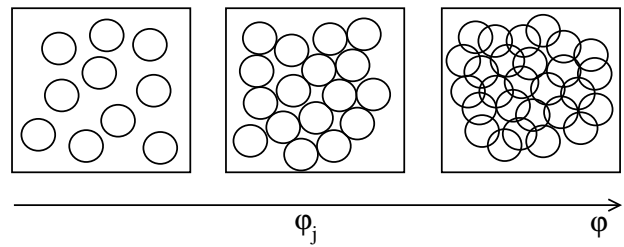}
\caption{
The jamming transition of harmonic spheres 
at zero temperature. Below $\ph_j$, particles do not 
overlap (as in hard sphere configurations) and the system flows. 
Above $\ph_j$, there is a finite density of overlaps,
and the energy and pressure are finite. In this region, 
hard spheres configurations cannot be found.}
\label{fig:jamming}
\end{figure}

Over the last decade, a large number of numerical observations was reported for this model~\cite{Vh09,LNSWW10}. A jamming transition is observed at $T=0$ at some critical volume fraction $\ph_j$, the density above which the packings carry a finite density of particle overlaps. This transition is pictorially represented in Fig.~\ref{fig:jamming}.
Numerically, the zero-temperature energy density, $e_{\rm GS}$, and pressure, $P$, are found to increase continuously from zero above $\ph_j$ as power laws~\cite{OLLN02}. The pair correlation function of the density fluctuations~\cite{hansen}, $g(r)$, develops singularities near $\ph_j$~\cite{DTS05,SLN06}, which are smoothed by thermal excitations~\cite{ZXCYAAHLNY09}. In particular, $g( 1 ) = \infty$ at $\ph_j$ and $T=0$. This behavior implies that the density of contacts between particles, $z$, jumps discontinuously from $0$ to a finite value, $z_j$, at $\ph_j$, and increases further algebraically above $\ph_j$~\cite{OLLN02,Vh09}. Thus, the jamming transition appears as a phase transition taking place in the absence of thermal motion, with a very peculiar critical behavior and physical consequences observable experimentally in a large number of materials.
Note however that in the unjammed state below $\ph_j$, the specific configurations sampled by the system might depend on the details of the preparation protocol. 
In the absence of fluctuations for instance~\cite{HB09,GBO06}, the contact number is not vanishing even below the jamming point, because particles remain in contact and the left cartoon in Fig.~\ref{fig:jamming} 
is not a faithful representation.

As discussed in detail by Krzakala and Kurchan~\cite{KK07a},
the jamming transition is very similar in 
nature to the so-called SAT-UNSAT transition of random constraint satisfaction problems~\cite{Mo07,MM09}. These are problems where a large set 
of constraints has to be simultaneously satisfied. Upon increasing the density of constraints per variable, a phase transition is found, from a phase where the problem is typically solvable (SAT) to a phase where it is typically not (UNSAT).
In the present context, the constraints are the non-overlapping conditions between spheres, and the control parameter is the density. At $\ph_j$, the system goes from a phase where the non-overlap conditions can
be satisfied to a phase where they cannot. In the context of random constraint satisfaction problems, it is very useful to introduce a soft version of the constraints, 
study the problem at finite temperature $T$, and finally take the $T\to 0$ limit~\cite{MP03,KMRSZ07,KZ08}. 
The reason is that introducing a temperature allows to use
powerful statistical mechanics tools, that can be used in a context where they are not a priori relevant~\cite{KK07a}.

\begin{figure}[t]
\includegraphics[width=.95\columnwidth]{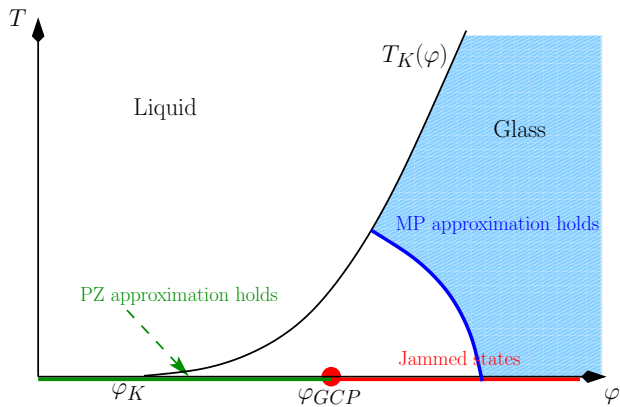}
\caption{(Color online)
A schematic phase diagram for the glassy states of the 
model defined by Eq.~(\ref{pot}) showing both the 
glass transition line at $T_K(\ph)$ and the jamming point 
at $T=0$ and $\ph = \ph_{GCP}$.
The Parisi-Zamponi (PZ) approach 
only holds at $T=0$ and $\ph < \ph_{GCP}$, while the 
M\'ezard-Parisi (MP) approach holds for large enough densities 
and temperatures. In this paper we develop a method to fill the gap between
the PZ and MP analytical schemes to explore the vicinity of 
the jamming transition.}
\label{fig:PDsch}
\end{figure}

To address the purely geometrical packing problem of soft spheres, we then suggest to study first the statistical mechanics of the model defined by Eq.~(\ref{pot}) 
at finite temperatures, before taking the $T \to 0$ limit where jamming occurs.  
Based on the analogy with constraint satisfaction problems, we have in mind the schematic phase diagram reported in Fig.~\ref{fig:PDsch}: 
see in particular the result for Potts glasses in~\cite{KZ08}, and see also~\cite{ricci10}. 
In this 
picture, that we will derive in the following, the liquid undergoes a glass transition at some temperature $T_{K}(\ph)$. The line $T_K(\ph)$ vanishes at a volume fraction $\ph_K$ which corresponds also to the ideal glass
transition for hard spheres~\cite{PZ10}. The point $\ph_K$ {\it is not the jamming transition}: indeed, the ground state energy and pressure remain zero across $\ph_K$. Above $\ph_K$, the system enters at zero temperature 
a {\it hard sphere glassy state}. 
In this state, particles vibrate near well-defined (but random) positions, 
and the system is not yet jammed~\cite{PZ10}. Jamming happens when
the glass reaches its close packing density, which we shall 
call ``glass close packing'' (GCP)~\cite{PZ10}. 
We identify the jamming transition with $\ph_{GCP}$. 
In Fig.~\ref{fig:PDsch} the density interval $[\ph_K, \ph_{GCP}]$
is simply the amorphous 
analog of the interval $\ph \in [0.54,0.74]$ for ordered 
states of hard spheres where a compressible crystalline structure
exists at thermal equilibrium.

The main theoretical difficulty in the study of the jamming 
transition at $\ph_{GCP}$ is that it happens deep inside the glass phase. 
Therefore, we need 
to develop first an accurate theoretical
description of the glass phase. 
As discussed in detail in Ref.~\cite{BJZ10}, 
previous theories of the glass phase fail (see Fig.~\ref{fig:PDsch}). 
The MP approach holds only at high enough temperature or density,
where a simple harmonic approximation for vibrations in the glass holds. The PZ approach holds only in the hard sphere region, $T=0$ and $\ph < \ph_{GCP}$.
The central theoretical achievement that we report in this paper is an approximation scheme that can naturally 
interpolate between those of MP and PZ and is correct in the entire
vicinity of the jamming transition, thus
allowing us to fully explore the phase diagram of Fig.~\ref{fig:PDsch},
and in particular the region of low temperature, 
$T \ll T_K(\ph)$
 and $\ph \sim \ph_{GCP}$. 
We obtain theoretically from a first principle calculation
a large number of the observed behaviors of harmonic 
spheres, 
and predict new results for the correlation functions 
of this system around the jamming transition. 
A short report of our results has been published in Ref.~\cite{JBZ11}.

\subsection{Discussion}

Our approach is very different from, but complementary to, 
recent theoretical works on jamming.

The aim of our work is to show, directly from the Hamiltonian, that the jamming transition exists, and determine from first principles its location and 
properties. Other approaches assume the existence
of jammed states and try to obtain geometrical informations on them~\cite{CCSB09,SWM08}.
A particularly interesting approach is to develop a scaling picture of the jamming transition 
by showing that the transition
is characterized by a diverging correlation length~\cite{WNW05,WSNW05,BW06}.
This approach allows to obtain a very detailed description of the jamming transition~\cite{Wyart},
that explains the anomalous scalings in shear~\cite{OLLN02,OSLN03} and transport~\cite{VXWLN10} properties of amorphous jammed packings.

Moreover, 
our method is crucially based on {\it equilibrium statistical mechanics}. In the glassy phase, on approaching $T_K(\ph)$ and below it, 
the equilibrium Gibbs distribution of the system is the superposition of many distinct ergodic components, each
representing a different glassy state; each of these ``basins'' (see Appendix~\ref{app:basins} for a more detailed discussion) is visited,
in our theory, according to its equilibrium weight. On the contrary, most of the numerical 
works on jamming focus on specific dynamic protocols, that often are athermal and therefore always out of equilibrium. It has been shown
that these protocols sample jammed states with highly non-uniform probabilities~\cite{GBO06,GBOS09}.
Even thermal protocols (see e.g.~\cite{DTS05}) inevitably fall out of equilibrium before $T_K(\ph)$ is reached, and therefore produce jammed
packings with probabilities that are strongly different from the equilibrium ones~\cite{KK07a,PZ10}.
Still, a crucial observation is that the critical properties around the (protocol dependent) jamming density $\ph_j$ are largely independent
of the protocol, and therefore of the probability with which jammed states are visited. This observation 
suggests that the critical properties are shared
by almost all jammed states, and therefore an equilibrium sampling of these states can appropriately describe these properties, as we showed
in this work.
Of course, some care will have to be taken when comparing with some
specific protocols, in which no fluctuations are present at all
(neither thermal, nor mechanical)~\cite{GBO06,HB09}. In those cases, as we already mentioned, 
the contact number is not vanishing even below the jamming
point. Once these effects are taken into account, we believe that the
generic picture we discuss in the following also applies to these
protocols.

Finally,
it is worth mentioning that if compression rates are small, partial crystallization takes place and the system can jam at any packing fraction
between $\ph_{GCP}$ and the close packing density corresponding to the most dense crystal~\cite{JM10,SOS11,JST11}. In our theory, crystallization
is not included by construction, as we will explain in the following, therefore this effect is absent. 
Extending this study to take into account
partial crystallization would be interesting but probably very hard.

The outline of the paper is the following. In Sec.~\ref{method}, 
we present the approximation scheme developed earlier, 
and describe our own approximation specifically developed
to study harmonic spheres. In Secs.~\ref{glass},  
\ref{jamming} and \ref{correlation} respectively, 
we present results concerning the glass transition, the jamming transition 
and the correlation functions.
In Sec.~\ref{comparison} we compare the theoretical 
results with numerical data.
We discuss and conclude the article in Sec.~\ref{conclusion}.

\section{The method} 
\label{method}

\subsection{The complexity}

In mean-field models of glasses, upon decreasing temperature, a number $\NN$ of metastable states, that scales as the exponential of $N$, appears~\cite{CC05}.
The dynamics of the glass is slowed down considerably: once the glass has fallen into one of these metastable states, 
the barrier it has to cross in order to jump to another metastable state is proportional to $N$, due to the mean-field nature of the models. 
In mean-field theory there are two distinct transitions upon lowering the temperature towards the glass state~\cite{CC05}: 
at a first transition $T_d$, the metastable states appear, giving rise to a finite {\it complexity} $\Si = N^{-1} \log \NN$, which is the entropic contribution
due to the exponential number of metastable states. Below this temperature, the dynamics of the system is arrested:
this temperature is indeed identified with the dynamical transition predicted by Mode-Coupling Theory~\cite{KT87a,Ca09}. 
At a lower temperature $T_K$, the complexity vanishes, and the number of metastable states is sub-exponential; this has been identified with the Kauzmann entropy
crisis transition~\cite{Ka48}. Therefore at the mean-field level the complexity is finite between $T_d$ and $T_K$.
Although we will not discuss this physics in the rest of the paper,
it is worth noting at this point that RFOT theory predicts that in a three (or finite) dimensional system the complexity is finite only locally.
This is because the system gains entropy by accessing many different states;
this creates an entropic driving force that allows the system to maintain its ergodicity by visiting all its metastable states. A nucleation theory of supercooled liquids has been developed based on these
ideas~\cite{LW07,Ca09,BB09}, which predicts that the transition at $T_d$ is avoided in finite dimension through activated events, and the dynamics is arrested only at $T_K$.
These ideas have been initially proposed on a phenomenological basis~\cite{LW07,Ca09,BB09} but are now starting to be confirmed by 
renormalization group computations in simple models~\cite{CDFMP10,CBTT11}.
While nucleation processes are surely important to describe activated dynamics below $T_d$, one expects that the local structural properties will not be affected by them and therefore mean-field
theory should provide precise predictions as long as static and local observables are concerned. 
For this reason, in this work we follow M\'ezard and Parisi \cite{MP96,MP99a} and directly apply the mean-field concepts described above to three dimensional glass-formers, neglecting activation. 

Let us denote by $\CC$ the microscopic configuration of the system.
The starting point of the discussion is the assumption that in the glassy region, phase space can be partitioned in {\it pure states}, 
indexed \mbox{by $\a = 1, \cdots, \NN$}, in such a way that a given configuration $\CC$ belongs to one state $\a$.
A practical implementation of this construction, which 
has been studied recently in Ref.~\cite{XFL11}, is discussed in Appendix~\ref{app:basins}.
Under this decomposition, the partition function can be written as:
\begin{align}\label{Cpart}
Z = \sum_\CC e^{-\b E(\CC)} = \sum_{\a=1}^\NN \sum_{\CC \in \a} e^{-\b E(\CC)}.
\end{align}
We define an effective free-energy of a pure state $f_\a$ by $f_\a = - \frac1{N\b} \ln \left( \sum_{\CC \in \a} e^{-\b E(\CC)} \right)$, 
and the complexity $\Si$ as the logarithm of the number $\NN(f)$ of pure states at a given free-energy $f$: $\Si(f) = \frac1N \log \NN (f) $. We assume
that, as in exactly solvable mean-field models~\cite{CC05}, the complexity $\Si(f)$ is non-zero in a finite interval $[f_{min}, f_{max}]$ of free energy,
and it vanishes linearly at $f_{min}$.
We find:
\begin{align}
\label{Zfonecopy}
Z = \int df \NN (f) e^{-\b f} =  \int df e^{-N \b (f - T \Si(f))}.
\end{align}
It is convenient for the following of the paper to consider, instead of the total free energy $F = -T N^{-1} \log Z$, a
``free entropy'' $\SS = N^{-1} \log Z = -\beta F$; the advantage is that this quantity has a finite
limit for $T\to 0$, which is the entropy of hard spheres.
Saddle-point evaluation at thermodynamic limit gives an expression for the free entropy:
\begin{align}
& \SS(T,\ph) \equiv \frac1N \ln Z = \Si(f^*(T,\ph)) - \b f^*(T,\ph) , \label{def_ssglass} \\
& \frac{\partial \Si}{\partial f}(f^*(T,\ph)) = \frac1T . \label{def_fstarglass}
\end{align}
The glass transition is reached at the Kauzmann temperature $T_K(\ph)$, when the complexity vanishes:
\begin{align}
\label{Sivanishes}
\Si(f^*(T_K(\ph),\ph)) = \Si(f_{min}) = 0.
\end{align}
When $T<T_K(\ph)$, the system has the free energy $f_{min}(T,\ph)$ and the complexity sticks to zero.
Therefore, detecting the glass transition and calculating the free energy of the glass phase amounts
to finding a way to compute the complexity as a function of the free energy. 

A schematic plot of $T_K(\ph)$ for harmonic spheres is reported in Fig.~\ref{fig:PDsch}. Although our
approach does not allow for a reliable calculation of $T_d(\ph)$~\cite{PZ10}, we expect that the latter
will be slightly bigger than $T_K(\ph)$ and follow a similar trend, as it can be shown in mean-field
lattice models (see e.g. Fig.2 in Ref.~\cite{FSZ11}).

\subsection{Replicating the system}

The need to replicate the system in order to compute the thermodynamics of the glass originates from the difficulty in defining an order 
parameter for the glass transition. Indeed, below $T_K$ the system freezes in an amorphous density profile which is structurally very
similar to a typical liquid configuration, therefore distinguishing the liquid and the glass by means of some observable related to structure
(like a correlation function) is extremely difficult. Edwards and Anderson proposed a way out of this difficulty by introducing
an order parameter defined as~\cite{MPV87}
\beq
q_{EA} = \frac1V \int d^3x \la \d \r(x) \ra^2 \ ,
\eeq
where $\la \d\r(x) \ra = \la \r(x) \ra -\r$ is the thermodynamic average of the fluctuation of the local density $\r(x)$ with respect
to its spatial average $\r = N/V$. This order parameter is clearly zero in the liquid phase while it is non-zero in a frozen amorphous
phase. The problem is that its computation is not easy since one has to compute the local density and average its square. A solution
is to introduce two identical copies of the system and compute
\beq
q_{EA} = \frac1V \int d^3x \la \d \r_1(x) \ra \la \d \r_2(x) \ra \ .
\eeq
The problem is that if the two replicas are independent, they will freeze in completely different profiles, and $q_{EA}$ will be zero
since the integrand will have random signs. To avoid this problem, we must
couple the two replicas via a small interaction, and then compute $q_{EA}$ by sending first $N\to\io$ and then switching off the
interaction. In a liquid phase, since ergodicity is maintained, the copy will eventually decorrelate from the original, and the order
parameter will go to zero. In an ergodicity-broken phase, the copy will stay trapped near the original and the order parameter 
will go to a finite value. 
This prescription allows one to detect the glass transition but does not give any information on the free-energy of the glass 
and thus the glass phase itself.

Monasson~\cite{Mo95} showed that replicas can provide a simple way to compute the complexity. 
He introduced an arbitrary number of replicas $m$, and assumed that a small coupling between each pair
of replicas is present, in such a way that all replicas are 
constrained to be in the same metastable state.
Since the number of metastable states does not change, the complexity is 
unaffected by this procedure. 
The partition function $Z_m$ of this replicated liquid then reads:
\begin{align}\label{rep_part_funct}
Z_m = \int df e^{-N\b (mf - T\Si(f))}.
\end{align}
The number $m$ of replicas is eventually taken to be non-integer, 
creating an additional $m$ dependence of the free-energy and complexity of the glass. 
Equations (\ref{def_ssglass}) and (\ref{def_fstarglass}) become now:
\begin{align}
& \SS(m;T,\ph) = \Si(f^*(m;T,\ph)) - m \b f^*(m;T,\ph) \label{def_ssglass_replicated} , \\
& \frac{\partial \Si}{\partial f}(f^*(m;T,\ph)) = \frac{m}{T} . \label{def_fglassreplicated}
\end{align}
This $m$-dependence of the free entropy allows one to compute the complexity as follows.
We can determine $\Si$ and $f^*$ as functions of $m$ at fixed $(\ph, T)$ using
\begin{align}
& \Si = -m^2\frac{\partial (\SS/m)}{\partial m} \  , \nonumber \\
& f^* = - T \frac{\partial \SS}{\partial m} \  . \label{def_sigma_replicated_parametric}
\end{align}
Then one can eliminate parametrically the dependence on $m$ and in this way reconstruct the
full curve $\Si(f)$.

We showed in the previous section that for $T < T_K$ the free energy of the glass is equal to $f_{min}(T)$, the point where $\Si(f)=0$. Therefore, to compute $f_{min}(T)$, we must impose that $\Si(f) = \frac{\partial (\SS/m)}{\partial m} =0$, which is equivalent to finding the minimum of $\SS/m$ with respect to $m$. Let us call $m^*(T,\ph)$ the point where $\SS/m$ assumes its minimum. Since $\Si=0$ at the minimum, using Eq.~(\ref{def_ssglass_replicated}), 
we get $\SS(m^* ; T,\ph) = - m^* \b f_{min}(T,\ph) $ which implies that:
\begin{align}
\SS_{glass}(T,\ph) & =  -\b f_{glass}(T,\ph) =  -\b f_{min}(T,\ph) \nonumber \\
& = \frac{\SS(m^*;T,\ph)}{m^*(T,\ph)} = \min_{m} \frac{\SS(m;T,\ph)}{m}.  \label{def_SS_glass}
\end{align}
Now we see that, in order to compute the thermodynamics of the ideal glass, one must be able to calculate the free entropy of a replicated liquid. In the following, we will present several approximation schemes that we applied and adapted, when necessary, for harmonic spheres.

\subsection{Effective potential approximation}
\label{sec:effpotapp}

Computing the free-energy and complexity of the glass amounts to computing the free entropy of a $m$-times replicated liquid. One can use a replicated version of standard liquid theory approximations such as the replicated Hyper-Netted-Chain (HNC) approximation~\cite{MP96,CFP98}, but these kind of approximations usually break down deep in the glass phase. 
For a discussion of the reasons for the failure of replicated HNC, 
see Ref.~\cite{PZ10}. 
One can also follow Refs.~\cite{MP99a,MP99b} and start 
at high density, and perform cage expansions: at high densities the copies stay close to the originals, forming molecules of size $A$. One can then expand the replicated free entropy with respect to this parameter $A$. However, to be able to explicitly compute the various integrals appearing in the computation, one traditionally has to suppose that the attraction between copies is harmonic, an assumption which breaks down for hard spheres, as discussed in~\cite{PZ10}.
To treat correctly the replicated hard sphere system, Parisi and Zamponi~\cite{PZ10} developed an effective potential approximation, which in the end amounts to performing an expansion in powers of $\sqrt{A}$ instead of $A$. We explain in the following the extension of the effective potential approximation to finite temperature harmonic spheres. By construction, we will see that we are bound to recover both high density MP results and zero temperature PZ results
in a unified treatment. 

\begin{figure}[t]
\includegraphics[width=.85\columnwidth]{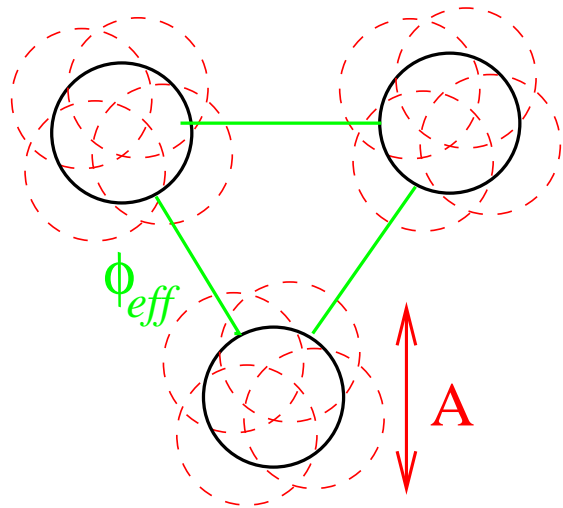}
\caption{(Color online) A schematic representation of the effective potential 
approximation. Each particle in the
original liquid is replicated $m$ times (dashed spheres). Assuming
that the replicated particles form a molecule of average cage
size $A$, we trace out in the partition sum the degrees of freedom
of ($m-1$) copies of the liquid to obtain an effective 
one-component liquid (black spheres) interacting with an effective
pair potential $\phi_{eff}$ (green lines).}
\label{fig:sketchveff}
\end{figure}

The starting point of all these approaches is the assumption that (due to the implicit coupling between replicas) the replicated system is composed of {\it molecules}
made of $m$ atoms, each belonging to a different replica. Making use of standard theory of molecular liquids, one can express the free energy as a functional
of the single molecule density and the interaction potential, see~\cite{PZ10} for details. In order to make the computation tractable, one then makes
a Gaussian {\it ansatz} for the probability distribution function of the positions $\bar x$ of the replicas within a molecule~\cite{MP99b,PZ10}:
\begin{equation}\label{rhoGauss}
\r (\vx_1 \cdots \vx_m) =  \int d^3 X \prod_{a=1}^m \frac{1}{(2 \p A)^{3/2}} e^{-\frac{(\vx_a-\vX)^2}{2 A}}
\end{equation}
We can now integrate out all replicas except one. The {\it effective 
interaction} between particles in replica 1, $\f_{eff}$, 
is obtained by averaging 
the full interaction over the probability distribution of the two molecules:
\beq\label{effpot_def}
\begin{split}
& e^{-\b \f_{eff}(\vec x_1-\vec y_1)} \equiv e^{-\b\f(\vx_1-\vec y_1)} \left\langle \prod_{a=2}^m e^{-\b \f(\vec x_a - \vec y_a)} \right\rangle_{\vec x_1,\vec y_1} \\
&  = \int d^3x_{2}\cdots d^3x_{m} d^3y_{2} \cdots d^3y_{m}  \times \\ 
& \times \r(\vx_1 \cdots \vx_m) \r(\vec y_1 \cdots \vec y_m) \prod_{a=1}^m e^{-\b \f(\vec x_a - \vec y_a)} \ ,
\end{split}\eeq
We get after some simple manipulation:
\begin{equation}
e^{-\b \f_{eff}(r)} = e^{-\b \f(r)} \int d^3r' \g_{2A}(\vr') 
q(A,T ; \vr-\vr')^{m-1},
\end{equation}
where $\g_{2A}$ is a normalized and centered Gaussian of variance $2A$ and $q(A,T;r)= \int d^3r' \g_{2A}(\vec r') e^{-\b \f(\vec r- \vec r')} $ 
is a function that can be explicitly computed for our harmonic potential and is given in Eq.~(\ref{def_qa}) of Appendix \ref{app_qA}.
This calculation is schematically represented in Fig.~\ref{fig:sketchveff}.

We can rewrite this result using bipolar coordinates:
\begin{align}
& e^{-\b \f_{eff}(r)}  = e^{-\b \f(r)} \frac{1}{r \sqrt{4 \p A}} \times \nonumber \\
& \times \int_0^\io du ~ u \left[ e^{-\frac{(r-u)^2}{4A}} \! - e^{-\frac{(r+u)^2}{4A}} \right] q(A,T;u)^{m-1} .
\label{def_effpot}
\end{align}

The integration of replicas $2 \cdots m$ induces also three-body (and more generally many-body) interactions on replica 1; if we neglect these, and just
keep the two-body effective potential, we obtain that the free entropy of the replicated liquid is the sum of an harmonic part $\SS_h(m,A)$ 
plus the free energy of a liquid interacting via the effective potential: $\SS = \SS_h + \SS_{liq}[\f_{eff}]$~\cite{PZ10}. 
Furthermore, we define
\begin{align}
& Q(r) = e^{-\b (\f_{eff}(r)-m\f(r))} -1  \nonumber \\
& e^{-\b \f_{eff}(r)} =e^{-\b m \f(r)} (1 + Q(r) ) \ ;
\end{align}
we note that $Q(r)$ is equal to zero when $A=0$, therefore for small $A$ we suppose that $Q(r)$ is a small perturbation, 
and we make use of equilibrium liquid perturbation theory. We finally obtain for the replicated free entropy $\SS(m,A;T,\ph) = \frac{1}{N} \log Z_m$:
\begin{align}
& \SS(m,A;T,\ph) = S_{h}(m,A) + \SS_{liq}(T/m,\ph) \nonumber \\
& \phantom{\SS(m,A;T,\ph)} + \frac{3 \ph}{\p} \int d^3r g_{liq}(T/m,\ph;r) Q(r) \label{S_of_Q} \ , \\
& \SS_h(m,A) = \frac{3}{2} (m-1) \ln (2\p A) + \frac{3}{2} (m-1 + \ln m) \ , \label{Sharm}
\end{align}
where $\SS_{liq}(T,\ph)=N^{-1} \log Z_{liq}$ is the free entropy of the liquid, $g_{liq}(T,\ph; r)$ its correlation function (to be computed at temperature $T/m$).

Equation~(\ref{S_of_Q}) allows for a full calculation of the replicated free entropy, and gives back both the MP small cage expansion in powers of $A$ at high density, and the PZ expansion in powers of $\sqrt{A}$ at zero temperature. Details about this can be found in the Appendix~\ref{app:SC} and \ref{app:HS}. The main theoretical goal of the paper is now achieved~\cite{JBZ11}: 
we obtained a set of equations that allow us to make a connection between zero temperature hard spheres and finite temperature harmonic spheres.
Note that the molecular liquid is assumed to be in a liquid state, therefore crystallization is excluded a priori from the theory~\cite{PZ10}.

\subsection{Low temperature approximation}

Our purpose in interpolating between zero and finite temperature was to be able to probe the vicinity of the jamming point in the $(\ph, T)$ phase diagram, thus we are interested mainly in the low temperature behaviour of the glass. We can exploit this by making an approximation that will allow us to push the analytical calculation much further. Defining the cavity distribution function $y$ by $g(T,\ph;r)=e^{-\b \f(r)} y(T,\ph;r)$, we make the approximation of taking the cavity function of the liquid as a constant and equal to its $T=0,r=1$ value, which we call $y^{HS}_{liq}(\ph)$. As $T$ goes to zero, we see that the exponential factor converges towards a step function around $r=1$, so that in all integrals that are cut at $r=1$ by the pair potential, we can safely evaluate $y$ at its $r=1$ value. Expanding $y$ in powers of $T$, it is easy to convince oneself that the temperature dependence leads to subdominant contributions in the integrals.
Thus we suppose:
\begin{align}
g_{liq}(T/m,\ph;r) \sim e^{-\b m \f(r)} y^{HS}_{liq}(\ph).  \label{y_approx_rep}
\end{align}
Plugging this approximation into Eq.~(\ref{S_of_Q}), we get (the details of the computation are in Appendix~\ref{app:A3}):
\begin{align}
&\SS(m,A;T,\ph) = S_{h}(m,A) + \SS_{liq}(T/m,\ph) \nonumber \\
& \phantom{\SS(m,A;T,\ph) =} + 4 \ph y^{HS}_{liq}(\ph) G(m,A;T), \label{defSS} \\
& G(m,A;T) = 3 \int_0^\io dr r^2 [q(A,T;r)^m - e^{-\b m \f(r)}],
\label{Sfirstorder}
\end{align}
where $\SS_{liq}(T/m,\ph)$ is the free entropy of the non-replicated liquid, evaluated at an effective temperature $T/m$.
These are the finite-temperature version of the replicated free entropy of hard-spheres obtained in Ref.~\cite{PZ10}. 

Many technical aspects presented in the following already appear in \cite{PZ10}, apart from the existence of a new control parameter, the temperature. The average cage radius $A^*(m;T,\ph)$ is obtained by maximization of $\SS$ against $A$. We find an implicit equation \mbox{for $A^*(m;T,\ph)$:}
\begin{align}
& J(m,A^*(m;T,\ph);T) = \frac{9}{4 \pi \rho y^{HS}_{liq}(\ph)} , \label{def_Amax} \\
& J(m,A;T) \equiv \frac{A}{1-m} \frac{\partial G(A,m;T)}{\partial A} \label{def_J} \\
& \phantom{J(m,A,T)} = 3 \frac{mA}{1-m} \int_0^\infty dr r^2 q(A,T;r)^{m-1} \frac{\partial q(A,T;r)}{\partial A} . \nonumber
\end{align}

\subsection{Liquid theory}

The last quantities that we need to compute are the free entropy $\SS_{liq}$ and 
the pair correlation function $g_{liq}(r)$ 
of the liquid. 
Given the pair correlation function, which represents the probability of finding a particle at distance $r$ from a particle fixed at the origin, we can express the internal energy $U$ as:
\begin{align}
U(T,\ph) = 12 \ph \int_0^{\io}dr \, r^2 g(T,\ph;r) \f(r) \ .
\label{def_U}
\end{align}
Plugging the low temperature approximation Eq.~(\ref{y_approx_rep}) for the liquid into this, we obtain:
\begin{align}
U_{liq}(T,\ph) & \underset{T \to 0}{\sim} 12 \ph y^{HS}_{liq}(\ph) \int_0^1 dr \,  r^2 (1-r)^2 e^{ - \b (1-r)^2} \ .
\label{y_approx}
\end{align}
We are interested in computing the free entropy $\SS_{liq}$ of the liquid, which is related to the liquid free-energy $F_{liq}$ by $\SS_{liq}(T,\ph)=-\b F_{liq}(T,\ph)$. The free entropy has a finite limit for hard spheres $T \to 0$, thus allowing to make the connection between hard spheres and soft spheres at finite temperature. Making use of the standard identity $U(T) = \frac{\partial (F/T)}{\partial (1/T)}$, we can derive the low temperature approximation for the free entropy of the liquid:
\begin{align}
& \SS_{liq}(T,\ph) = S_{liq}^{HS}(\ph) + 6 \ph y^{HS}_{liq}(\ph) \nonumber \\
& \times \left[ \frac{\sqrt{\pi}}{2} \sqrt{T} \left( 2 + T \right) \text{erf} \left( \frac{1}{\sqrt{T}} \right) \right.  \left.+ T \left(e^{-1/T} - 2 \right) \right] ,
\label{dev_energy}
\end{align}
where $y^{HS}_{liq}(\ph)$ and $\SS_{liq}^{HS}(\ph)$ are short-hand notations for $y_{liq}(T=0,\ph;r=1)$ and $\SS_{liq}(T=0,\ph)$.

At this level of approximation, it is clear that the only input that is needed from liquid theory is the equation of state of the hard sphere liquid.
From any given equation of state one can easily deduce
the hard sphere free entropy $\SS_{liq}^{HS}$ and cavity function $y^{HS}_{liq}$.
The most reasonable choice would be to use the phenomenological Carnahan-Starling (CS) equation of state, as in~\cite{PZ10}, that provides
the best fit to the hard sphere pressure. 

In this paper we use instead the Hyper-Netted Chain (HNC) approximation: the HNC equations are non-linear integro-differential equations 
that require a numerical solution. 
Although HNC is known to be less accurate for the hard sphere system, using HNC allows us to also compare the present directly our results to the ones obtained from the MP approach valid at large density and finite temperatures. 
The procedure for solving these equations is described in full details in~\cite{BFJS10}.
The HNC approximation overestimates
$y^{HS}_{liq}$ by $20\%$ in the relevant range of volume fraction $\ph \sim 0.64$. This has the effect of reducing the glass transition density
obtained from the theory, from the value $\ph_K = 0.62$ obtained from CS~\cite{PZ10} to $\ph_K = 0.58$ obtained with HNC. 
Therefore the reader should keep in mind that the glass densities reported in the following are lower than the correct ones.
In any case, here we are more interested in the low-temperature scaling 
in the glass phase than to the actual value of the glass
transition density. We also checked that the scaling results 
are insensitive to the precise choice of the equation
of state of the hard sphere liquid.

\section{Thermodynamics of the glass} 
\label{glass}

\subsection{Complexity and phase diagram}

As discussed above, the glass transition is signalled by the point where the saddle point in
Eq.~(\ref{Zfonecopy}) reaches the minimum $f_{min}$ at which the complexity $\Si(f)$ vanishes.
In the replica formalism, the equilibrium complexity $\Si(f^*)$ in Eq.~(\ref{Sivanishes}) corresponds
to the complexity in Eq.~(\ref{def_sigma_replicated_parametric}) evaluated at $m=1$. We call this
quantity the ``equilibrium'' complexity of the liquid $\Si_{eq}(T,\ph)$. The latter is easily computed
by expanding the equations around $m=1$:
\begin{align}
& \Si_{eq}(T,\ph) = S_{liq}(T,\ph) -\frac{3}{2} \ln(2 \pi A^*(m=1;T,\ph) - 3  \nonumber \\
& \phantom{\Si(T) =} - 12 \ph y^{HS}_{liq}(\ph) H(m=1,A^*(m=1;T,\ph);T) , \\
& H(m,A;T) \equiv \frac{1}{m} \frac{\partial G(m,A;T)}{\partial m} \nonumber \\
& \phantom{H(m,A,T)}= \frac{1}{m} \int_0^\infty dr r^2 q(A,T;r)^m \ln q(A,T;r) \nonumber \\
& \phantom{H(m,A,T) =} - \frac{\beta}{m} \int_0^1 r^2 \f(r) e^{- m \beta \f(r)} .
\label{phi_expanded}
\end{align}
We see that the only unknown in the previous equation is the optimal cage radius $A^*$. It is computed with Eq.~(\ref{def_Amax}) evaluated at $m=1$. 

\begin{figure}
\includegraphics[width=.45\textwidth]{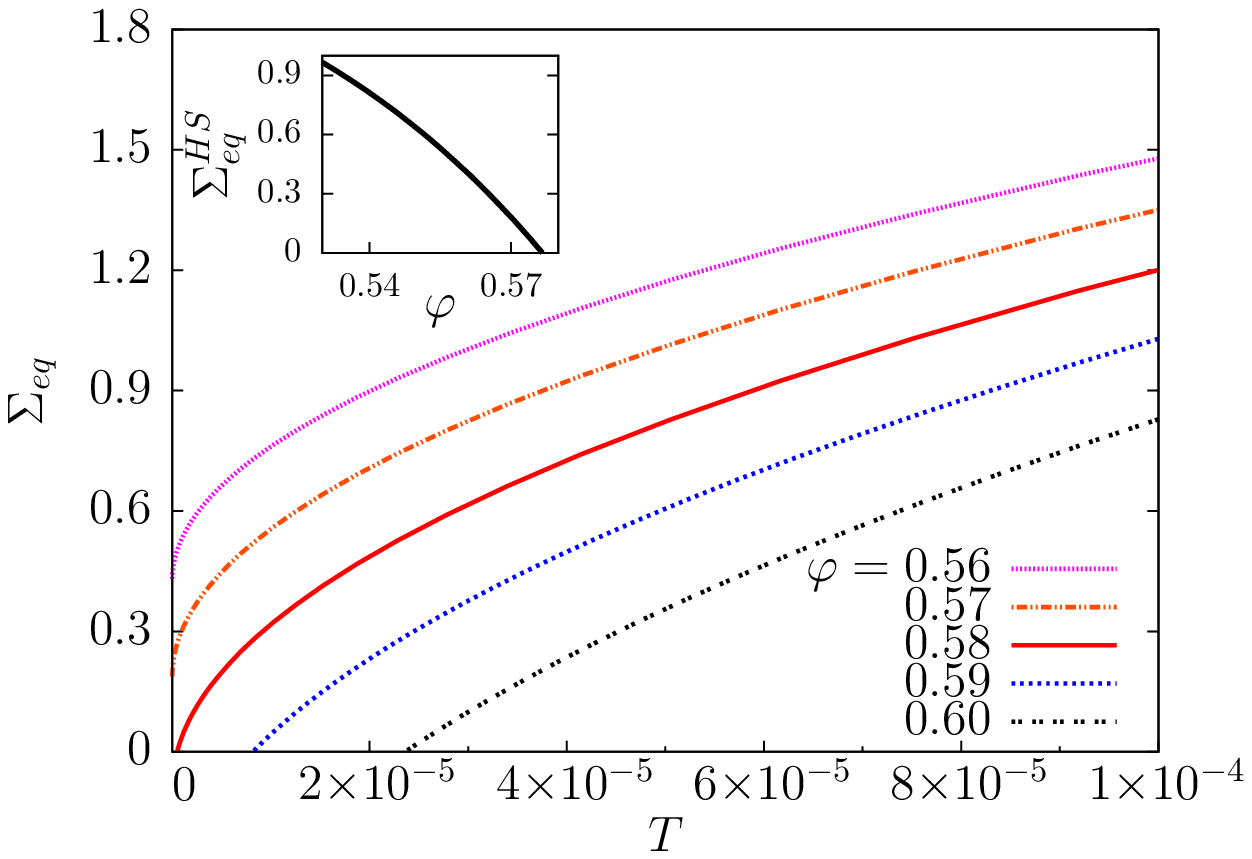}
\caption{(Color online) 
Equilibrium complexity against temperature for several volume fractions around $\ph_K$, calculated within the effective potential approximation. Inset: complexity of hard spheres, obtained as the $T \to 0$ limit of the complexity $\Si_{eq}$, plotted against the volume fraction
}
\label{fig:sigma}
\end{figure}

Since our computation makes connection with the hard spheres calculation of PZ, we expect to retrieve all their results in 
the $T=0$ limit for $\ph < \ph_{GCP}$. Thus for $\ph$ smaller than $\ph_K$ (the Kauzmann transition of hard spheres), the system should be in a liquid phase, and thus its complexity should not vanish, and converge to the hard sphere complexity $\Si^{HS}_{eq}(\ph)$ found by PZ. In practice, since PZ performed a $\sqrt{A}$ expansion, whereas we did not, our numerical values for $\ph_K$ and $\Si^{HS}_{eq}(\ph)$ could differ slightly. 
Nevertheless, $A$ is typically found between $10^{-3}$ and $10^{-6}$, so that the difference is indeed very small. For example $\ph_K$ calculated by PZ differs from our value by less than $6 \cdot 10^{-3}$.

For $\ph \ge \ph_K$, the complexity vanishes at a temperature $T_K(\ph)$, called the Kauzmann temperature, that increases with the density. These results are summarized in Fig.~\ref{fig:sigma}, where we show the complexity as a function of temperature for several volume fractions around $\ph_K$. In the inset we show the zero temperature limit of the complexity $\Si_{eq}^{HS}$ as a function of the density, that vanishes at $\ph_K$.

\begin{figure}
\includegraphics[width=.45\textwidth]{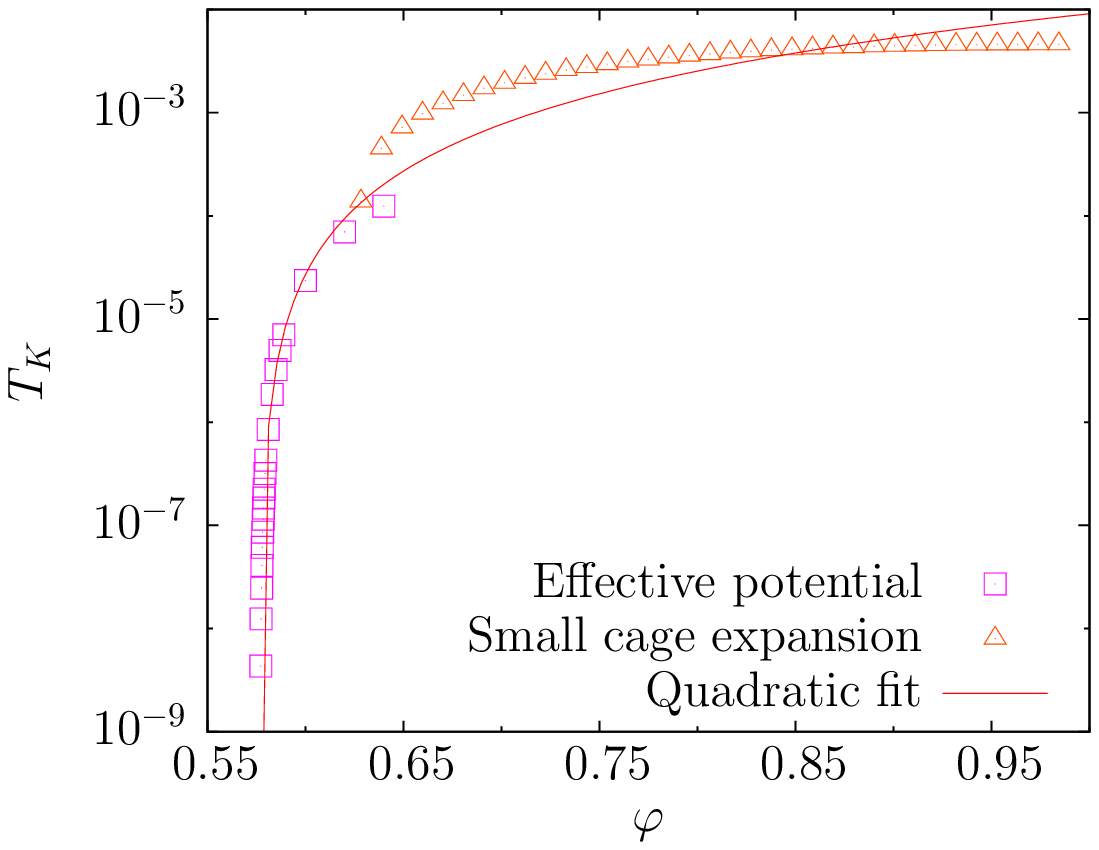}
\caption{(Color online)
Kauzmann temperature against volume fraction, within the
the effective 
potential approach developed in this work (open squares). 
The quadratic fit (red line) gives an estimated 
Kauzmann transition for hard spheres at $\ph_K = 0.576898$.
The small cage M\'ezard-Parisi approach breaks down 
at low temperatures and low densities, as explained in Ref.~\cite{BJZ10}.}  
\label{fig:tk}
\end{figure}

From the complexity we deduce the phase diagram 
shown in Fig.~\ref{fig:tk}. 
In this figure we report $T_K(\ph)$, as obtained in the framework of the effective potential approximation developed 
in this paper, and we compare it with the result obtained
from the M\'ezard-Parisi small cage expansion~\cite{BJZ10}.
In the effective potential case we find that the Kauzmann temperature goes to zero at $\ph_K  \approx 0.5769$, quadratically in $\ph - \ph_K$. 
Conversely, the result from the small cage expansion is a finite $T_K(\ph)$ which jumps to zero abruptly at a value of $\ph$ which
is unrelated to hard sphere results. This is due to fact that 
the small cage expansion is valid only in the region indicated 
in Fig.~\ref{fig:PDsch}.
On the other hand, our effective potential computation becomes
inaccurate when the temperature is too high because of the 
approximation in Eq.~(\ref{y_approx_rep}). 
Still we obtain a reasonable matching of the two approximation schemes for intermediate densities around $\ph = 0.64$.
It would be easy, in principle, to reconcile the two approximations at all volume fractions, including the crossover regime, 
by avoiding the low temperature approximation made in Eq.~(\ref{y_approx_rep}), but these computations would require a much heavier numerical treatment.

\subsection{Relaxation time and glass fragility}

The calculation of the liquid complexity, apart from signalling the glass transition, can be related to the relaxation time of the liquid by making use of the Adam-Gibbs scenario \cite{Gi56,GD58,AG65}, which relates the relaxation time $\t$ of a liquid to its configurational entropy. In the case of our mean-field vision of the phase space of the liquid, the configurational entropy can be identified with the equilibrium 
complexity $\Si_{eq}$. In that case the Adam-Gibbs relation yields:
\begin{align}
\t(T,\ph) = A \exp \left( {\frac{c}{T \Si_{eq}(T,\ph)}} \right),
\label{adam-gibbs}
\end{align}
where $c$ is a constant.
We can push the correspondence between complexity and relaxation time a bit further, and use it to extract informations on the fragility of the glass. The fragility of a glass quantifies how different the relaxation processes are from usual Arrhenius relaxation processes, and can be observed 
by representing the logarithm of the relaxation time 
against the inverse temperature. A linear curve is then the signature of an Arrhenius relaxation, while steeper curves indicate a greater fragility, i.e. a faster increase of the relaxation time.

\begin{figure}
\includegraphics[width=.45\textwidth]{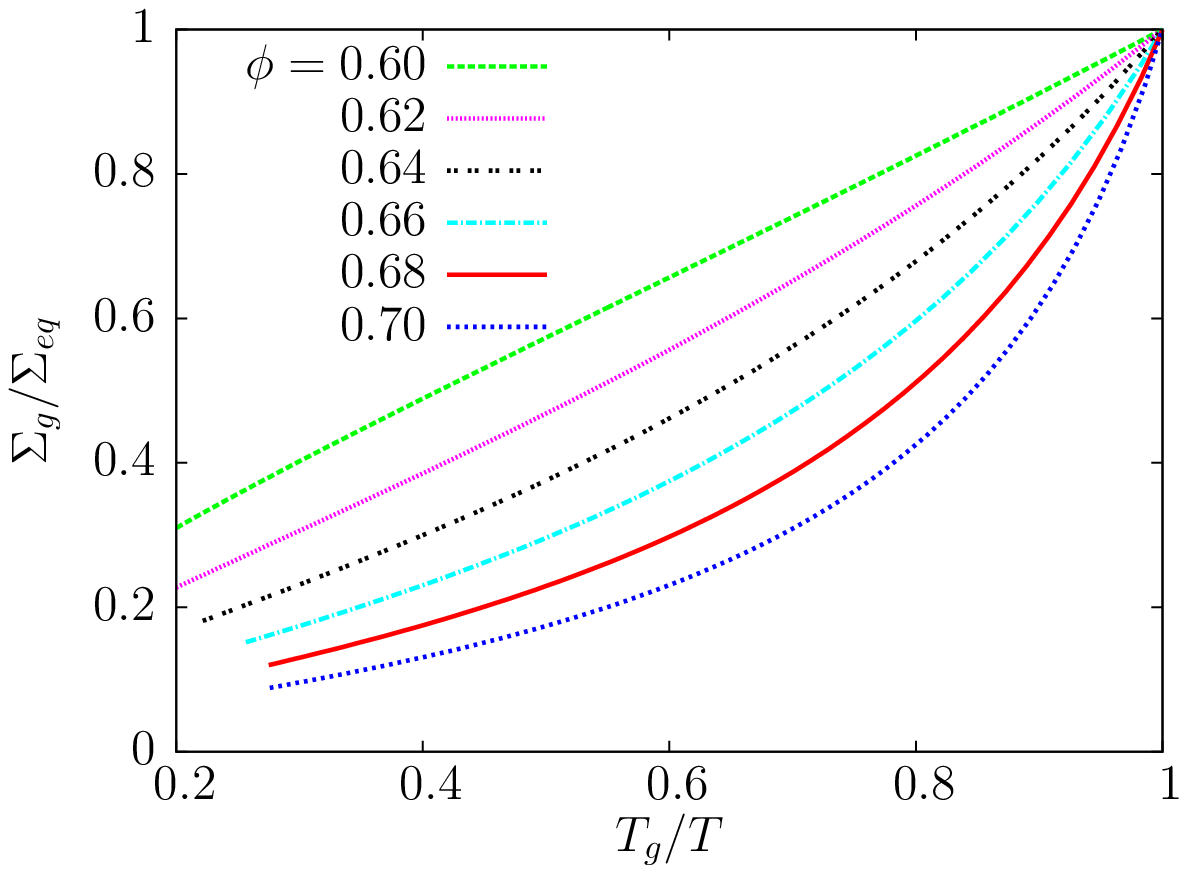}
\caption{(Color online)
A thermodynamic Angell plot of $\Si_g/\Si_{eq}$, with $\Si_g=1$ as a definition of the glass transition, plotted against $T_g/T$, for different volume fractions. The thermodynamic fragility is predicted to increase 
rapidly with volume fraction, in agreement with numerical 
simulations~\cite{BW09b}.}
\label{fig:angell}
\end{figure}

In Refs.~\cite{BW09b,BW09a}, Berthier and Witten showed that, for harmonic spheres, compressing the system leads to a significant
increase of the fragility. Using our results for the complexity, we are able to check qualitatively whether replica theory can reproduce this feature. 
Indeed, thanks to Eq.~(\ref{adam-gibbs}), the fragility can be extracted equivalently from the relaxation time or from the complexity.
Since the fragility is usually evaluated at the conventionally defined laboratory glass transition,
we arbitrarily define the glass transition temperature $T_g(\ph)$ as the temperature at which the equilibrium complexity is equal to one,
$\Si(T_g(\ph),\ph) = 1$,
which is a typical value of the 
configurational entropy at the glass transition in
most numerical simulations and experiments (its precise value is immaterial
for our purposes).
Using these values of $T_g$ and $\Si_g=1$, we can construct an Angell plot for the complexity following Ref.~\cite{MA01}. 
In Fig.~\ref{fig:angell} we show the inverse of the complexity, 
linked to the logarithm of the relaxation time by Eq.~(\ref{adam-gibbs}), plotted against $T_g/T$, for several densities. 
The fragility is the slope of the curves in $T_g/T=1$~\cite{MA01}.
One can clearly see that increasing the density drastically increases the fragility of the glass-former.

\subsection{Free-energy of the glass}

We turn now to the calculation of the free entropy of the glass, 
$\SS_{glass}$. We have seen in Sec.~\ref{method} that in order to compute $\SS_{glass}(T,\ph)$, one needs to optimize the replicated free entropy $\SS(m,A;T,\ph)/m$ defined in Eq.~(\ref{defSS}) with respect to $A$ first, via Eq.~(\ref{def_Amax}), then with respect to $m$, via Eq.~(\ref{def_SS_glass}). Calling $A^*(T,\ph)$ and $m^*(T,\ph)$ the optimal values of $A$ and $m$, we obtain:
\begin{align}
\SS_{glass}(T,\ph) = \frac{\SS(m^*(T,\ph),A^*(T,\ph);T,\ph)}{m^*(T,\ph)}.
\label{eqSSglass_final}
\end{align}

\begin{figure}
\includegraphics[width=8.5cm]{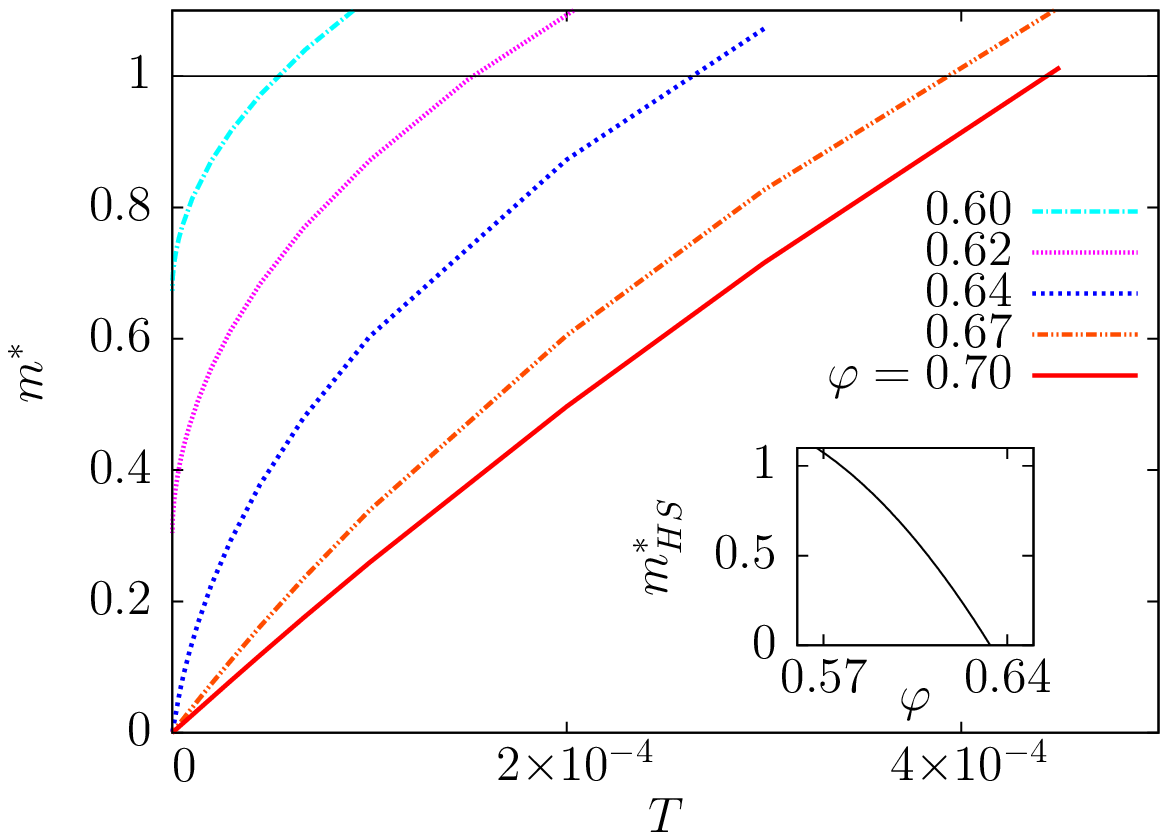}
\caption{(Color online)
Optimal number of replicas $m^*$ as a function of $T$, for several volume fractions. The inset shows the $T \to 0$ limit of $m^*$ as a function of the volume fraction.
}
\label{fig:mstar}
\end{figure}

Starting from Eq.~(\ref{eqSSglass_final}), we can deduce all quantities of interest, for instance the pressure 
and energy of the glass and its specific heat. The derivations are reported in Appendix~\ref{app:Uglass}. 
In particular, our approximations imply that the energy 
of the glass $U_{glass}$ can be computed directly from the knowledge of the effective potential $\f_{eff}$ 
via the following relation:
\begin{align}
U_{glass}(T,\ph) = 12 \ph y^{HS}_{liq}(\ph) \int_0^1 dr ~ r^2 (1-r)^2 e^{-\b \f_{eff}(r)}.
\label{def_U_glass}
\end{align}
Similarly, we find that the reduced pressure (or ``compressibility factor'') $p = \b P/\r$ is given
by
\beq
\label{def_p_glass}
\begin{split}
p_{glass}&(T,\ph) 
 = \frac{1}{m^*} \, p_{liq}(T/m^*,\ph) \\ 
&- \frac{4 \ph}{m^*} \left[ y^{HS}_{liq}(\ph) + \ph \frac{d  y^{HS}_{liq}(\ph)}{d\ph} \right]
G(m^*,A^*;T).
\end{split}\eeq

In Fig.~\ref{fig:mstar}, we show the behavior of the optimal number of replicas $m^*$ as a function 
of the temperature, for different volume fractions.
We see that the $T \to 0$ limit of $m^*(T,\ph)$ converges 
when $\ph$ is not too large to a finite value, which we call $m^*_{HS}(\ph)$, and which is shown in the inset. The replica parameter of hard spheres vanishes at a density $\ph_{GCP}$, the glass close packing. This point is the equivalent, in our mean-field picture, of the jamming point of harmonic spheres. The behavior of the cage radius $A^*$ is similar to that of $m^*$. We will study in greater detail the behavior of the system around $\ph_{GCP}$ in the next section.

\begin{figure}
\includegraphics[width=8.5cm]{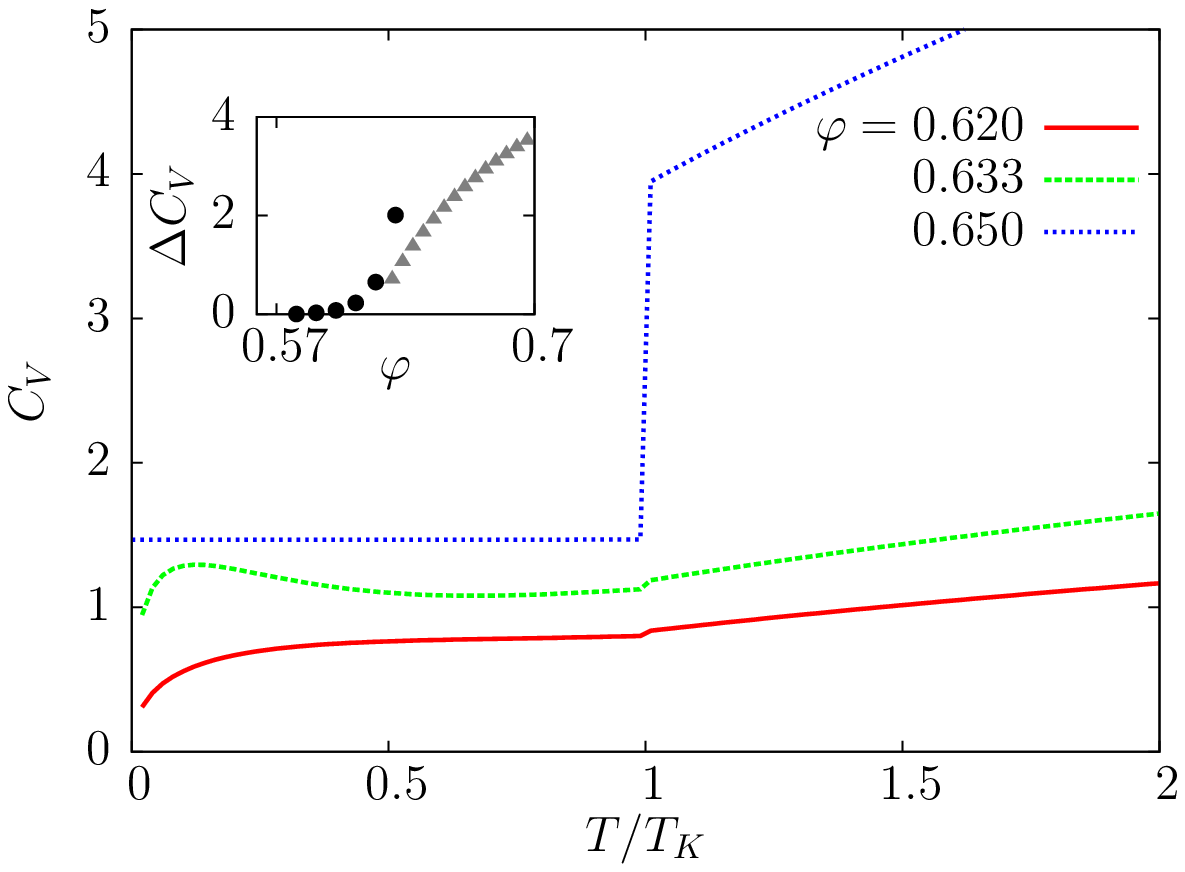}
\caption{(Color online)
Specific heat $C_V$ of the system plotted against $T/T_K$, for several volume fractions.
The curves for $\ph = 0.620, 0.633$ are obtained with the effective potential method developed in this
paper. The curve for $\ph = 0.650$ is obtained with the M\'ezard-Parisi small cage expansion~\cite{MP99b},
since at this density $T_K(\ph)$ is too high and the effective potential method is not reliable. {Inset:}~Specific heat jump at $T_K$; the black dots
are obtained with the effective potential approximation, the gray triangles are obtained with the
M\'ezard-Parisi small cage expansion.
}
\label{fig:cv}
\end{figure}

From the knowledge of $m^*$ and $A^*$, we can deduce the energy and specific heat of the glass. We show in Fig.~\ref{fig:cv} the temperature evolution of the 
specific heat for three densities, one below $\ph_{GCP}$, one very close to it, and one above. 
The temperature dependence of the specific heat is qualitatively in good
agreement with the numerical data shown in Ref.~\cite{BW09b}, the only difference
being that the abrupt jump predicted at $T_K$ using the theory is replaced
by a smoother evolution and a small maximum at the numerical glass
temperature.

Upon crossing the ideal glass transition, the specific 
heat undergoes a finite jump. We find that the amplitude of this jump 
increases continuously with $\ph$ from $\ph_K$. 
A qualitatively similar result was obtained in numerical 
simulations~\cite{BW09b}, 
where the jump of specific heat was studied at the 
numerical glass transition temperature. The behavior of the specific heat 
correlates well with the evolution of the thermodynamic fragility discussed 
in Fig.~\ref{fig:angell}.
Finally, we note that the $T \to 0$ limit of the specific heat jumps discontinuously from $0$ to $3/2$ at $\ph_{GCP}$, which reveals that the ground state 
properties of the glass phase change abruptly at the jamming transition, as
we now study in more detail.

\section{Jamming point of harmonic spheres}
\label{jamming}

We now turn to the study of the region of the phase diagram deep in the glass
phase, inside the line $T_K(\ph)$ and close to the jamming point $\ph_{GCP}$, see Fig.~\ref{fig:PDsch}.
In this region, $m^*(T,\ph)$ is very small, as we discussed in the last section (see Fig.~\ref{fig:mstar}). In principle, one could just compute $m^*(T,\ph)$ and then take the limit $T\to 0$ and $\ph \to \ph_{GCP}$ (``jamming limit''). However, both numerically and analytically, it is much more convenient to exchange the optimization with respect to $m$ and $A$ with the jamming limit, take the latter first, and then optimize the free energy. This is because, in the jamming limit, many of the integrals that appear in the free energy simplify considerably. However, the scaling of $m$ and $A$ in the jamming limit is different depending on the order of the limits, which emphasizes the asymmetry that exists between both sides of the jamming point. In this section we discuss how to exchange the jamming limit with the optimization procedure.

\subsection{Zero temperature, below close packing}
\label{sec:IVA}

We consider first the limit $T\to 0$ for fixed $\ph < \ph_{GCP}$. In this case, we are bound to recover the results of \cite{PZ10} for hard spheres. Thus, we expect the optimal number of replicas $m^*(T,\ph)$  and cage radius $A^*(T,\ph)$ to tend to finite values $m^*_{HS}(\ph)$ and $A^*_{HS}(\ph)$ when $T \to 0$. Therefore, to recover hard spheres results, one has to take the limit $T \to 0$ at fixed $A$ and $m$, and optimize the resulting free entropy over $A$ and $m$. The corresponding expression for the replicated free entropy is given in Appendix~\ref{app:HS} and coincides with that of \cite{PZ10}. We refer to this paper for further details on the hard sphere case. 

We solved numerically the optimization equations; the result for $m^*_{HS}(\ph)$ is plotted in the inset of Fig.~\ref{fig:mstar}. Approaching $\ph_{GCP}$ by the left, $m^*_{HS}$ tends to zero linearly:
\beq\label{eq:mstHS}
m^*_{HS} = \widetilde \mu \, (\ph_{GCP} - \ph),
\eeq
and $A^*_{HS}$ also goes to zero with $m$ as $\a m$. Thus the close packing limit for hard spheres can be computed by taking first the limit $T \to 0$, and then $m \to 0$ with $A = \a m$. Optimization on $A$ will now become an optimization on $\a$. From the inset of Fig.~\ref{fig:mstar} we find 
\begin{equation}
\ph_{GCP} = 0.633353, 
\end{equation}
and 
$\wt \m = 20.7487$, see Appendix~\ref{app_numbers} for more details.
Therefore, we find that the location of the jamming transition 
is different from the density of the hard sphere 
glass transition, which confirms the distinct nature 
of both phenomena.

In this limit ($T \to 0$ at finite $m$ and $A$), it is easy to see that $\phi_{eff}$ has an hard-core of the same size of the original potential.
Therefore, from Eq.~(\ref{def_U_glass}), it is straightforward to see that the energy of the glass vanishes: the system becomes indeed a system
of hard spheres. From Eq.~(\ref{def_p_glass}), since both $p_{liq}(T=0,\ph)$ and $G(m,A;T=0)$ are finite, we find that the reduced pressure goes to a finite value $p_{glass}(T=0,\ph)$ which is the reduced pressure of the hard sphere glass. The latter diverges at $\ph_{GCP}$ since
$m^* \to 0$ at that point. One can show that for $\ph \to \ph_{GCP}^-$, since $m^*$ vanishes linearly, one obtains~\cite{PZ10}
\beq
p_{glass}(T=0,\ph) \sim \frac{3.03430 \, \ph_{GCP}}{\ph_{GCP} - \ph}
\eeq
where the prefactor is only $1\%$ different from the correct value which is the space dimension $d=3$, predicted by free volume theory~\cite{SW62}
and by the small cage expansion of Ref.~\cite{PZ10}.

\subsection{Zero temperature, above close packing}
\label{sec:IVB}

\begin{figure}
\includegraphics[width=8.5cm]{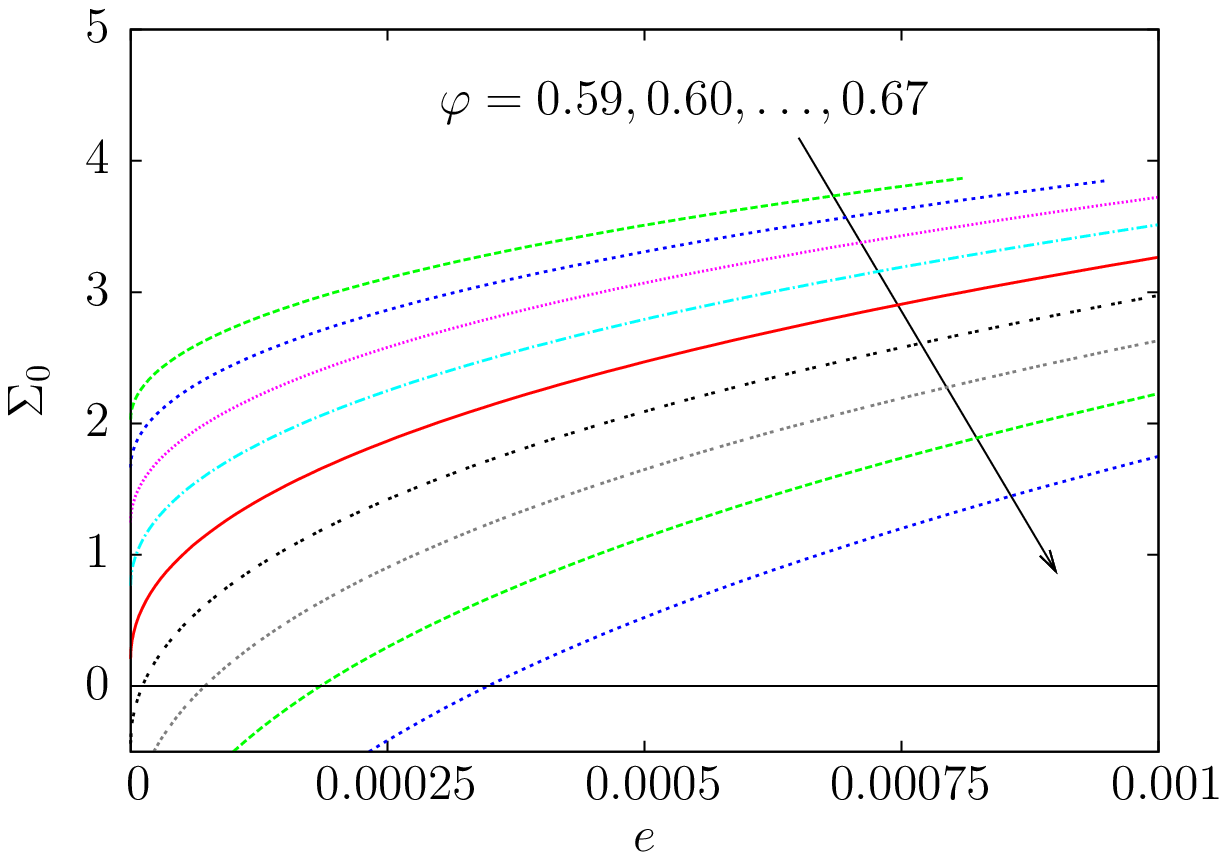}
\caption{(Color online)
Complexity of the energy minima $\Si_0(e,\ph)$ as a function of energy for several densities.
}
\label{fig:sigmae}
\end{figure}

For $\ph > \ph_{GCP}$, the harmonic spheres have a finite energy at $T = 0$, due to overlaps. Then, because of the repulsion the spheres are jammed, and $A^*$ vanishes at $T=0$. It turns out from the numerical calculation of $m^*$ and $A^*$, both in the small cage expansion \cite{MP99b} and in the potential approximation (see the previous section) that the optimal number of replicas $m^*$ tends to zero when $T \to 0$. One finds that $m = T/\t$ and $A = m \a$ with constants $\a$ and $\t$. Therefore for $\ph > \ph_{GCP}$ one has to take the $T\to 0$ limit with $\a = A/m$ and $\t=T/m$ kept constant. Optimization over $m$ and $A$ is replaced by an optimization over $\a$ and $\t$.

The replicated free entropy (\ref{defSS}) simplifies considerably in this limit. The details of the calculations are discussed in Appendix~\ref{app:SS}. The expression for $\SS_0(\a,\t;\ph)$ in this limit is given in Eq.~(\ref{eq:def_Psi_alpha_mu}). We first calculate the optimal cage radius $\a$ with Eq.~(\ref{optim_A_T=0}), and replace the solution $\a^*(\ph)$ in Eq.~(\ref{eq:def_Psi_alpha_mu}). We obtain the replicated free entropy as a function of $\t$, $\SS_0(\t;\ph) = \SS_0(\a^*,\t;\ph)$. To understand the physical meaning of this quantity, we have to go back to Eq.~(\ref{def_ssglass_replicated}) and take the limit $T\to 0$ with $m = T/\t$. Since the free energy $f^*$ reduces to the energy $e^*$ for $T=0$, we get
\beq
\SS_0(\t;\ph) = \lim_{T\to 0} \SS(T/\t ; T,\ph) = \Si_0(e^*) - e^*/\t \ .
\eeq
Following the reasoning that led to Eq.~(\ref{def_sigma_replicated_parametric}), we get
\beq\label{eq:zeroTLeg}
\begin{split}
& \Sigma_0(\t,\ph) = \frac{\partial \left(\t \SS_0 \right)}{\partial \t},  \\
& e(\t,\ph) = \t^2 \frac{\partial \SS_0}{\partial \t},
\end{split}\eeq
and therefore 
\begin{align}
e_{GS}(\ph) & = U_{glass}(T=0,\ph) =- \min_{\t} [ \t \SS_0(\a^*,\t;\ph) ] \nonumber \\
& =-\t^*(\ph) \SS_0(\a^*,\t^*(\ph);\ph) ,
\end{align}
since the ideal glass energy coincides with the amorphous ground state energy.
We obtain the scaling of $\t^*$ and of the energy around $\ph_{GCP}$ by means 
of a development around $\ph_{GCP}$. The calculation 
is presented in Appendix~\ref{app:SS},
and the result is:
\begin{align}
\begin{split}
& \t^*(\ph) \sim \wt\t (\ph -\ph_{GCP}) , \\
& e_{GS}(\ph) \sim \wt e (\ph-\ph_{GCP})^2 ,
\end{split}
\label{eq:tauGCP}
\end{align}
for $\ph \to \ph_{GCP}^+$, so that the jamming limit corresponds to $\t \to 0$, as expected. 
We get $\ph_{GCP} = 0.633353$, $\wt\t = 0.00835535$ and $\wt e =0.263017$ 
(see Appendix~\ref{app_numbers}). In addition to the scalings in Eq.~(\ref{eq:tauGCP}), 
we obtain the following scaling for the complexity against the ground-state energy for small $e$:
\begin{equation}
\Sigma_0(e,\ph) = \Sigma_0^{HS}(\ph) + A(\ph) \sqrt{e } + \cdots
\end{equation}
The function $\Sigma_0(e,\ph)$ is reported in Fig.~\ref{fig:sigmae}.
The energy of the ground state $e_{GS}(\ph)$ corresponds to the vanishing of the complexity. We know that $\Sigma^{HS}_0(\ph) \sim \ph_{GCP} - \ph$. Since the coefficient $A$ is constant around $ \ph_{GCP}$, one obtains the quadratic scaling of the energy.

Finally, from Eq.~(\ref{def_p_glass}) we see that the pressure $P_{glass}(T=0,\ph) = \lim_{T\to 0} (\r T p_{glass})$ 
is proportional to $\t^*$ and therefore
\beq
P(\ph) \sim \wt P (\ph -\ph_{GCP})  
\label{scalp}
\eeq
with $\wt P = 0.403001$ (see Appendix~\ref{app_numbers}). 
This result is indeed consistent with the exact 
relation $P_{glass}(T=0,\ph) = \frac{6 \ph^2}\pi \frac{de_{GS}}{d\ph}$.

In this section we have thus derived the scaling 
laws (\ref{eq:tauGCP}) and (\ref{scalp}) first observed numerically
above the jamming transition at zero temperature~\cite{OLLN02,OSLN03}, which 
indicate that solidity emerges continuously at the jamming density. 
These results also confirm the correspondance between the jamming 
transition observed numerically and the glass close packing density
defined within our theoretical approach.

\subsection{Scaling around jamming}

We have shown in Secs.~\ref{sec:IVA} and \ref{sec:IVB} 
that all the thermodynamic quantities are singular at $\ph_{GCP}$
and $T=0$;
for instance $m^*$ is finite below $\ph_{GCP}$ while it vanishes proportionally to $T$ above $\ph_{GCP}$.
The reduced pressure is also finite below $\ph_{GCP}$, while it diverges at $\ph_{GCP}$ and is formally infinite
above $\ph_{GCP}$.
Both the energy and the pressure vanish below $\ph_{GCP}$ while they are finite above $\ph_{GCP}$.

\begin{figure}
\includegraphics[width=8cm]{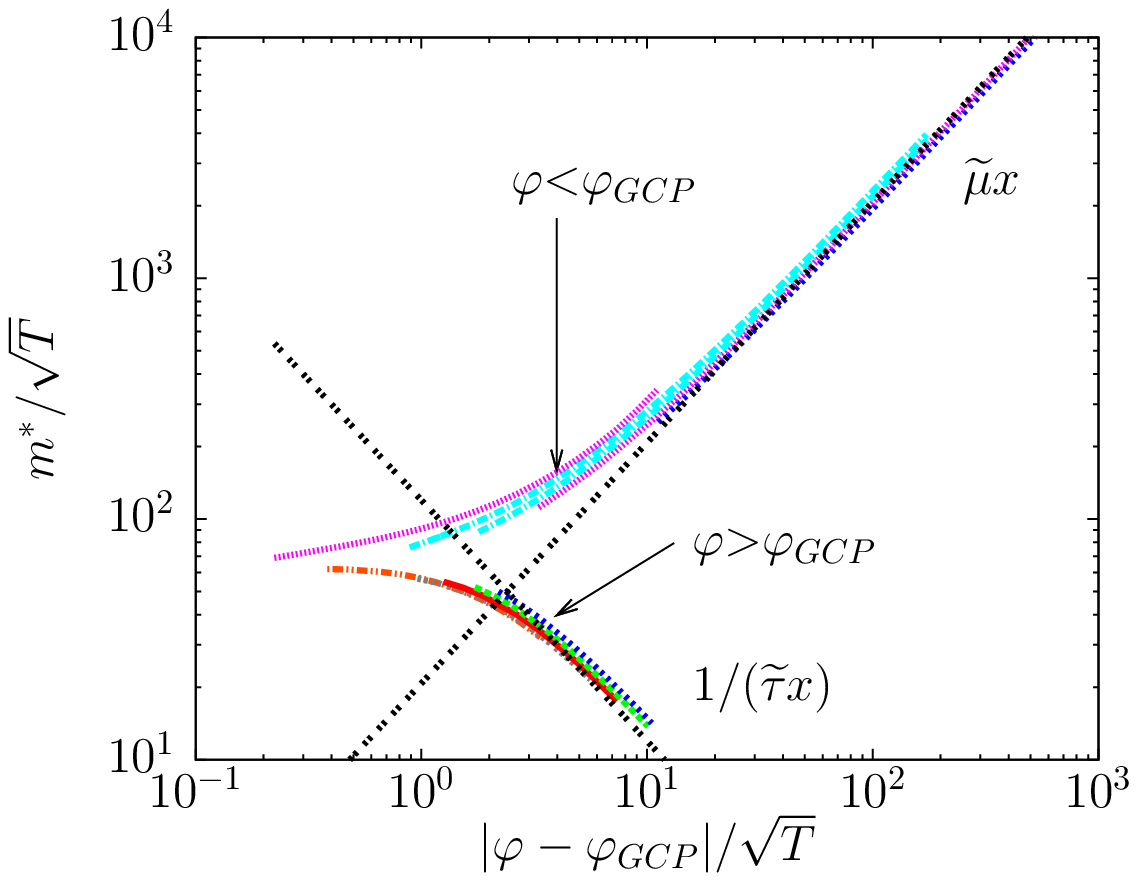}
\caption{(Color online)
Scaling of $m^*(T,\ph)/\sqrt{T}$ as a function of the rescaled temperature 
$|\ph-\ph_{GCP}|/\sqrt{T}$ for $\varphi=0.58 , 0.59 , \ldots , 0.70$. The asymptotic 
forms corresponding to Eq.~(\ref{eq:mstarscaling}) are also reported.
}
\label{fig:scaling_m}
\end{figure}

From this observation, and since all quantities are analytic at finite $T$, 
it follows that all these quantities must satisfy scaling relations if $T$ is small
enough and $\ph$ is close enough to $\ph_{GCP}$~\cite{OT07}. 
We discuss explicitly the case of $m^*$ for which we
assume a scaling relation of the form
\beq
m^*(T,\ph) = T^\g \widetilde m_\pm\left( \frac{|\ph - \ph_{GCP}|}{T^\n} \right) \ ,
\eeq
where the two scaling functions correspond to the two sides of the transition.
In the hard sphere limit $T\to 0$ and $\ph < \ph_{GCP}$, to recover Eq.~(\ref{eq:mstHS})
we need $\wt m_-(x \to \io) = \wt \mu x$ and $\g = \n$. For $\ph > \ph_{GCP}$ instead,
$m^* = T/\t^*(\ph)$ and to recover Eq.~(\ref{eq:tauGCP}) we need $\wt m_+(x \to \io) = 1/(\wt\t x)$
and $\g = \n = 1/2$.
Finally, we are led to the scaling
\beq\label{eq:mstarscaling}
\begin{split}
& m^*(T,\ph) = \sqrt{T} \, \widetilde m_\pm\left( \frac{|\ph - \ph_{GCP}|}{\sqrt{T}} \right) \ , \\
& \wt m_-(x \to \io) = \wt \mu x \ , \\
& \wt m_+(x \to \io) = \frac{1}{\wt\t x} \ .
\end{split}\eeq
This scaling is successfully tested in Fig.~\ref{fig:scaling_m}.

Qualitatively similar scaling forms (with different exponents) 
apply to thermodynamic quantities such 
as the energy, the pressure, or the complexity. 
Although we do not show the corresponding scaling plots explicitly, 
the energy and the pressure are plotted and
compared with numerical data in the following (Fig.~\ref{fig:ep}).
Note finally that a scaling plot for the complexity 
is related, in the spirit of the Adam-Gibbs relation in 
Eq.~(\ref{adam-gibbs}), to the scaling plot of the 
numerically measured relaxation time 
reported in Ref.~\cite{BW09b,BW09a}.

\section{The pair correlation function}
\label{correlation}

The pair correlation function is a key quantity that gives a lot of insight into the microstructure of dense fluids or packings.
In this section we derive an expression for the correlation function of the glass and use it to discuss the scaling
of the contact peak of $g(r)$ close to the jamming transition.

\subsection{General expression of the correlation function of the glass}

It has been shown in \cite[Eqs.(67)-(68)]{PZ10} that one can compute the correlation function of the glass starting from the replicated free energy. This is done by adding a small perturbation $v(r)$ to the potential of replica 1 and taking the derivative of the free energy with respect to $v(r)$. The glass correlation function is then calculated as:
\begin{align}
g_{glass}(r) = - \frac2\r \frac{\d \SS}{\b \d v(r)}.
\end{align}
 We start from Eq.~(\ref{S_of_Q}) and we use approximation (\ref{y_approx_rep}). Since the perturbation acts only on replica 1, we have
\begin{align}
\SS(m,A; & T,\ph) = S_{h}(m,A) + \SS_{liq}\left[\ph, e^{-\b m \phi(r) -\b v(r)} \right] \nonumber \\
& + \frac{3 \ph}\pi y^{HS}_{liq}(\ph) \int d^3r e^{-\b m \phi(r) -\b v(r)}  Q(r),
\end{align}
and $Q(r)$ depends only on the potentials of replicas 2 to $m$, so it is independent of $v$. Taking the derivative with respect to $v(r)$ we have, using again Eq.~(\ref{y_approx_rep}):
\begin{align}
g_{glass}(r)&  = g_{liq}(T/m,\ph; r) +  y^{HS}_{liq}(\ph) e^{-\b m \phi(r)} Q(r) \label{gGlass} \\
& =   y^{HS}_{liq}(\ph) e^{-\b m \phi(r)} [1 + Q(r)]  \nonumber \\
& = y^{HS}_{liq}(\ph) e^{-\b \phi_{eff}(r)} . \nonumber
\end{align}
Therefore the correlation function of the glass is directly related to 
the effective potential within the low temperature 
approximation we used to compute the thermodynamics of the system. Of course, an improved expression for this correlation could be obtained by using the HNC approximation to describe the effective liquid. In that case, $g_{glass}(r)$ would be the HNC correlation of the effective liquid. Still, the much simpler expression (\ref{gGlass}) is enough to capture the scaling of $g_{glass}(r)$ around jamming, at least close to the first peak. In the rest of this section we discuss the scaling form that is obtained starting from Eq.~(\ref{gGlass}).

A very interesting quantity related to $g(r)$ is the number of contacts, defined as the number overlaps per particle.
Within our low temperature approximation it is given by:
\begin{align}
z(T,\ph) & = 24 \ph \int_0^1 dr \, r^2 g_{glass}(r)   \nonumber \\
& = 24 \ph y^{HS}_{liq}(\ph) \int_0^1 dr \, r^2 e^{-\b \f_{eff}(r)}  \ .
\label{eq:zdef}
\end{align}
Using the definition of the effective potential, Eq.~(\ref{effpot_def}), we can rewrite this in a simpler way:
\begin{align}
z(T,\ph) = 24 \ph y^{HS}_{liq}(\ph) \int_0^{\io} du \, u^2 q(A,T;u)^{m-1} \widehat q(A,T;u) \ , 
\end{align}
with a function $\widehat q(A,T;u)$ which is very similar to $q(A,T;u)$; its form and the details of the derivation can  be found in Appendix \ref{app_z}.

The starting point of the analysis of the correlation function of the glass are Eq.~(\ref{gGlass}) and
Eq.~(\ref{def_effpot}) where the effective potential is defined. In the following we keep the notation $e^{-\b\f_{eff}}$ since this function has a finite limit for $T\to 0$ even if $\b$ is formally infinite. 

\subsection{Scaling of the peak of the pair correlation at finite temperature}

\begin{figure}
\begin{flushleft}
\includegraphics[width=8cm]{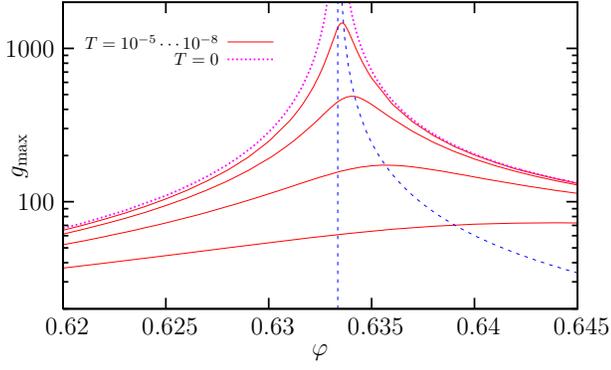}
\end{flushleft}
\caption{(Color online)
Evolution of the maximum of the glass pair correlation function with $T$ and $\ph$. While $g_{max}$ diverges on both sides of the transition at $T = 0$ as $g_{max} \sim |\ph - \ph_{GCP}|^{-1}$, this divergence becomes a smooth maximum at finite $T$ near the transition, whose position shifts with temperature. This behavior compares very well with numerical \cite{ZXCYAAHLNY09,DTS05,SLN06} and experimental observations \cite{ZXCYAAHLNY09,Ch10a}.
}
\label{fig:gmax}
\end{figure}

Here we discuss the effect of thermal fluctuations on 
the maximum of $g_{glass}(r)$ near the 
the $T=0$ jamming transition at $\ph_{GCP}$. This situation was studied 
numerically and experimentally in Refs.~\cite{ZXCYAAHLNY09,JB10}. 
We focus on the scaling of this maximum 
in a region of small $T$ and for $\ph \sim \ph_{GCP}$.
In Fig.~\ref{fig:gmax} we show the behavior of the maximum 
of $g_{glass}(r)$, which we call $g_{max}$, 
as a function of the density, for temperatures ranging from $10^{-5}$ to $0$. 
This figure demonstrates that 
we are able to derive analytically all behaviors 
reported in Refs.~\cite{ZXCYAAHLNY09,Ch10a,DTS05,SLN06}. 
Namely, 
the density at which the pair correlation function reaches its 
maximum shifts towards higher values when the temperature departs 
from $0$ and increases as $\sqrt{T}$, 
while the maximum $g_{max}$ diverges as $|\ph - \ph_{GCP}|^{-1}$
on both sides of the transition.

As discussed in the previous section, the zero temperature limit is 
very different for $\ph < \ph_{GCP}$ 
or $\ph > \ph_{GCP}$. Thus we now discuss these two limits separately.

\subsection{Zero temperature below jamming: hard spheres}

In this case we send $T \to 0$ before taking the jamming limit $m \to 0$ and $A = \a m$, corresponding to hard spheres approaching jamming from below. Taking the limit $T\to 0$ we get the expression of \cite{PZ10}:
\begin{align}
e^{-\b\f_{eff}(r)} = & \th(r-1)  \frac{1}{r \sqrt{4 \p A}} \nonumber \\
& \times \int_0^\io du ~ u \left[ e^{-\frac{(r-u)^2}{4A}} \!\!\! - 
e^{-\frac{(r+u)^2}{4A}} \right] q(A;u)^{m-1}, \nonumber \\
\end{align}
with $q(A;r)$ defined in (\ref{qAHS}). In the jamming limit $m\to 0$ and $A = \a m$ one can show (see Appendix~\ref{app:HSpeak}) that this has the scaling~\cite{PZ10} in the region $r = 1^+$:
\begin{align}
e^{-\b\f_{eff}(r)} & =  \frac{1}{m} \D_\a \left( \frac{r-1}{m \sqrt{4 \a}} \right) \ , 
\end{align}
where the function $\D_\a$ is given in Eq.~(\ref{DeltaExact}) and is very well approximated by 
$\D_0(\l) = 1 - \sqrt{\pi} \l e^{\l^2} (1 - \erf(\l))$.
For $\ph \to \ph_{GCP}^-$ one has $m \sim \wt \mu \, (\ph_{GCP} - \ph)$ and \mbox{$A^*=\a^* m$} with a finite $\a^* = \a^*(\ph_{GCP})$:
then the peak of the correlation function for $r=1^+$ has a simple scaling form:
\beq
g_{glass}(r) \sim \frac{y^{HS}_{liq}(\ph)}{\wt\mu \, (\ph_{GCP} - \ph)} \D_{\a^*}
\left[ \frac{r-1}{\sqrt{4 \a^*} \wt \m (\ph_{GCP} - \ph) }   \right] \ ;
\eeq
the numerical values of the parameters can be found in Appendix~\ref{app_numbers}.
The height of the peak scales as \mbox{$g_{glass}(r=1^+,\ph \to \ph_{GCP})\sim 1.09922 / (\ph_{GCP} - \ph)$} and its width as $\ph_{GCP} - \ph$. One can obtain scaling
functions:
\begin{align}
\label{eq:HSscaling}
& g_{glass}(r) \, (\ph_{GCP} - \ph) =  f\left[ \frac{r-1}{\ph_{GCP} - \ph} \right]   , \\
& \text{with } f(\l) =  \frac{y^{HS}_{liq}(\ph)}{\wt\mu} \D_{\a^*}\left[ \frac{\l}{\wt \m \sqrt{4 \a^* }  }   \right]  ,
\end{align}
or alternatively:
\begin{align}
& \frac{g_{glass}(r)}{g_{glass}(1^+)}  =  f\left[ (r-1) (1+4 \ph g_{glass}(1^+)) \right]  , \\
& \text{with }  f(\l) =  \D_{\a^*}\left[ \frac{\l}{8 \ph y^{HS}_{liq}(\ph) \sqrt{ \a^* }  }   \right] .
\end{align}
The number of contacts is the integral of the peak at $r = 1^+$ which is given by
\begin{align}
z & \sim 24 \ph \int_{peak} dr \, g_{glass}(r) \nonumber \\
& \sim 24 \ph y^{HS}_{liq}(\ph) \sqrt{4 \a^* } \int_0^\io d\l \, \D_{\a^*}(\l)   \label{eq:zHS}
\end{align}
Replacing the values given in Appendix~\ref{app_numbers}, we get \mbox{$z = 6.13720$} for $\ph=\ph_{GCP}$, 
which is slightly above the expected $z=2d=6$. 

Note that it has been shown in \cite{PZ10} that within a systematic expansion in $\sqrt{\a}$, one gets $z=2d$ at first order. Here instead, in order to match with the soft sphere computation, we are using an approximation scheme which is not a consistent expansion in $\sqrt{\a}$. This could explain the fact that $z \neq 2 d$. It would be interesting to have a full computation of the next term in the small $\a$ expansion to check whether the equality $z=2 d$ survives at this order. Unfortunately this seems to be a quite hard task.

\subsection{Zero temperature above jamming}

For $\ph > \ph_{GCP}$, the appropriate limit is to send $T\to 0$ with $m = T/\t$ and $A = \a m$. Starting from Eq.~(\ref{effpot_big}) and taking this limit, the effective potential takes the following form (the calculation is 
presented in Appendix~\ref{app_effpot_soft}):
\begin{align}
& e^{-\b \f_{eff}(r)} = \th(r-1) + \th(1-r) \th\left( r - \frac{4 \a}{\t+4\a} \right) \nonumber \\ 
& \times (1+4\a/\t)  \left( 1 + \frac{4\a (r-1) }{\t r} \right)^2  e^{-\frac{\t+4\a}{\t^2} (r-1)^2}.
\label{phieffGCPabove}
\end{align}
Again, since $\t \sim \ph-\ph_{GCP}$, close to $\ph_{GCP}$ this function develops a delta peak close to $r=1$ with height $(\ph-\ph_{GCP})^{-1}$ and width $\ph-\ph_{GCP}$. The main difference we observe with the hard sphere limit is that, despite the fact that both cases give rise to a delta function at $r=1$, the mechanism is different since for hard spheres the support is concentrated on $r=1^+$ while for soft spheres it is concentrated on $r=1^-$. Another interesting observation is that the theory predicts that $g_{glass}(r)$ should vanish exactly for $r < \frac{4 \a}{\t+4\a}$; this means that no pair of spheres can have an overlap larger than $\frac{\t}{\t+4\a}$. Indeed in the limit $\t \to 0$ we recover the $GCP$ at which no overlaps are present. Unfortunately, a direct numerical test of this predicion is impossible: this is because when $1-r = \frac{\t}{\t+4\a}$, the exponential term in Eq.~(\ref{phieffGCPabove}) is equal to $\exp (-1/(\t + 4 \a))$. Since both $\t$ and $\a$ are already extremely small (of the order of $10^{-3}$ at best), $g_{glass}(r)$ is extremely small in the region where we predict it to vanish, and numerically the signal to noise ratio becomes too large in that region.

Again in the limit $\ph \to \ph_{GCP}^+$ we obtain the scaling forms
\begin{align}
\label{eq:SSscaling}
& g_{glass}(r) \, (\ph-\ph_{GCP}) =  f\left[ \frac{1-r}{\ph - \ph_{GCP}} \right] \ , \\
& f(\l) = \frac{y^{HS}_{liq}(\ph) 4 \a^*}{\wt\t} \left( 1 - \frac{4 \a^* \l}{\wt\t} \right)^2 e^{- \frac{4 \a^* \l^2}{\wt\t^2}} \ ,
\end{align}
or alternatively:
\begin{align}
& \frac{g_{glass}(r)}{g_{glass}(1^+)}  =  f\left[ (r-1) (1+4 \ph g_{glass}(1^+)) \right] \ , \\
& f(\l) =  \left( 1 - \frac{\l}{4 \ph y^{HS}_{liq}(\ph)} \right)^2 e^{- \frac{\l^2}{64 \a^* (\ph y^{HS}_{liq}(\ph))^2}} .
\end{align}

The number of contacts is easily obtained from Eq.~(\ref{eq:zdef}) and Eq.~(\ref{phieffGCPabove}) by a change of variable:
\begin{align}
z_0(\ph) & = z(T=0,\ph)= 24 \ph y^{HS}_{liq}(\ph) \int_0^1 dr \, r^2 e^{-\b \f_{eff}(r)} \nonumber \\ & = 24 \ph y^{HS}_{liq}(\ph) \int_0^1 dr \, r^2 e^{-\frac{(r-1)^2}{\t + 4 \a^*}} \nonumber \\
& = 8\ph y^{HS}_{liq}(\ph) G_0^{HS}(\a^* + \t/4) ,
\end{align}
with $G_0^{HS}(\a)$ given in Eq.~(\ref{eq:Galfadef}). The expression above is linear in $\t$ at small $\t$. Therefore, for $\ph \to \ph_{GCP}^+$, since $\t \sim (\ph - \ph_{GCP})$, we have $z_0(\ph) = z_0(\ph_{GCP}) + C (\ph - 
\ph_{GCP})$. This result is at odds with the numerical finding 
of a square root behavior~\cite{DTS05,SLN06}. 
We comment further on this discrepancy 
in Sec.~\ref{conclusion}. At $\ph_{GCP}$, we have
\beq
z_0(\ph_{GCP}) = 8\ph y^{HS}_{liq}(\ph) G_0^{HS}(\a + \t/4) = 6.13720 \ ,
\eeq
which is differs slightly from the 
expected isostatic value $z=2d$, for the technical reasons 
already discussed above in the hard sphere case.

Note that this value of $z_0(\ph_{GCP})$ is identical to the one obtained on the hard sphere side (as it can be easily proven analytically), despite the fact that the two definitions of $z$ are not equivalent. In the soft sphere case we defined $z$ as the number of overlaps. On the hard sphere side, this number is strictly zero, therefore we defined $z$ as the integral of 
the delta peak at $r=1^+$.

\section{Scaling around jamming: comparison with numerical data}
\label{comparison}

In this section, we compare the prediction of the theory with numerical data, in particular to test the
scaling in temperature and in $\d \ph = \ph - \ph_{GCP}$ around the glass close packing point. 

\subsection{Details of the numerical procedure}

We investigated
a system of $N=8000$ identical harmonic spheres. The system was prepared using the standard procedure of 
Ref.~\cite{OLLN02}. 
We started from a random configuration at high density; we then minimized the
energy, and then reduced slowly the density, at each step minimizing the energy again, until a jammed
configuration of zero energy and volume fraction $\ph_j$ was found. This jammed configuration corresponds, according
to the mean-field interpretation of~\cite{PZ10}, to the $T=0$, $p \to \io$ limit of a given hard sphere glass,
or equivalently to the $T\to 0$, $P\to 0$ limit of a corresponding soft sphere glass. 
Starting from that configuration, 
we prepared configurations at 
different $T$ and $\ph$ by using molecular dynamics simulations
as in Ref.~\cite{BW09a} to thermalize the 
system inside the glass state selected by the initial jammed configuration. 
Since we always used very small $T$ and $\ph \sim \ph_j$, the system is not able to escape from that 
glass state, so it always remains in the vicinity of the original 
jammed configuration. This is a crucial requirement to numerically obtain 
scaling behavior near $\ph_j$ as microscopic rearrangments would 
directly affect the location of $\ph_j$ in the simulations~\cite{CBS09}.  
We tested the absence of such an effect in our simulations 
by repeating the energy minimization starting from some equilibrated
configurations at different $T$ and $\ph$, 
and checking that we always found the same value of $\ph_j$ (within numerical errors).
For the particular series of run described below, 
we estimated $\ph_j = 0.643152\pm 0.000020$ by fitting the zero 
temperature energy using $e(\ph) \propto (\ph-\ph_j)^2$.

\subsection{Difficulties in the comparison with the theory}

When comparing the theory with the numerical data, we face two difficulties. First of all, the value
of $\ph_j$ depends on the numerical protocol and on its particular realization we used, 
and therefore cannot be directly compared with $\ph_{GCP}$.
Note that $\ph_{GCP}$ should represent an upper bound for $\ph_j$, whatever the protocol, but the value
$\ph_{GCP}=0.633353$ we report seems to contradict this statement.
The reason is simply that our theoretical computations are
based on the HNC equation of state for liquid hard spheres, which (as 
discussed above), underestimates
the value of $\ph_{GCP}$:
a better result is obtained using the Carnahan-Starling equation of state,
which gives $\ph_{GCP}=0.683$, consistently with the value of $\ph_j$ found above~\cite{PZ10}. 
To avoid confusion, here we stick to the use of the HNC equation of state;
the theoretical calculations can be easily repeated for any other equation of state, and the scaling
around $\ph_{GCP}$ is unaffected by this choice. 
Once again we stress that the absolute values
of density reported in this paper should always 
be taken with this caveat in mind.
Since we are interested here mainly in the relative
distance to jamming, which is $\ph - \ph_j$ for the numerical data, and $\ph - \ph_{GCP}$ for the theory,
the absolute values of density are not crucial for our purposes. 

The second 
difficulty is the following. For hard spheres, the theory predicts that
the reduced pressure diverges as $p^{HS}_{glass} \sim 3 \ph_{GCP}/(\ph_{GCP}-\ph)$, while
$y^{HS}_{glass}(\ph) \sim 1.1/(\ph_{GCP}-\ph)$ for $\ph \to \ph_{GCP}^-$, see Appendix~\ref{app_numbers}.
As one can easily check, these scalings imply that the theory violates the exact thermodynamic relation 
$p^{HS}_{glass} = 1 + 4 \ph y^{HS}_{glass}(\ph)$, as already noticed 
in~\cite{PZ10}. In particular, the divergence of $p^{HS}_{glass}$ is correct, while the
coefficient of the divergence of $y^{HS}_{glass}(\ph)$ is overestimated by a factor
of $\sim 1.46562$. We will comment on the origin of this problem in Sec.~\ref{conclusion}. This error affects the
prefactors in the scalings of $g_{glass}(r)$ and of $z$ around $\ph_{GCP}$. 
To correct for this error, we rescaled the $\d \ph$ of the numerical data in such a way that
the coefficient of the divergence of $y^{HS}_{glass}$ is the same as 
the theoretical one. 

To summarize, when comparing numerical data with the theory, 
we consider $\d \ph = \ph -\ph_{GCP}$ for the theory,
and $\d \ph = 1.46562 (\ph - \ph_j)$ for the numerical data such that
the values of $y^{HS}_{glass}(\d \ph)$ for theory and simulations are 
exactly the same
close to $\d \ph = 0$. Note that this adjustment is based solely on 
the analysis of the hard sphere side of the jamming 
transition ($\d \ph<0$), but we find that it allows
to describe the full scaling also for positive $\d \ph$. 

If the reader is uncomfortable with the above reasonings, 
another way of thinking is that we need to adjust two free parameters
(the position of the transition and the amplitude of the 
rescaling of $\d\ph$) 
to obtain the best comparison of theory and numerical simulations.
While the location is $\ph_j$ was always adjusted in previous
work~\cite{DTS05,SLN06,OLLN02,Vh09}, we need to adjust also one
prefactor as we seek to compare theoretical predictions to simulations 
in absolute values, and not only at the level of leading 
diverging contributions.

\subsection{Energy, pressure, average coordination}

\begin{figure}[t]
\includegraphics[width=8cm]{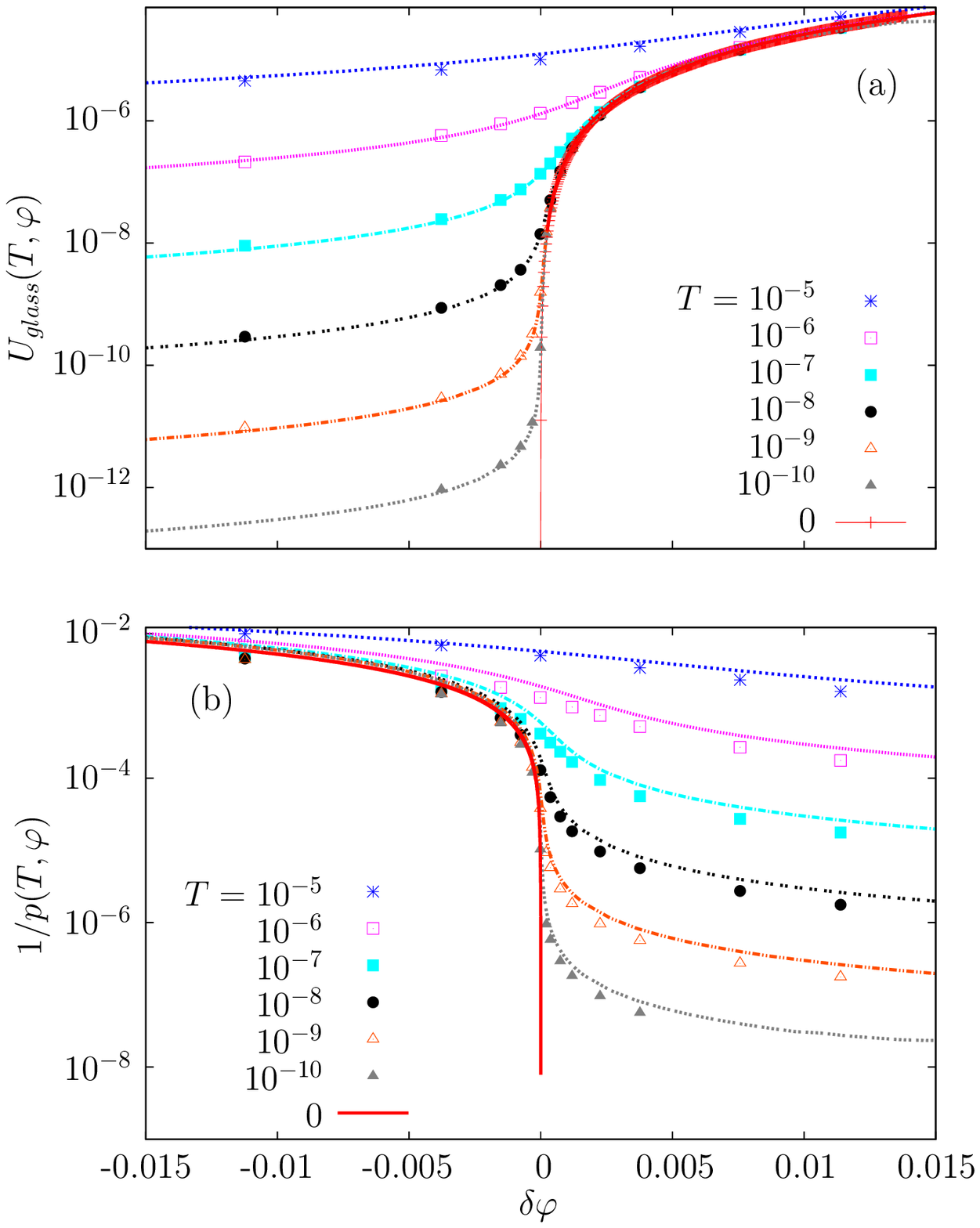}
\caption{(Color online)
Energy $U_{glass}(T,\ph)$ (top panel) and inverse reduced pressure $1/p_{glass}(T,\ph)$ (bottom panel)
as functions of distance from jamming $\d \ph$, for several temperatures.
We define $\d \ph = \ph -\ph_{GCP}$ for the theory,
and $\d \ph = 1.46562 (\ph - \ph_j)$ for the numerical data.
}
\label{fig:ep}
\end{figure}

We start our discussion by the simplest observables. In Fig.~\ref{fig:ep} we plot the average energy $U_{glass}(T,\ph)$ 
and the inverse reduced pressure $1/p_{glass}(T,\ph)$ as functions
of $\d\ph$ for several temperatures. The agreement between theory and numerical data, with the rescaling of $\d\ph$ discussed
above, is nearly perfect. The scaling around jamming is clearly visible in the figures. For instance, $U_{glass}(T,\ph)$ tends
to a finite value for $\d\ph>0$, while for $\d\ph < 0$ it goes to zero as a power law, since in this case the system becomes a hard
sphere glass. Similarly, the reduced pressure is finite for $\d\ph<0$, while it diverges proportionally to $\b$ for $\d\ph>0$, since
in this case the pressure is finite at zero temperature. 
At finite temperature, the curves interpolate between the two regimes.
Scaling functions similarly to the ones shown in 
Fig.~\ref{fig:scaling_m} for $m^*$ can easily 
be constructed.

\begin{figure}[t]
\includegraphics[width=8cm]{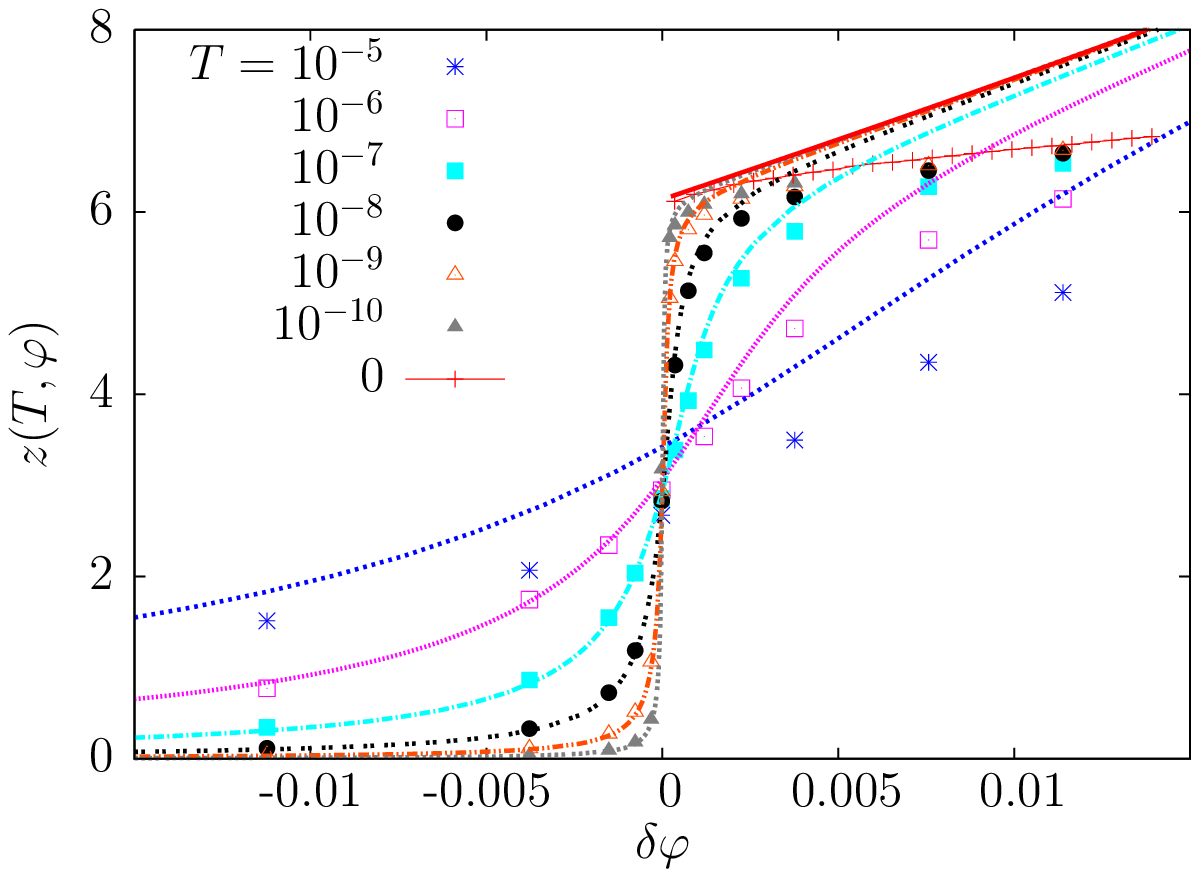}
\caption{(Color online)
Average number of contacts $z$ as a function of distance from jamming $\d \ph$, for several temperatures.
We define $\d \ph = \ph -\ph_{GCP}$ for the theory,
and $\d \ph = 1.46562 (\ph - \ph_j)$ for the numerical data.
}
\label{fig:z}
\end{figure}

In Fig.~\ref{fig:z} we report the average coordination number $z(T,\ph)$, as a function
of $\d\ph$ for several temperatures. At $T=0$, the average coordination jumps
from 0 to $6$ (for numerical data) or to $6.13720$ (for the theory). 
For $\d\ph>0$, the average 
coordination grows linearly in $\d\ph$ for the theory, while it grows 
as $\d\ph^{1/2}$ for the numerical
data. Therefore, the theory fails to capture the correct
evolution of this structural property at $\d\ph>0$ and $T=0$. 

A more
detailed description of what happens is obtained by looking to the data 
at finite $T$.
When temperatures become too large, e.g. at $T=10^{-5}$, 
the theory eventually fails because of the low-$T$ approximation we 
made on the liquid theory in
Eq.~(\ref{y_approx_rep}). 
For any fixed temperature $T \leq 10^{-6}$, we observe that the 
theory describes perfectly the numerical data for $\d\ph < 0$ and also 
for positive $\d\ph$ up to some crossover value  $\d\ph_h(T)$. 
Around $\d\ph_h(T)$ the
theory starts deviating from the numerical data. 
The data in Fig.~\ref{fig:z} suggest that $\d\ph_h(T)$ is an increasing
function of $T$, \ie that for higher temperatures the theory performs better, if the temperature is low enough
that approximation (\ref{y_approx_rep}) makes sense. Indeed, we 
expect that for $\d\ph_h(T) \to 0$ for
$T\to 0$, since at $T=0$ the theory fails 
to capture the behavior of the contact number at $\d\ph>0$. 
We will come back to this point below.

\subsection{Correlation function}

\begin{figure}[t]
\includegraphics[width=8cm]{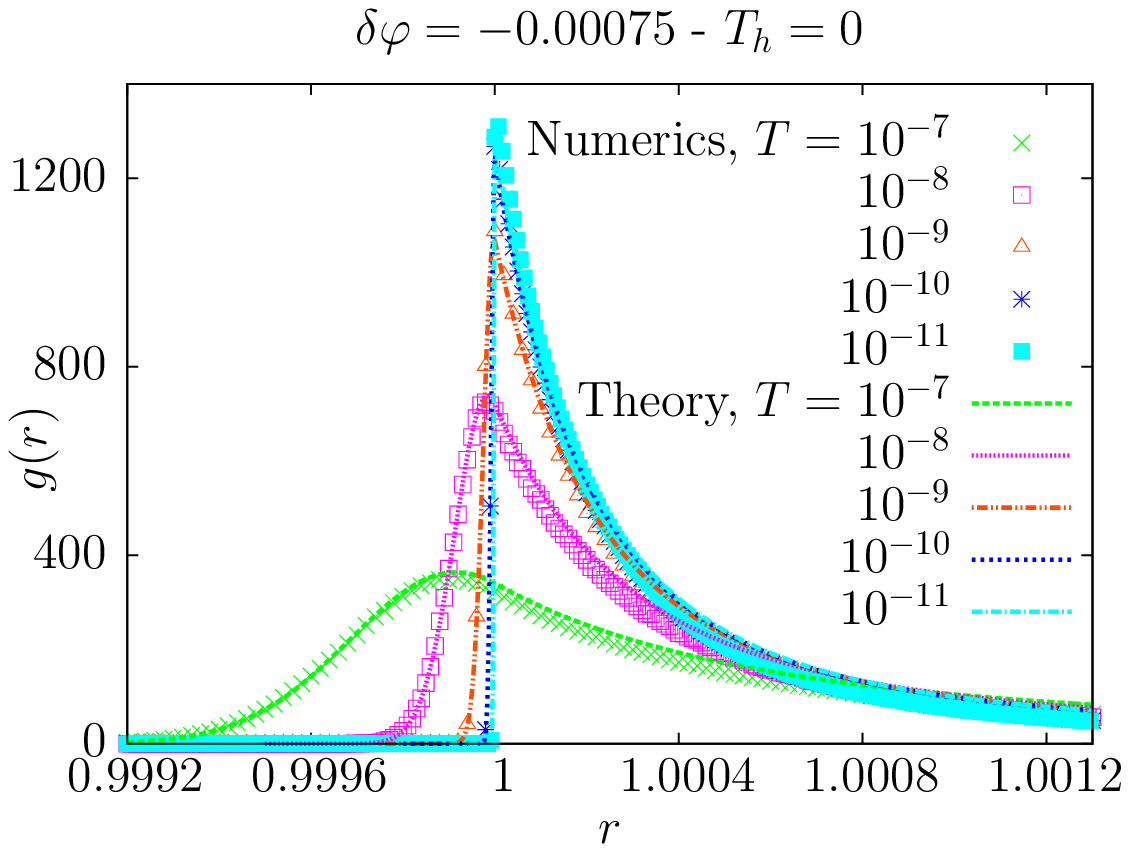}
\caption{(Color online)
Pair correlation $g_{glass}(r)$ just below jamming predicted by theory 
(full lines) and measured in numerical simulations (symbols).
Here $\d\ph = -0.00075$.
}
\label{fig:grbelow}
\end{figure}

\begin{figure*}
\begin{flushleft}
\includegraphics[width=12cm]{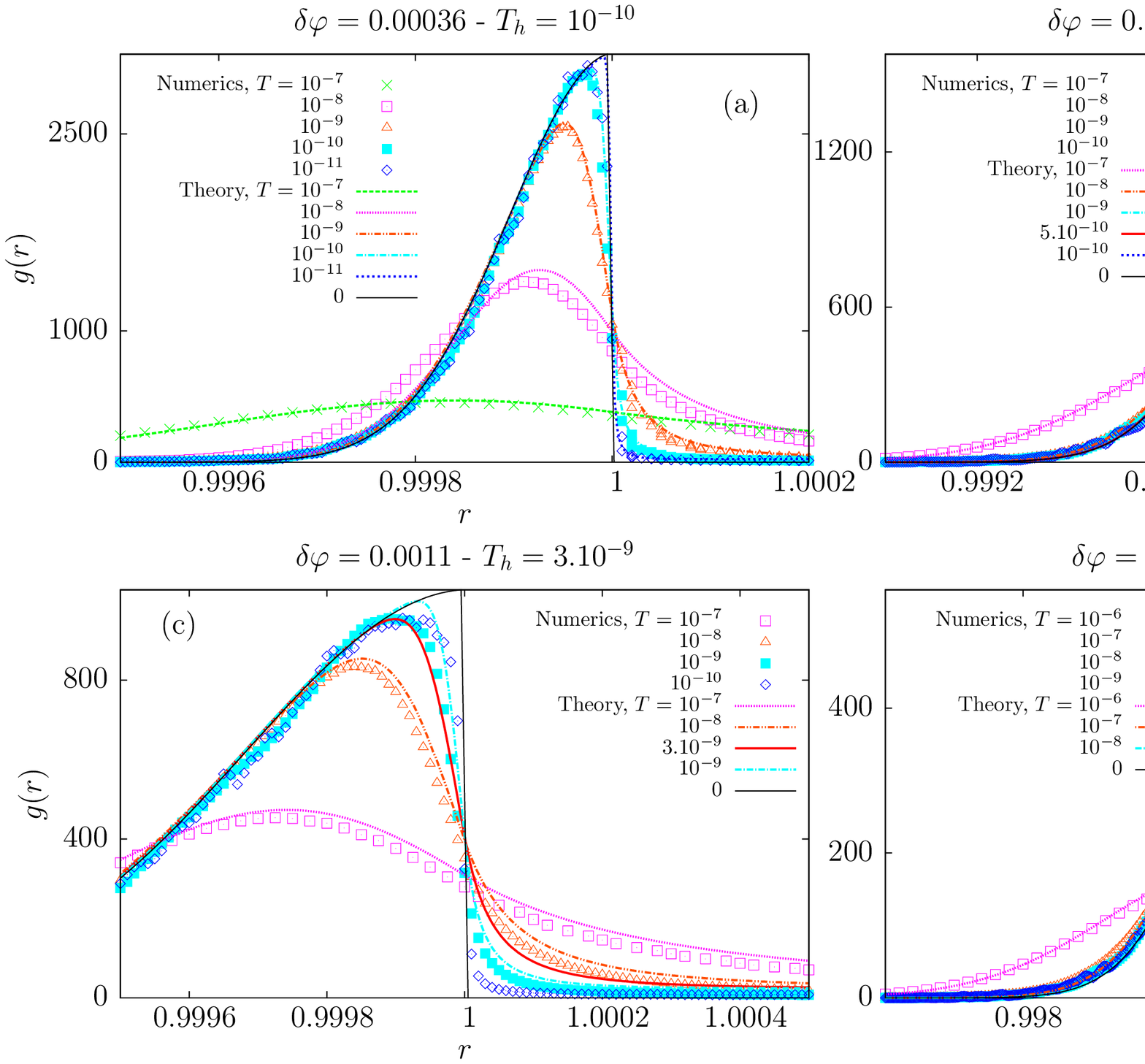}
\end{flushleft}
\caption{(Color online)
Pair correlations $g_{glass}(r)$ above jamming predicted by theory (full lines) and measured in numerical simulations (symbols).
The agreement is very good for $T > T_h(\d\ph)$, the values of $T_h(\d\ph)$ are reported above the figures. Below $T_h(\ph)$,
the numerical curves do not evolve anymore, while the 
theoretical curves continue to evolve towards the $T=0$ theoretical 
limit.}
\label{fig:grabove}
\end{figure*}

We now report data for the scaling of the contact peak of 
$g_{glass}(r)$ near jamming.
We first consider a fixed $\d\ph < 0$, 
and change the temperature. In Fig.~\ref{fig:grbelow} we report data for
$\d\ph = -0.00078$ for several temperatures. In this regime the theory works very well down to $T=0$; this is expected
since we know that the theory works at $T=0$ for hard spheres~\cite{PZ10}.

Next, we consider a fixed $\d\ph > 0$ and plot $g_{glass}(r)$ for several temperatures. This is done in Fig.~\ref{fig:grabove} for
four different values of $\d\ph$. Here, we observe that, like for the average coordination (Fig.~\ref{fig:z}), the theory works
very well at moderately large $T$. However, on lowering $T$, at some point we observe that the numerical $g_{glass}(r)$ saturates
to a limiting curve which has a maximum at $r<1$. 
By contrast, the theoretical curves slowly evolve towards the 
$T=0$ theoretical limit, which is a half-Gaussian centered 
in $r=1$, with its $r>1$ part removed.

\begin{figure}
\includegraphics[width=8cm]{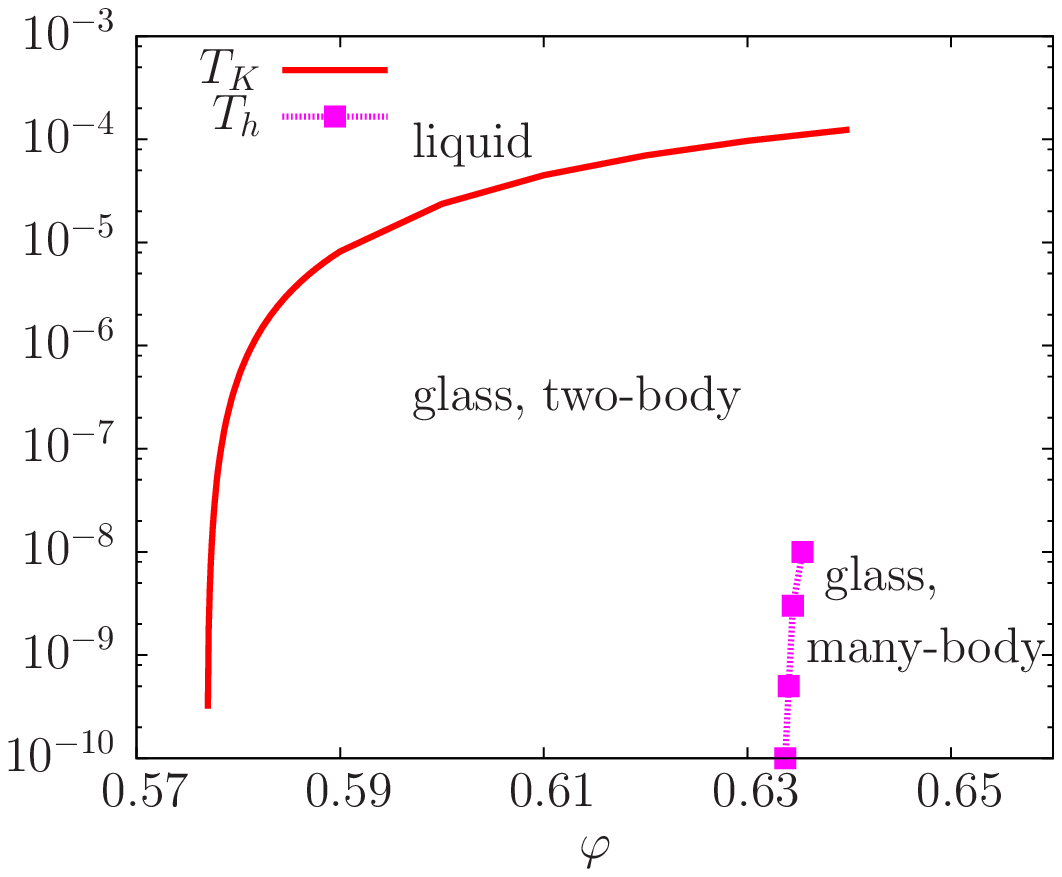}
\caption{(Color online)
The crossover temperature $T_h(\ph)$, 
as determined from Fig.~\ref{fig:grabove}, 
and compared with the Kauzmann temperature from Fig.~\ref{fig:tk}.}
\label{fig:Th}
\end{figure}

Therefore, below a certain temperature $T_h(\d\ph)$ 
(the inverse of the function $\d\ph_h(T)$ mentioned above), a deviation between
theory and numerical data is observed close to the maximum of $g_{glass}(r)$. Clearly, the deviation between data and theory is a
smooth crossover, so determining $T_h(\d\ph)$ is not easy.
Here we choose the following procedure. The numerical
data accumulate on a master curve as $T\to 0$. Therefore we define $T_h(\d\ph)$ as the value of temperature at which the
theoretical curve best fits the $T\to 0$ numerical curve. As an example, in the upper left panel of Fig.~\ref{fig:grabove},
we see that the theoretical curve for $T=10^{-10}$ fits very well the numerical curves for both $T=10^{-10},10^{-11}$, while
the $T=10^{-11}$ theoretical curve is quite different. We fix $T_h=10^{-10}$ for this value of $\d\ph$. 
Clearly the ambiguity on the precise numerical value of 
$T_h$ is rather large. Despite this, we are able to qualitatively confirm 
that $T_h(\d\ph)$ is an increasing function of $\d\ph$, as already 
suggested in the previous section. 
The function $T_h(\d\ph)$ determined 
from the pair correlation functions is 
reported in Fig.~\ref{fig:Th}. It emerges continuously from zero above 
$\ph_{GCP}$. 
A comparison with the temperature scale given by the glass 
temperature shows that it is orders of magnitude smaller
than $T_K$. Thus it is only in the small regime of $T<T_h$ 
and $\ph > \ph_{GCP}$ that our the effective potential 
approach misses some important physics, as we discuss below 
in Sec.~\ref{subdisc}.

\begin{figure*}
\begin{flushleft}
\includegraphics[width=12cm]{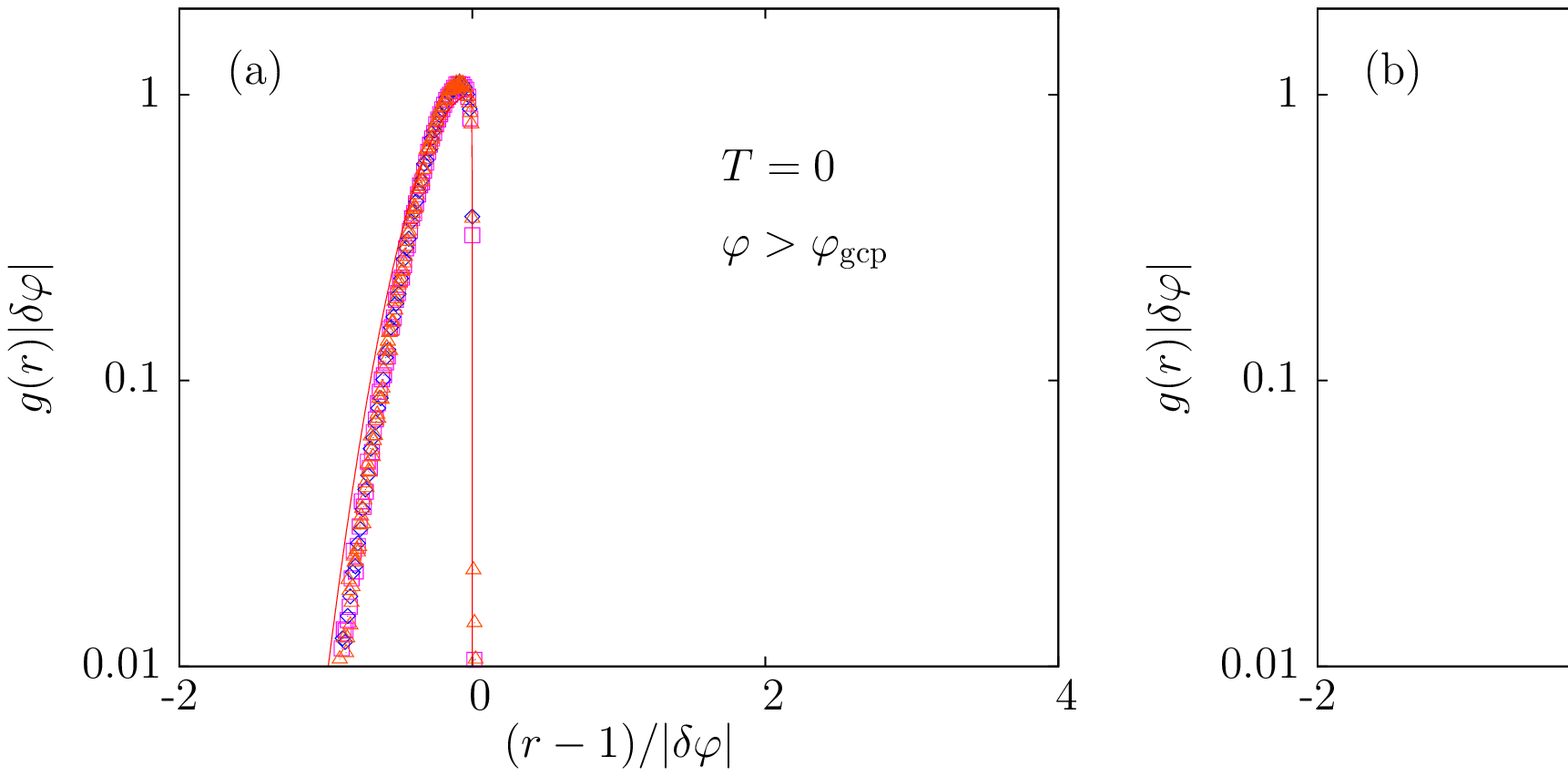}
\end{flushleft}
\caption{(Color online)
Scaling functions at $T = 0$ above (left) and below (right) the jamming transition showing the convergence of the first peak near $r = 1$ to a delta function with asymmetric scaling functions on both sides of the transition. 
To ease visualization, we show the evolution of $g(r)/g_{max}$, where $g_{max}$ can be read from Fig.~\ref{fig:gmax}.
}
\label{fig:grTzero}
\end{figure*}

Finally, in Fig.~\ref{fig:grTzero} we show that $g_{glass}(r)$, at $T=0$, follows the scaling relations predicted by the theory, Eqs.~(\ref{eq:HSscaling}) and (\ref{eq:SSscaling}), for 
$| \d\ph | \to 0$, with different scaling functions on the two sides of the transition. While the scaling is perfect for $\d\ph<0$~\cite{PZ10}, for $\d\ph>0$ there is a small deviation
around the maximum as already discussed, 
which is however very small in the scaling regime for $\d\ph \to 0$.

\subsection{Discussion}
\label{subdisc}

The main discrepancy between the theory and numerical data is in the region $\d\ph>0$ and very small temperature $T<T_h(\d\ph)$.
Here, the average coordination scaling $z \propto \d\ph^{1/2}$ at $T=0$ is not obtained. Furthermore, the shape of the peak of $g_{glass}(r)$ is
only partially captured by the theory.
We tentatively attribute these discrepancies to the particular nature of the vibrational modes at $T=0$ and $\d\ph>0$. It has been
shown that in this regime the low-frequency spectrum is dominated by spatially 
correlated vibrational modes. These modes
cannot be described by our approach, in which vibrations are assumed to 
be Gaussian and only two-body correlations are taken into account.
Therefore it is to be expected that the particular features of 
the jamming transition that are related to these correlated soft modes
are not well reproduced by our theory.

On the hard sphere side, the soft modes induce a square-root divergence of the pair correlation function 
$g_{glass}(r) \propto 1/\sqrt{r-1}$. Yet, for $\d\ph \to 0^-$, 
this square root singularity is well separated from the contact 
peak~\cite{PZ10};
therefore the shape of the scaling function for the contact peak 
is not affected by the soft 
modes and it is indeed correctly reproduced by the theory (see \cite{PZ10}
and Fig.~\ref{fig:grTzero}).

On the other side of the transition and for small $\d\ph > 0$,
it is commonly assumed that both 
the contact peak and the square root singularity are shifted
 by $\d\ph$.
Under this assumption, the square root contribution should become 
$g_{glass}(r) \sim 1/\sqrt{r-1+ a \d\ph}$. 
Inserting this form of $g_{glass}(r)$ into Eq.~(\ref{eq:zdef}) yields
the result that $z = z_0 + c \d\ph^{1/2}$, where $z_0$ is 
the contribution of the contact peak. While $z_0$ is clearly 
unaffected by the shift, the scaling of $z$ with $\delta \phi$
directly stems from the square root singularity of the pair correlation
function.
Therefore, the scaling of $z$ is dominated 
by the soft modes contribution, which might explain why it is not 
well captured by our theory. 

On the other hand, inserting the same expression in the general formula for the internal energy, Eq.~(\ref{def_U}), one finds
that $e_{GS} \sim \wt e \d\ph^2 + d \d\ph^{5/2}$, the first term being the contribution of the contact peak, the second being the one of the square root
singularity. Therefore, for the energy the contribution of the anomalous modes is subdominant, and for this reason we observe that the theory reproduces
well the numerical data in this case. The anomalous correction $\d\ph^{5/2}$ is of course not reproduced by the theory, that gives an expansion of $e_{GS}$
in integer powers of $\d\ph$, but these are subdominant terms. 

We also suspect that the shape of $g_{glass}(r)$ near the peak 
is also influenced by a mixing of the contact peak 
with the square root singularity. While both contributions are 
well separated for $\ph < \ph_j$, this is not necessarily the case
above jamming, which likely explains the small deviations
observed in Figs.~\ref{fig:grabove} and \ref{fig:grTzero}, but we could not 
find an empirical or scaling argument to disentangle both contributions.

Finally, it is interesting to remark that these deviations
between theory and simulations are washed out by a finite temperature 
$T > T_h(\d\ph)$. We suspect that $T_h(\d\ph)$ represents a temperature
below which the many-body direct interactions induced by the presence of
many replicas become relevant (see the discussion in section~\ref{sec:effpotapp}).
Above this crossover temperature, vibrations are probably
much less correlated and our approximations become correct. 

A possible interpretation is that the low-temperature correlated vibrations
that are relevant below $T_h(\d\ph)$ are induced by the presence of strongly
correlated soft vibrational modes 
at $T=0$~\cite{LNSWW10,Vh09,OLLN02,WNW05,WSNW05,BW06,Wyart,OSLN03}.
Interestingly, the fact that $T_h(\d\ph) \to 0$ for $\d\ph \to 0^+$ gives a hint 
that the correlations of these modes
at $\d\ph=0$ are extremely fragile, 
such that an arbitrary small temperature is able to 
destroy them and restore the agreement between 
theory and numerical data. This observation is consistent with the numerical finding that
around the jamming point, strong anharmonicity is present~\cite{XVLN10,SBOS11}
and the Hessian matrix of the potential does not capture the correct density of states
of the vibrations~\cite{SBOS11}.

\section{Conclusions}
\label{conclusion}

In this final section we summarize our results and we discuss some 
perspectives for future work.

\subsection{Summary of results}

Understanding theoretically the physical properties
of dense packings of soft repulsive particles is a fully 
nonequilibrium problem, because the 
jamming transition occurs in the absence of thermal fluctuations 
deep inside a glassy phase where phase space is dominated by the 
existence of a large number of metastable states. 
Athermal dynamical protocols sample packings with highly non-uniform 
weights~\cite{GBO06,GBOS09}, while thermal protocols fall out of equilibrium
at the glass transition~\cite{DTS05}.

Nevertheless, we showed that it can be 
successfully addressed using 
equilibrium statistical mechanics tools.
We have developed a mean-field replica theory of the jamming
transition of soft repulsive spheres which satisfactorily  
derives, from first principles, 
the existence and location of a jamming transition. 
This transition, that happens only at $T=0$, is exactly 
the same as the SAT/UNSAT transition of random constraint satisfaction 
problems~\cite{Mo07,MM09,KZ08}:
it is the point where the Parisi parameter $m^*$ that characterizes 
the 1RSB glass phase goes to zero, the reduced pressure of
the hard sphere glass diverges, or equivalently its pressure and 
energy become finite. 
Within our approach, the jamming transition occurs deep into the glassy 
phase, and is thus a phenomenon which is physically distinct from 
the glass transition itself~\cite{PZ10,MKK09}. 
Finally, we have shown that the average coordination jumps from zero to 
$\sim 6$ at the transition, and we derived scaling functions that
describe well the contact peak of the pair correlation function.

We want to stress here again that the absolute values
of density reported in this paper are affected by the fact that 
we used the HNC equation of state for the 
liquid~\cite{PZ10}, and thus quantitative agreement with 
simulations cannot be expected. Still, we found that
the scaling around jamming is almost insensitive to the particular equation of state that is chosen for the liquid.
We also investigated the region of small $T$ around the jamming transition. We showed that the theory reproduces quite well
the scaling of numerical data with $T$ and $\d\ph$, the distance from the jamming transition. On the hard spheres side
of the transition ($\d\ph < 0$) the theory works well down to $T=0$. On the soft sphere side ($\d\ph > 0$) 
we have identified a temperature $T_h(\d\ph)$ below
which, according to our interpretation, correlated vibrations~\cite{WNW05,BW06} 
that are neglected in our theory become relevant for some observables, 
and produce a series
of interesting scalings that are not captured by our 
theory~\cite{Wyart,Vh09,LNSWW10}.
The most striking of these is the scaling of the number of contacts
which our theory fails to reproduce.
It is a major  open problem to try and include 
in the theory a 
correct description of the divergent correlations in the vibrational modes~\cite{WNW05,BW06} 
that are relevant
for $\d\ph >0$ and $T<T_h(\d\ph)$. 

\subsection{Technical aspects}

There are several more technical 
aspects on which the theory could be improved, even in the region 
where it works quite well.

Our theory falls within the general framework of the replica method, using the molecular liquid formulation of \cite{MP99a,MP99b}.
However, to be able to describe the jamming transition, we had to develop a new approximation scheme, that is based on the effective potentials
formulation of~\cite{PZ10}, and allows to write the molecular liquid as a simple liquid with effective potentials 
involving an arbitrary number of particles.
Our approximation is based on the following steps:
\begin{enumerate}
\item As in~\cite{PZ10}, we neglect the three (and more)-body interactions, and just keep the two-body effective potential.
\item On top of this, we observe that the effective potential is given by the original potential (rescaled by a factor $m$) plus a small perturbation, 
therefore
we use liquid perturbation theory to write the free energy of the replicated liquid as in Eq.~(\ref{S_of_Q}).
\item Finally, we performed the low temperature approximation Eq.~(\ref{y_approx_rep}), in order to simplify the numerical calculations.
\end{enumerate}
The last two approximations have been done mainly for practical reasons. However, we found that they are related to the first one. Indeed, if we get rid of 
approximation 3, we introduce an explicit dependence of $y_{liq}$ on temperature, hence on $\t$ in the limit $T\to 0$ and $\d\ph>0$. 
It is easy to realize that this dependence affects the scaling close
to jamming. An additional term $\sqrt{\t}$ appears in the small $\t$ expansion (see Appendix~\ref{app_smalltau}), 
which implies that $\t \propto \d\ph^{2}$ and $e \propto \d\ph^{3}$. An inspection
of the small $\t$ expansion reveals that if $y_{liq}$ does not depend on $\t$, then a delicate cancellation 
happens to eliminate the $\t^{1/2}$ term and produce
the correct scalings. The crucial observation is that the anomalous term contains the derivative of $y_{liq}$ 
with respect to temperature, which is related to a three-body
correlation function. Therefore we suspect that in a full treatment this term is cancelled by a term 
coming from the three body interaction potential. Therefore, getting
rid of the third step of approximation might require taking into account three-body correlations as well, 
therefore getting rid of all steps at once. 

This is of course an
extremely interesting direction for future research, mainly because we suspect that getting rid of approximation 2 
should allow to eliminate the violation of the exact thermodynamic relation 
$p_{glass}^{HS}=1 + 4 \ph y_{glass}^{HS}(\ph)$ that
happens in the theory and forces us to rescale $\d\ph$ in order to compare with numerical data.
It is also possible that taking into account interactions involving several particles one could capture 
at least partially the long range correlations at jamming.

Finally, our theory neglects the existence of rattlers (particles with less than 4 contacts, which are therefore
not mechanically blocked in jammed packings) since the cages are assumed to be identical for all particles~\cite{PZ10}.
Therefore the rattlers are approximated as jammed particles.
This should not influence the shape of $g(r)$ close to the contact peak
since typically rattlers are not in contact with other particles, so they do not contribute
in this region, which was the main focus of this work. The theory can be in principle extended to include
cage heterogeneity, and therefore describe rattlers. The computations will be more involved because a distribution
of cage radii $P(A)$ should be included.

\subsection{Some general directions for future work}

Our approach is general
enough that it can be systematically improved and generalized to various
models (\eg Hertzian potentials, or truncated Lennard-Jones potential). 
Moreover, recent work reported on a method to compute the shear modulus within the replica approach~\cite{MY10,SF11}. 
This could allow to add the external drive to the theory, and obtain
a complete theoretical picture of the jamming transition in the 
three dimensional 
temperature, density, stress jamming phase diagram
originally proposed by Liu and Nagel~\cite{LN98}.

Finally, we believe that in the long term these studies could open a way 
towards a quantitative renormalization group treatment of structural glasses, 
following recent studies that formulated the renormalization group for lattice models~\cite{CDFMP10,CBTT11}. 
If successful, this program could lead to a correct treatment of
long-range correlations in the glass phase~\cite{WNW05,BW06}, and in particular at 
the jamming point, and could also allow us to address the 
behavior of pair correlations over a broader range of distances, 
since pair correlation functions also exhibit singular behavior
beyond the contact peak~\cite{DTS05,SLN06}, while the 
structure factor~\cite{DTS05b} and isothermal compressibility~\cite{BCCDS11} 
also display intriguing anomalies at large scale which remain to 
be predicted theoretically.

\acknowledgments

We wish to thank D.~Frenkel, A.~Liu, G.~Parisi, 
A.~Sicilia, M.~Wyart, and N.~Xu 
for many useful discussions and comments on our paper
and for sharing with us the results of their work.
We thank F.~van~Wijland for giving us the opportunity to 
pursue our collaboration on this project.
H. Jacquin PhD work is funded by a Fondation CFM-JP Aguilar grant.
L. Berthier is partially funded by 
R\'egion Languedoc-Roussillon.

\appendix

\section{Complexity and the free-energy of basins}
\label{app:basins}

The replica calculation presented in this paper is based on the assumption that phase space can
be decomposed into a set of ``pure states'', which is the basis of Eq.~(\ref{Cpart}).
The precise definition of these states in statistical mechanics is however tricky and not very practical
for numerical implementation. 
Although less rigorous, 
a more intuitive presentation in terms of the minima of the potential energy function, sometimes
called ``inherent structures''~\cite{SW85} or ``basins'',
is very enlightening and allows an explicit connection with many numerical procedures
encountered in the litterature to study the jamming transition~\cite{OLLN02,XFL11}. 
For the sake of completeness, we 
include this discussion below.
It is worth noting, however, that a partitioning of phase space in basins
always depends on the details of the way these basins are defined. We expect
that the particular definition we use below will be a good approximation of
the equilibrium decomposition, at least for small enough temperatures.

\subsection{Definition of basins and their internal free energy}

\begin{figure}
\includegraphics[width=.33\textwidth]{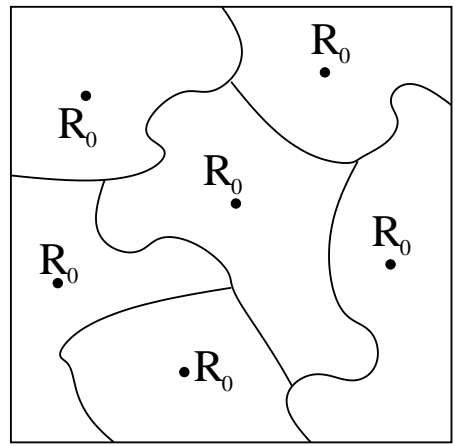}
\caption{A partitioning of phase space into basins, 
labeled by a representative 
configuration $\vec R_0$, which we assume
to have minimal energy in the basin.
Of course the shape of the basins is much more complex
due to the high-dimensional nature of phase space. The above
is just a very schematic representation.
}
\label{f1}
\end{figure}

We suppose that the phase space of all configurations of the spheres 
$\vec{R}$ can be unambiguously
partitioned in basins. Each basin is labeled by a 
representative configuration, $\vec R_0$, see Fig.~\ref{f1}.
For instance, one might think that a basin represents
the set of configurations that end up in $\vec R_0$
after an energy minimization ($\vec R_0$ is an energy minimum in this case). Or, for hard spheres,
we can assume that a basin is the set of configurations that end up in $\vec R_0$ after a fast 
compression.
One can, alternatively, minimize the energy first, then if the energy is zero (the configuration is not 
jammed), perform a fast compression, in such a way that $\vec R_0$ is always a jammed 
configuration. In any case, $\vec R_0$ will have minimal energy in the basin at the end of the 
procedure. Note that the partition defined in this way is, a priori, dependent on the precise procedure
used to perform the energy minimization or the compression. It is reasonable to expect that the ambiguity
due to this is small and negligible in most cases.

The potential energy of the harmonic spheres is $V[\vec R] = \sum_{i<j} \phi( | r_i - r_j |)$, and we define 
the free energy of a basin as
\beq
e^{-\b N f[\vec R_0]} = \int d\vec R \, G(\vec R,\vec R_0) \, e^{-\b V[\vec R_0]}
\eeq
where $G(\vec R, \vec R_0) = 1$ if and only if configuration $\vec R$ belongs to basin $\vec R_0$.
Note that for $\beta \to \io$ we have
\beq
e^{-\b N f[\vec R_0]} \to 
\begin{cases} 
v_b(\vec R_0) = e^{N s[\vec R_0]} \hskip10pt \text{if $e[\vec R_0]=0$} \\
e^{-\b N e[\vec R_0]}  \hskip41pt \text{if $e[\vec R_0] > 0$}
\end{cases}
\eeq
where $v_b$ is the volume of the configurations of the basin that are 
compatible with the hard sphere constraint 
(its logarithm being the internal entropy
of the basin). Indeed, if the basin contains zero-energy configurations ($e[\vec R_0]=0$), 
then at zero temperature its free energy is
dominated by the number of those configurations. In the opposite case, if there are 
no hard sphere configurations ($e[\vec R_0]>0$),
it is very reasonable to assume that the minimum of the energy is unique and
if we assume that $\vec R_0$ is the configuration
of minimal energy inside the basin, then the weight is just dominated by its energy 
$e[\vec R_0] = V[\vec R_0] / N$.

\subsection{The replica method and the complexity}

The replica method allows to compute the following object (for a given temperature $T$ and 
volume fraction $\ph$), see Eq.~(\ref{rep_part_funct}):
\beq
Z_m = \sum_{\vec R_0} e^{-\b m N f[\vec R_0]} \ .
\eeq
We can then compute a configurational entropy (or complexity) $\Si(m;T,\ph)$ according to 
Eq.~(\ref{def_sigma_replicated_parametric}):
\beq\label{Sibasins}
\begin{split}
\Si(m; T, \ph ) &= -m^2 \frac{d}{dm} \left[ \frac1m \log Z_m \right] \\
& =  - \sum_{\vec R_0} \frac{e^{-\b m N f[\vec R_0]}}{Z_m} 
\log \left[  \frac{e^{-\b m N f[\vec R_0]}}{Z_m}  \right] \\
& = - \sum_{\vec R_0} p_{\vec R_0} \log  p_{\vec R_0}
\end{split}\eeq
where 
\beq
p_{\vec R_0} = \frac{e^{-\b m N f[\vec R_0]}}{Z_m}  \ .
\eeq
Therefore, the complexity is {\it the entropy of basins, weighted by the weight $p_{\vec R_0}$}. The 
weight is constructed by assigning an ``energy'' $N f[\vec R_0]$ to each
basin, and then constructing a Boltzmann weight at the effective temperature $\t = T/m$.

\subsection{Special limits}

Equation (\ref{Sibasins}) allows to give simple interpretation to the different limits we considered in 
Sec.~\ref{jamming}, as follows:
\begin{enumerate}
\item
Consider $m=1$. Then the basins are weighted by their equilibrium free energy at temperature $T$. 
The associated complexity $\Si_{ eq}(T, \ph) = \Si(m=1;T,\ph )$
corresponds to the number of basins that {\it dominate the equilibrium partition sum}. 
In other words, if we take an equilibrium configuration at given $(T,\ph)$, this 
configuration belongs, with probability 1 for $N\to\io$, to one among $\exp[ N \Si_{ eq} ]$ 
basins (all the other basins having a negligible probability at equilibrium).
This equilibrium complexity is reported in Fig.~\ref{fig:sigma}.
\item
Consider taking $\b \to \io$ first, then $m=1$. In this case we get, at low enough density 
where hard sphere configurations exist (omitting the normalization)
\beq
p_{\vec R_0} \propto v_b[\vec R_0] \ .
\eeq
The basins are weighted by their hard sphere volume. Therefore $\Si^{HS}_{ eq}(\ph) = \Si(m=1;T=0, \ph)$ 
is the {\it equilibrium complexity}
of hard spheres: if one takes one hard sphere configuration at equilibrium at volume fraction $\ph$, the 
latter will belong with probability 1 to one among $\exp [ N \Si^{HS}_{ eq} ]$ basins.
This quantity is reported in Fig.~16 
of Ref.~\cite{PZ10} and in the inset of Fig.~\ref{fig:sigma}.
\item 
Consider taking $m=0$. In this case, the basins are not weighted. 
Therefore, $\Si(m=0;T,\ph)$ is just the logarithm of the {\it total} number of basins, of any energy and entropy.
\item
Consider taking $\b \to \io$ first, again in the region of $\ph$ where hard sphere configurations exist. 
Then, the weight becomes
\beq
p_{\vec R_0} \propto v_b[\vec R_0]^m \ ,
\eeq
which means that basins are weighted by their volume of hard spheres configurations raised to the power $m$. 
In particular, those basins that do not contain any hard sphere configuration
(those of positive energy) have zero weight and {\it are not counted}. Now, take $m \to 0$. The resulting weight is
\beq
p_{\vec R_0} \propto \theta\{ v_b[\vec R_0] > 0 \}  \ ,
\eeq
in other words we give uniform weight to all basins that contain at least one hard sphere configuration. 
The corresponding complexity is
\beq
\Si_0^{HS}(\ph) = \lim_{m\to 0} \lim_{T\to 0}  \Si(m;T,\ph )
\eeq
and it counts {\it the total number of basins for hard spheres}.
\item
Finally, consider taking the limit $\b \to \io$, with the effective temperature 
$\tau = T/m$ held constant (meaning that $m\to 0$ at the same time).
In this case, the weight becomes
\beq
p_{\vec R_0} \propto e^{-N e[\vec R_0]/\t} \ .
\eeq
Therefore, $\Si_0(\t,\ph) = \lim_{T \to 0} \Si(m=T/\t ; T, \ph)$ counts the basins 
{\it weighted by their minimal energy at 
temperature $\t$, irrespective of their entropy} 
(in particular basins that contain hard sphere configurations have weight 1 in this case).
Obviously, $\lim_{\t \to 0} \Si_0(\t,\ph)  = \Si_0^{HS}(\ph)$ for $\ph \leq \ph_{ GCP}$. 
Moreover, one can perform a Legendre transformation
with respect to $\t$ to eliminate $\t$ in favor of its conjugate variable (the basins minimal energy) and therefore
obtain $\Si_0(e,\ph)$, which counts the basins of minimal energy $e$ at density $\ph$. 
Again, $\lim_{e \to 0} \Si_0(e,\ph) = \Si_0^{HS}(\ph)$  for $\ph \leq \ph_{ GCP}$,
while for $\ph > \ph_{ GCP}$, one finds that $\Si_0(e,\ph)$ 
vanishes at a positive energy $e_{ GS}(\ph)$ which is the global minimum energy for that density and
increases as $e_{ GS}(\ph) \propto (\ph - \ph_{ GCP})^2$~\cite{JBZ11}.
The quantity $\Si_0(e,\ph)$ is reported in Fig.~\ref{fig:sigmae}. A 
similar object was recently measured numerically~\cite{XFL11}.
\end{enumerate}

\begin{widetext}

\section{Some calculations and simplications}

\subsection{The function $q$}
\label{app_qA}

The function $q(A,T;r)$ that appears in the effective potential Eq.~(\ref{def_effpot})
is given, using the definition and introducing bipolar coordinates, by:
\beq\label{qAdef}
q(A,T;r)  = \int d^3r' \g_{2A}(\vec r') e^{-\b \f(\vec r-\vec r')} =
 \frac{1}{r \sqrt{4 \p A}} \int_0^\io du ~ u \left[ e^{-\frac{(r-u)^2}{4A}} - e^{-\frac{(r+u)^2}{4A}} \right]
e^{-\b \f(u)} \ .
\eeq
The Gaussian integral can be explicitly computed and the result is
\begin{align}
q(A,T;r) &= \frac12 \left(2+\text{Erf}\left[\frac{r-1}{2 \sqrt{A}}\right]- \text{Erf}\left[\frac{r+1}{2 \sqrt{A}}\right] \right) +\frac{4 A^{3/2} \b}{(1+4 A \beta)\sqrt{\pi} r }
\left( e^{-\frac{(r-1)^2 }{4 A }} -e^{-\frac{(r+1)^2}{4 A }} \right) \nonumber \\
&+ e^{-\frac{(r-1)^2 \beta }{1+4 A \beta }} \frac{ (r+4 A \beta )}{2r (1+4 A \beta )^{3/2} }  \left(\text{Erf}\left[\frac{r+4 A \beta }{2 \sqrt{A (1+4 A \beta )}}\right] + \text{Erf}\left[\frac{1-r}{2 \sqrt{A (1+4 A \beta )}}\right]\right) \nonumber \\
&- e^{-\frac{(r+1)^2 \beta }{1+4 A \beta }}  \frac{(r-4 A \beta )}{2 r (1+4 A \beta)^{3/2}} \left( \text{Erf}\left[\frac{r-4 A \beta }{2 \sqrt{A (1+4 A \beta )}}\right] - \text{Erf}\left[\frac{1+r}{2 \sqrt{A (1+4 A \beta )}}\right] \right). 
\label{def_qa}
\end{align}

\subsection{Link with the M\'ezard-Parisi small cage expansion}
\label{app:SC}

Here we show that our effective potential approximation reproduces the result of \cite{MP99b} 
in a small $A$ expansion at fixed $\b$. Starting from Eq.~(\ref{qAdef}), we note that for small $A$,
$\vec r'$ is small, and we expand
the potential $\f(\vec r-\vec r')$ (which is assumed to be differentiable) for small $\vec r'$:
\begin{align}
q(A,T;r) & = \int d^3r' \g_{2A}(\vec r') e^{-\b \f(\vec r-\vec r')} , \\ 
& \sim e^{-\b \f(r)} [ 1 -\b A \D \f(r)
+ \b^2 A (\nabla \f(r))^2 + O(A^2)] . \nonumber
\label{qAMP}
\end{align}
Plugging this in (\ref{Sfirstorder}) and using an integration by parts, one easily obtains:
\beq
G(m,A;T) = 3 \b A (1-m) \int_0^1 dr r^2 e^{-\b m \f(r)} \D \f(r) \ .
\eeq
This result, together with Eq.~(\ref{defSS}), reproduces the result of~\cite{MP99b}, 
with the additional approximation we made for the correlation function of the liquid,
Eq.~(\ref{y_approx_rep}). Note that without the latter approximation, we would not reproduce
the result of~\cite{MP99b} exactly; this would require an expansion around the center of mass
of the molecule and not around replica 1 as we did in Eq.~(\ref{effpot_def}).

\subsection{Simplification of the replicated free entropy in the low-temperature approximation}
\label{app:A3}

We show here the details of the simplification of 
the replicated free entropy in the low-temperature approximation, Eq.~(\ref{effpot_def}).
Starting from Eq.~(\ref{S_of_Q}) and using Eq.~(\ref{effpot_def}), we get
\begin{align}
\int d^3r g_{liq}(T/m,\ph; r) Q(r) & 
\sim y^{HS}_{liq}(\ph) \int dr e^{-\b m \f(r)} Q(r) \equiv y^{HS}_{liq}(\ph) \frac{4 \p}{3} G(m,A;T) \ ,
\end{align}
where:
\begin{align}
G(m,A;T) & = \frac{3}{4 \p} \int d^3r ~e^{-\b m \f(r)} Q(r) = 3 \int_0^\io dr \, r^2 [e^{-\b \f_{eff}(r)}-e^{-\b m \f(r)}] .
\end{align}
Then we observe that, using Eqs.~(\ref{rhoGauss}) and (\ref{effpot_def}), we get
\begin{align}
& \int d^3r \, e^{-\b \f_{eff}(\vec r)} = \frac1V \int d\vec x_1 d\vec y_1 e^{-\b \f_{eff}(\vec x_1-\vec y_1)} \nonumber \\ 
& = \frac1V \int d^3x_{1}\cdots d^3x_{m} d^3y_{1} \cdots d^3y_{m} \r(\vx_1 \cdots \vx_m) \r(\vec y_1 \cdots \vec y_m)  
\prod_{a=1}^m e^{-\beta \f(\vec x_a - \vec y_a)} , \nonumber \\
& \equiv  \frac1V \int d^3X d^3Y q(A,T;\vec X-\vec Y)^m = \int d^3r \, q(A,T;\vec r)^m \ ,
\end{align}
therefore we obtain
\beq
\label{Gdef_app}
G(m,A;T) = 3 \int_0^\io dr \, r^2 [q(A,T;r)^m - e^{-\b m \f(r)}] \ ,
\eeq
which leads to the desired result, Eqs.~(\ref{defSS}) and (\ref{Sfirstorder}).

\subsection{Energy and pressure of the glass}
\label{app:Uglass}

Here we show how to compute the energy and pressure of the glass, which can be computed using
the general thermodynamic relations
$U = - d\SS/d\b$ and $p = \b P/\r = -\ph d\SS/d\ph$, where $p$ is the so-called ``reduced pressure''
or ``compressibility factor''. 
From expression 
(\ref{def_SS_glass}), together with (\ref{defSS}) and (\ref{Sfirstorder}), we have that
$\SS_{glass}(T,\ph) = \SS(m^*,A^*;T,\ph)/m^*$, where 
$m^*$ and $A^*$ are determined by optimization of
$\SS(m,A;T,\ph)/m$. Therefore, we do not need to take explicit derivatives with respect to $m$ and $A$.

We get for the energy:
\beq\nonumber
\begin{split}
U_{glass}(T,\ph) & = - \frac{1}{m^*} \frac{\partial \SS(m^*,A^*;T,\ph)}{\partial \b} \\ 
&= -\frac{\partial \SS_{liq}(T/m^*,\ph)}{\partial (\b m^*)} - 12 \ph y^{HS}_{liq}(\ph) \int_0^\io dr \, 
r^2 \left[ q(A^*,T;r)^{m^*-1} \frac{\partial q(A^*,T;r)}{\partial \b} + \phi(r) e^{-\b m^* \phi(r)} \right]
 \\
&= U_{liq}(T/m^*,\ph) - 12 \ph y^{HS}_{liq}(\ph) \int_0^\io dr \, 
r^2  \phi(r) e^{-\b m^* \phi(r)} - 12 \ph y^{HS}_{liq}(\ph) \int_0^\io dr \, 
r^2 q(A^*,T;r)^{m^*-1} \frac{\partial q(A^*,T;r)}{\partial \b} \\
&= - 12 \ph y^{HS}_{liq}(\ph) \int_0^\io dr \, 
r^2 q(A^*,T;r)^{m^*-1} \frac{\partial q(A^*,T;r)}{\partial \b} ,
\end{split}\eeq
where we made use of Eq.~(\ref{y_approx}) for the liquid energy.
Now we make use of the definition of $q$, Eq.~(\ref{qAdef}), to get
\beq
- \frac{\partial q(A,T;r)}{\partial \b} = 
\frac{1}{r \sqrt{4 \p A}} \int_0^\io du ~ u \f(u) \left[ e^{-\frac{(r-u)^2}{4A}} - e^{-\frac{(r+u)^2}{4A}} \right]
e^{-\b \f(u)} .
\eeq
We plug this in the equation above, and we get (exchanging the names of $u$ and $r$ in the integral):
\beq\begin{split}
U_{glass}(T,\ph) & = 12 \ph y^{HS}_{liq}(\ph) \int_0^\io dr \, 
r^2 \f(r) \ e^{-\b \f(r)} \frac{1}{r \sqrt{4 \p A}} \int_0^\io du \ u \   
 \left[ e^{-\frac{(r-u)^2}{4A^*}} - e^{-\frac{(r+u)^2}{4A^*}} \right] q(A^*,T;u)^{m^*-1} \\
& = 12 \ph y^{HS}_{liq}(\ph) \int_0^\io dr \, 
r^2 \f(r) \ e^{-\b \f_{eff}(r)} 
\end{split}\eeq
where we used the definition (\ref{def_effpot}). This is the desired result, Eq.~(\ref{def_U_glass}).

For the pressure, we get the simple result
\beq\begin{split}
p_{glass}(T,\ph) & = -\frac{\ph}{m^*} \frac{\partial \SS(m^*,A^*;T,\ph)}{\partial \ph} \\
& = \frac{1}{m^*} p_{liq}(T/m^*,\ph) - \frac{4 \ph}{m^*} \left[ y^{HS}_{liq}(\ph) + \ph \frac{d  y^{HS}_{liq}(\ph)}{d\ph} \right]
G(m^*,A^*,T) .
\end{split}\eeq

\subsection{Simplified expression of the number of contacts and of the energy}
\label{app_z}

Here we derive a simplified expression for the number of contacts. Using Eq.~(\ref{effpot_def}) we obtain
\beq\begin{split}
z(T,\ph) & =  24 \ph y^{HS}_{liq}(\ph) \int_0^1 dr \, r^2 e^{-\b \f_{eff}(r)}  \\
& = 24 \ph y^{HS}_{liq}(\ph) \int_0^\io du \, u^2 q(A,T;u)^{m-1}
 \frac{1}{u \sqrt{4 \p A}} \int_0^1 dr \, r \left[ e^{-\frac{(r-u)^2}{4A}} - e^{-\frac{(r+u)^2}{4A}} \right]
e^{-\b (1-r)^2} \\
& = 24 \ph y^{HS}_{liq}(\ph) \int_0^\io du \, u^2 q(A,T;u)^{m-1} \, \widehat q(A,T;u),
\end{split}\eeq
where we defined the function
\beq
\widehat q(A,T;u) = \frac{1}{u \sqrt{4 \p A}} \int_0^1 dr \, r \left[ e^{-\frac{(r-u)^2}{4A}} - e^{-\frac{(r+u)^2}{4A}} \right]e^{-\b (1-r)^2}.
\eeq
Note that $\widehat q(A,T;u)$ is equal to the part of $q(A,T;u)$
that vanishes in the zero temperature limit. This allows to compute
it starting from Eq.~(\ref{def_qa}):
\beq\begin{split}
\widehat q(A,T;r) & = q(A,T;r) - \lim_{T\to 0} q(A,T;r)  \\
&= - \sqrt{\frac{A}{\pi}} \frac{1}{(1+4 A \beta) r }
\left( e^{-\frac{(r-1)^2 }{4 A }} -e^{-\frac{(r+1)^2}{4 A }} \right)  \\
&+ e^{-\frac{(r-1)^2 \beta }{1+4 A \beta }} \frac{ (r+4 A \beta )}{2r (1+4 A \beta )^{3/2} }  \left(\text{Erf}\left[\frac{r+4 A \beta }{2 \sqrt{A (1+4 A \beta )}}\right] + \text{Erf}\left[\frac{1-r}{2 \sqrt{A (1+4 A \beta )}}\right]\right)  \\
&- e^{-\frac{(r+1)^2 \beta }{1+4 A \beta }}  \frac{(r-4 A \beta )}{2 r (1+4 A \beta)^{3/2}} \left( \text{Erf}\left[\frac{r-4 A \beta }{2 \sqrt{A (1+4 A \beta )}}\right] - \text{Erf}\left[\frac{1+r}{2 \sqrt{A (1+4 A \beta )}}\right] \right). 
\end{split}
\eeq

Similarly, for the energy of the glass we get
\beq\begin{split}
U_{glass}(T,\ph) & =  12 \ph y^{HS}_{liq}(\ph) \int_0^1 dr \, r^2 (1-r)^2 e^{-\b \f_{eff}(r)}  \\
& = 12 \ph y^{HS}_{liq}(\ph) \int_0^\io du \, u^2 q(A,T;u)^{m-1} 
 \frac{1}{u \sqrt{4 \p A}} \int_0^1 dr \, r (1-r)^2 \left[ e^{-\frac{(r-u)^2}{4A}} - e^{-\frac{(r+u)^2}{4A}} \right]
e^{-\b (1-r)^2} \\
& = 12 \ph y^{HS}_{liq}(\ph) \int_0^\io du \, u^2 q(A,T;u)^{m-1} \, \left( -\frac{\partial \widehat q(A,T;u)}{\partial \b} \right).
\end{split}\eeq

\section{The hard sphere limit}
\label{app:HS}

\subsection{Free energy of hard spheres}

Here we discuss how Eq.~(\ref{Sfirstorder}) is simplified in the limit $T \to 0$, with fixed $m$ and $A$, that is the appropriate one 
below $\ph_{GCP}$.
Taking the limit $T \to 0$ at fixed $A$, Eq.~(\ref{def_qa}) 
gives back the results of Parisi and Zamponi~\cite[Eqs.~(C4)]{PZ10}: 
\begin{align}
q(A;r) = \lim_{T \to 0} q(A,T;r)
 =& \frac{1}{2} \left( 2 + \text{Erf}\left[ \frac{r-1}{2 \sqrt{A}} \right] - \text{Erf}\left[ \frac{r+1}{2 \sqrt{A}} \right] \right) + \sqrt{\frac{A}{\pi}} \frac{1}{r} \left( e^{-\frac{(r-1)^2}{4A}} - e^{-\frac{(r+1)^2}{4A}} \right).
\label{qAHS}
\end{align}
In this limit we obtain from Eq.~(\ref{Sfirstorder}) the following result:
\beq\label{SSHSfin}
\begin{split}
& \SS(m,A;\ph) = S_{h}(m,A) + \SS^{HS}_{liq}(\ph) + 4 \ph y^{HS}_{liq}(\ph) G(m,A) \\
& G(m,A) = 3 \int_0^\io dr r^2 [q(A;r)^m - \th(r-1)],
\end{split}
\eeq
These equations coincide with the ones reported in~\cite[Eq.~(72)]{PZ10}.

\subsection{Jamming limit from below}

The jamming limit from the hard sphere side $\ph \to \ph_{GCP}^-$ is obtained by
taking $m\to 0$ and $A = \a m$ \cite{PZ10}.
In this limit, the function $q(A;r)$ goes to $1$ for $r>1$ and to $0$ for $r<1$.
Using the asymptotic expansions of the error function we get:
\beq\label{qAHSsmallA}
q(A;r) \sim \frac{1}{2} \left( 1 + \text{Erf}\left[ \frac{r-1}{2 \sqrt{A}} \right] 
\right) + \sqrt{\frac{A}{\pi}} \frac{1}{r}  e^{-\frac{(r-1)^2}{4A}} 
\sim \theta(r-1) +
\sqrt{\frac{A}{\pi}} e^{-\frac{(r-1)^2}{4A}} \left[ \frac{1}{r} + \frac1{1-r} R\left(\frac{|1-r|}{\sqrt{A}}\right) 
\right] ,
\eeq
where the function $R(x)$ goes to $1$ for $x\to \io$ and is proportional to $x$ for $x\to 0$,
in such a way that there is no singularity at $r=1$.

Based on this result, we can evaluate the integral that defines $G(m,A)$ in Eq.~(\ref{SSHSfin}).
For $r>1$, we have $q(A;r) \to 1$ and therefore
\beq
\int_1^\io dr r^2 [ q(m \a ;r)^m - 1] = \int_1^\io dr r^2 \{ m [ q(m \a;r) -1] + \cdots \}
= S_1 m^{2} + S_2 m^{5/2} + \cdots 
\eeq
For $r<1$, we have instead
\beq
\int_0^1 dr r^2 q(m \a ;r)^m \sim \left( \sqrt{\frac{\a m}{\pi}} \right)^m \int_0^1 dr r^2
 e^{-\frac{(r-1)^2}{4\a}} \left[ \frac{1}{r} + \frac1{1-r} R\left(\frac{1-r}{\sqrt{A}}\right) \right]^m .
\eeq
The last term in square brackets in the integral gives rise to a regular expansion in powers
of $m$ for small $m$. Collecting all these results together we get
\beq\begin{split}
&G(m,\a m) \sim m^{m/2} G^{HS}_0(\a) (1 + R_1 m + R_2 m^2 + \cdots) +  S_1 m^{2} + S_2 m^{5/2} + \cdots  \ ,\\
&G_0^{HS}(\a) = 3 \int_0^1 r^2 e^{-\frac{(1-r)^2}{4 \a}} dr \ .
\end{split}\eeq

Then, in the limit $m \to 0$, Eqs.~(\ref{SSHSfin}) reduce to:
\beq\label{eq:Galfadef}
\begin{split}
& \SS^{HS}_0(\alpha;\ph) = -\frac{3}{2} \left( \ln (2 \pi \alpha) + 1 \right) + \SS^{HS}_{liq}(\ph) + 4 \ph y^{HS}_{liq}(\ph) G_0^{HS}(\a)  \ , \\
& G_0^{HS}(\a) = 3 \int_0^1 r^2 e^{-\frac{(1-r)^2}{4 \a}} dr = 3 \big[ \sqrt{\pi \alpha} (1+2\alpha) \text{Erf}\left(\frac{1}{2 \sqrt{\alpha}}\right) + 2 \alpha e^{-\frac{1}{4 \alpha}} - 4 \alpha \big]  \ . \\
\end{split}
\eeq
Optimization over $\a$ yields an equation for $\a^*_{HS}(\ph)$:
\begin{align}
& J^{HS}_0(\a^*_{HS}) = \frac{3}{8 \ph y^{HS}_{liq}(\ph)} ,
\hskip20pt \text{with} \hskip20pt
 J^{HS}_0(\a) = \a \frac{d G^{HS}_0(\a)}{d \a}.
\end{align}

Next we calculate the leading order in the small $m$ expansion around the previous result. We have
\beq
\SS(m,\a;\ph) = \SS^{HS}_0(\alpha;\ph) + \frac{1}{2} \big[ 3 + 4 \ph y^{HS}_{liq}(\ph) G_0^{HS}(\a) \big] \, m \log m  
+ O(m, (m \log m)^2, \cdots).
\eeq
Therefore for the complexity we get at the leading order (recall that we do not need to add the
explicit derivative with respect to $\a$, 
since $\a$ is defined by the condition that it vanishes):
\beq
\Si(m,\a;\ph) = -m^2 \frac{d}{dm} [ \SS(m,\a;\ph)/m ] = 
\SS^{HS}_0(\alpha;\ph) - \frac{m}{2} \big[ 3 + 4 \ph y^{HS}_{liq}(\ph) G_0^{HS}(\a) \big]  \ .
\eeq
The expression above has to be computed in $\a = \a^*_{HS}(\ph)$.
Therefore, $\SS_0^{HS}\left( \a^*(\ph);\ph \right) = \Sigma^{HS}_0(\ph)$ 
is the complexity of hard-spheres at $m=0$ \cite{PZ10}, which vanishes linearly
at $\ph = \ph_{GCP}$.
Since $m^*_{HS}(\ph)$ is defined by $\Si(m,\a^*_{HS}(\ph);\ph) =0$, we get
the leading term of $m^*_{HS}(\ph)$ for $m \to 0$ or $\ph \to \ph_{GCP}^-$:
\beq\label{mstarHSsmall}
m^*_{HS}(\ph) = \frac{2 \,\Sigma^{HS}_0(\ph) }{ 3 + 4 \ph y^{HS}_{liq}(\ph) G_0^{HS}(\a^*_{HS}(\ph)) }.
\eeq

\subsection{Scaling of the effective potential on approaching jamming from below}
\label{app:HSpeak}

Here we discuss the emergence of the delta peak at $r=1^+$ in the effective potential of hard
spheres for $\ph \to \ph_{GCP}^-$~\cite{PZ10}, where $A = m \a$ and $m\to 0$ as discussed
above. For small $A$, we have
\beq
e^{-\b \f_{eff}(r)} = \th(r-1) \frac{1}{r \sqrt{4 \p A}} \int_0^\io du ~ u  e^{-\frac{(r-u)^2}{4A}}  q(A;u)^{m-1} \ , 
\eeq
with $q(A;r)$ given in (\ref{qAHSsmallA}). First we note that if $r - 1 \gg \sqrt{A}$, then the integral is
peaked at $u > 1$, the function $q \to 1$ and the integral goes to 1 too. 
Therefore $e^{-\b \f_{eff}(r)} \to 1$ for $r - 1 \gg \sqrt{A}$.

We have then to study the region $r - 1 \sim \sqrt{A}$. In this region one can easily show that the part
of the integral at $u>1$ gives a finite contribution. If we want to study the divergent part, 
we can truncate the integral at $u=1$.
We then introduce the scaling variable $\l = (r-1)/\sqrt{4 m A} = (r-1)/(m \sqrt{4 \a}) > 0$ and
in the integral we change variable to $u = 1-x \sqrt{4 A/m} = 1- x \sqrt{4 \a}$. 
Using the first equation in (\ref{qAHSsmallA}), and defining
\beq\label{Thdef}
\Th(s) = \frac12 \left[ 1 + \text{Erf}(s) \right] \ ,
\eeq
we get for the divergent part
\beq
e^{-\b \f^{div}_{eff}(\l)} =  \frac{1}{\sqrt{\p m}} \int_0^{1/\sqrt{4 \a}} dx \, (1- x \sqrt{4 \a}) \, e^{-  (\l \sqrt{m} + x/\sqrt{m}    )^2   } 
\left[ \Th\left( -\frac{x}{\sqrt{m}} \right) 
 + \sqrt{\frac{m}{\pi}} \frac{\sqrt{\a}}{1- x \sqrt{4 \a}}  e^{-\frac{x^2}{m}} \right]^{m-1}
 \ .
\eeq
Making use of the asymptotic large $s$ expansion of the error function, we have
$\Th(-s) \sim e^{ -s^2}/(2 \sqrt{\pi} s)$, and we get
\beq\begin{split}
e^{-\b \f^{div}_{eff}(\l)} & =  \frac{e^{-m \l^2}}{\sqrt{\p m}} \int_0^{1/\sqrt{4 \a}} dx \, (1- x \sqrt{4 \a}) \, e^{-  2 \l x - x^2 } 
\left[ \sqrt{\frac{m}{\pi} }\frac{1}{2 x}
 + \sqrt{\frac{m}{\pi}} \frac{\sqrt{\a}}{1- x \sqrt{4 \a}}   \right]^{m-1} \\
 & \sim \frac{1}{m} \int_0^{1/\sqrt{4 \a}} dx \, (1- x \sqrt{4 \a}) \, e^{-  2 \l x - x^2 } 
\left[  \frac{1}{2 x}
 +  \frac{\sqrt{\a}}{1- x \sqrt{4 \a}}   \right]^{-1} \\
& = \frac1m \D_\a(\l)  \ ,
\end{split}\eeq
where we introduced the scaling function that describes the delta peak:
\beq\label{DeltaExact}
\D_\a(\l) = 2 \int_0^{1/\sqrt{4 \a}} dx \, x \, (1- x \sqrt{4 \a})^2 \, e^{-  2 \l x - x^2 } 
\eeq
Note that at $\ph_{GCP}$ we have $\a \sim 10^{-4}$; therefore an excellent approximation of the scaling
function is obtained by setting $\a=0$:
\beq
\D_0(\l) = 2 \int_0^{\io} dx \, x \, e^{-  2 \l x - x^2 } = 1 - \sqrt{\pi} \l e^{\l^2} (1 - \text{Erf}(\l))  \ ,
\eeq
that allows to recover the results of~\cite[Eq.~(C40)]{PZ10}.

\section{The zero temperature soft sphere limit}
\label{app:SS}

\subsection{Free energy of soft spheres at $T=0$}

When $\ph > \ph_{GCP}$, the appropriate $T\to 0$ limit is taken with
$m = T/\t$ and $A = m\a$ for constants $\a$ and $\t$.
We start by looking at the asymptotics of $q(\a m, \t m ; r)$ when $m$ goes to 0. 
From the definition (\ref{def_qa}) we find that:
\beq\label{qjabove}
q(\a m, \t m ; r) \underset{m \rightarrow 0}{\sim}  
\left\{
\begin{array}{ll}
 e^{-\frac{\f(r)}{m (\t + 4 \alpha)}} \frac{r + 4 \alpha / \t}{r ( 1 + 4 \alpha / \t )^{3/2}} & \hskip20pt r<1 \\
1 & \hskip20pt r>1 \\
\end{array}
\right.
\eeq
Therefore the part of the integrals in $G(m,A;T)$ for $r>1$ vanishes in this limit.
Plugging Eq.~(\ref{qjabove}) in Eq.~(\ref{Gdef_app}) we get:
\begin{align}
& G(m , \a m ; \t m) \underset{m \to 0}{\sim} 3 \int_0^1 r^2 \left[ e^{-\frac{\f(r) / \t}{1 + 4 \alpha / \t }} - e^{- \f(r) / \t} \right] dr \ ,
\end{align}
which can be written, using the definition (\ref{eq:Galfadef}), as:
\begin{align}
G(m,\a m; \t m) \underset{m \to 0}{\sim} G_0(\a,\t) = G_0^{HS}(\alpha + \t / 4) - G_0^{HS}(\t / 4) \ .
\end{align}
The free entropy is given by Eq.~(\ref{defSS}). 
Replacing $T = \t m$ and $A = \alpha m$ and taking the $m \rightarrow 0$ limit, we obtain:
\begin{align}
\SS_0(\alpha,\t;\ph) & = -\frac{3}{2} \left( \ln (2 \pi \alpha) + 1 \right) + \SS_{liq}(\t,\ph) + 4 \ph y^{HS}_{liq}(\ph) G_0(\a,\t).
\label{eq:def_Psi_alpha_mu}
\end{align}
As before, optimization over $\a$ yields an equation for $\a^*(\t,\ph)$ for each $\t$ and $\ph$:
\begin{align}
& J_0(\a^*(\t,\ph),\t) = \frac{3}{8 \ph y^{HS}_{liq}(\ph)} , \nonumber \\
& J_0(\a,\t) = \a \frac{d G_0}{d \a}(\a,\t).
\label{optim_A_T=0}
\end{align}
In this limit, the free energy is much simplified and the optimization over $\a$ and $\t$ is easily carried out.

\subsection{Small $\t$ expansion: jamming from above}
\label{app_smalltau}

A very important observation is that, taking the $\t \rightarrow 0$ limit, we recover the hard-sphere value 
of $\SS^{HS}_0(\a;\ph)$ given in Eq.~(\ref{eq:Galfadef}). 
Indeed we expect that $\t$ tends to zero when approaching $\ph_{GCP}$ from above. Therefore, to expand around the jamming point, we must
perform a small $\t$ expansion of the free entropy (\ref{eq:def_Psi_alpha_mu}).

We start from (\ref{eq:def_Psi_alpha_mu}) and make use of the expressions (\ref{dev_energy}) and (\ref{eq:Galfadef}).
We note that some remarkable cancellations happen, that eliminate a dangerous term in $\sqrt{\t}$.
Keeping only the lowest orders in the expression (\ref{eq:def_Psi_alpha_mu}) of the free entropy, we get:
\begin{align}
& \SS_0(\a,\t;\ph) = \SS_0^{HS}(\a;\ph) +  \t\SS_1(\a,\ph) + o(\t), \nonumber \\
& \SS_1(\a,\ph) = \ph y^{HS}_{liq}(\ph) \left[ \frac{1}{ \a} J_0^{HS}(\a) - 6 \right].
\label{eq:dev_gcp}
\end{align}
The optimal value of $\a$ is defined by :
\begin{equation}
\frac{d \SS_0^{HS}}{d \a}\left(\a^*(\t,\ph);\ph\right) + \t \frac{\partial \SS_1}{\partial \a} \left( \a^*(\t,\ph),\ph \right) = 0 \ ,
\end{equation}
and it has the form
\begin{equation}
\a^*(\t,\ph) = \a^*_{HS}(\ph) + \t b(\ph) + o(\t).
\end{equation}
Recalling that $\a^*_{HS}$ maximizes $\SS_0^{HS}(\a;\ph)$, therefore $\partial_\a \SS_0^{HS}(\a;\ph)=0$ in $\a=\a^*_{HS}$,
we see that the $\t$-dependence of $\a^*$ does not contribute at lowest order. Thus:
\begin{equation}
\SS_0 \left(\a^*(\t,\ph),\t;\ph \right) = \SS_0^{HS}\left( \a^*_{HS}(\ph);\ph \right) + \t \SS_1\left(\a^*_{HS}(\ph),\ph \right) + o(\t).
\end{equation}
Remember that $\SS_0^{HS}\left( \a^*(\ph);\ph \right) = \Sigma^{HS}_0(\ph)$
is the complexity of hard-spheres at $m=0$ (see Appendix~\ref{app:HS}).
We define $\SS_1(\ph) = \SS_1 \left(\a^*_{HS}(\ph),\ph \right) $ and we obtain
\begin{equation}
\SS_0 \left(\t;\ph \right) = \Sigma_0^{HS}(\ph) + \t \SS_1 (\ph) + o(\t) \ .
\end{equation}
Using Eq.~(\ref{eq:zeroTLeg}), we obtain
\beq\label{eq:tausmall}
\begin{split}
& \Sigma_0(\t,\ph) = \Sigma_0^{HS}(\ph) + 2\t \SS_1(\ph) , \\
& e(\t,\ph) = \t^2 \SS_1(\ph).
\end{split}\eeq
Finally, reconstructing this parametric equation 
defining $\Sigma_0$, we obtain :
\begin{equation}
\Sigma_0(e,\ph) = \Sigma_0^{HS}(\ph) + 2 \sqrt{e \SS_1(\ph)}.
\end{equation}

\subsection{Effective potential in the zero temperature limit for soft spheres}
\label{app_effpot_soft}

We start from the definition (\ref{def_effpot}) of the effective potential and we take the limit $T\to 0$ with $m = T/\t$ and $A = \a m$.
A direct inspection of the different terms shows that, since $A$ is very small, one can discard a number of terms to obtain:
\begin{align}
\label{effpot_big}
& e^{-\b \f_{eff}(r)} = e^{-\b \f(r)} \frac{1}{r \sqrt{4 \p A}} \int_0^\io du ~ u  e^{-\frac{(r-u)^2}{4A}}  q(A,T;u)^{m-1} \ , \\
& q(A,T;u) \sim \Th\left( \frac{u-1}{\sqrt{4A}} \right) + e^{-\frac{\b(u-1)^2}{1+4A\b}} \frac{u+4A\b}{u(1+4A\b)^{3/2}} \Th\left( \frac{1-u}{\sqrt{4A(1+4A\b)}} \right) ,
\end{align}
with $\Th(s)$ defined in Eq.~(\ref{Thdef}).
We should now make a separate analysis for $r>1$ and $r<1$. In both cases, the integral is dominated by $u \sim r$. Therefore,
for $r>1$ also $u>1$ and in this case $q(A,T;u)\sim 1$, so we get $e^{-\b \f_{eff}(r)} \to 1$.

The more interesting case is for $r<1$. We have
\begin{align}\label{appC2effpot}
e^{-\b\f_{eff}(r)}=e^{-\b (1-r)^2}  \int_0^\io du \frac{u e^{-\frac{(r-u)^2}{4A}}}{r \sqrt{4 \p A}} \left[ \Th\left( \frac{u-1}{\sqrt{4A}} \right) + e^{-\frac{\b(u-1)^2}{1+4A\b}} \frac{u+4A\b}{u(1+4A\b)^{3/2}} \Th\left( \frac{1-u}{\sqrt{4A(1+4A\b)}} \right) \right]^{m-1}  \ ,
\end{align}
For $r<1$, the integral over $u$ is dominated by the decay of the function away from $u=1$ (for $u<1$).
We recall that $\Th(s \to -\io) \sim e^{-s^2}$, while $\Th(s \to\io) \to 1$.
Therefore, inside the square brackets we can 
neglect the first term $\Th\left( \frac{u-1}{\sqrt{4A}} \right)$ whose decay is faster.
We get, neglecting all the terms that are small for $m\to 0$
\begin{align}
e^{-\b\f_{eff}(r)}=e^{-\b (1-r)^2}  \frac{1}{r \sqrt{4 \p \a m}} \int_0^\io du ~ u e^{\frac1m \left\{ - \frac{(r-u)^2}{4 \a} + \frac{ (u-1)^2}{\t+4\a} \right\} } e^{-\frac{(u-1)^2}{\t+4\a}} \frac{u(1+4\a/\t)^{3/2}}{u+4\a/\t} \ . 
\end{align} 
The term in the exponential proportional to $1/m$ dominates. Rearranging terms we obtain:
\begin{align}
e^{-\b\f_{eff}(r)} & = e^{- \frac{(1-r)^2}{m \t}}  e^{\frac{(r-1)^2}{m \t}} \frac1r \int_0^\io du ~ u \frac{e^{-\frac{[u+4\a/\t - r(1+4\a/\t)]^2}{4 \a m (1+4\a/\t)}}}{\sqrt{4 \p \a m (1+4 \a/\t)}} e^{-\frac{(u-1)^2}{\t+4\a}} \frac{u(1+4\a/\t)^2}{u + 4\a/\t}  \\
& = \frac1r \int_0^\io du ~ u ~ \d [ u-1-(r-1)(1+4\a/\t) ] e^{-\frac{(u-1)^2}{\t+4\a}} \frac{u(1+4\a/\t)^2}{u+4\a/\t} \\
& =  \th\left( r - \frac{4 \a}{\t+4\a} \right)  \frac{(1+4\a/\t)[1+(r-1)(1+4\a/\t)]^2}{r^2}  e^{-\frac{\t+4\a}{\t^2} (r-1)^2},
\end{align}
since the delta function forces $u=1+ (r-1)(1+4\a/\t)$, but the integral vanishes
unless this value of $u$ is positive. Adding the contribution for $r>1$, we get the final result:
\begin{align}
e^{-\b \f_{eff}(r)} = \th(r-1) + \th(1-r) \th\left( r - \frac{4 \a}{\t+4\a} \right)  \frac{(1+4\a/\t)[1+(r-1)(1+4\a/\t)]^2}{r^2}  e^{-\frac{\t+4\a}{\t^2} (r-1)^2}.
\end{align}

Note that the above expression is discontinuous in $r=1$, while at any finite $T$ the effective potential is continuous.
We can derive the scaling function that describes the emergence of the discontinuity for $T\to 0$ 
as follows. We start from Eq.~(\ref{appC2effpot}) and
we perform a change of variable $u = 1 + x \sqrt{4A}$ in the integral, and define $\l = (r-1)/\sqrt{4A}$. Then we get
in the zero temperature limit
\beq\begin{split}
&e^{-\b \f_{eff}(r)} = e^{-\frac{4 \a}{\t} \l^2 \th(-\l)} \frac{1}{\sqrt{\pi}} \int_{-\io}^\io dx e^{-(\l -x )^2} \frac{1}{q_0(\a,\t;x)}, \\
& q_0(\a,\t;x) = \Th(x) + \frac{1}{\sqrt{1+4\a/\t}} \Th\left(-\frac{x}{\sqrt{1+4 \a/\t}}\right) e^{-\frac{4\a}{\t+4 \a} x^2},
\end{split}\eeq
which provides the scaling function in the region where $\l$ is of order one, and $r-1 \sim \sqrt{A}$.
It is quite easy to check that the function above tends to $1$ for $\l\to\io$, while it tends to 
$1+4\a/\t$ for $\l\to-\io$, therefore interpolating between the two regimes $r>1$ and $r<1$.

\section{Numerical calculations}
\label{app_numbers}

In this Appendix we collect all the numerical results of the various calculations presented in the paper.

\subsection{Hard spheres}

We start from the determination of $\ph_{GCP}$, which is done in the most precise way by looking at the point
where $\Si_0^{HS}(\ph)=0$, see Appendix~\ref{app:HS} and \ref{app_smalltau}. We get
\beq\begin{split}
& \ph_{GCP}=0.633353 \ , \\
& \a_0^{HS}(\ph_{GCP}) = 9.72187 \cdot 10^{-5} \ , \\
\end{split}\eeq
with $y^{HS}_{liq}(\ph_{GCP}) = 23.6238$,
and from Eq.~(\ref{mstarHSsmall}), using $G_0^{HS}(\a^*_{HS}(\ph_{GCP}))= 0.0170908$, we get
\beq
m^*_{HS}(\ph) = \wt \mu (\ph_{GCP} - \ph) = 20.7487 (\ph_{GCP} - \ph).
\eeq
From this and from Eq.~(\ref{def_p_glass}) 
we can get the scaling of the reduced pressure for $\ph \to \ph_{GCP}^-$:
\beq\begin{split}
p_{glass} & \sim \frac{1}{\wt \mu (\ph_{GCP} - \ph)} \left[ p^{HS}_{liq}(\ph_{GCP}) - 4 \ph_{GCP}  
\left( y^{HS}_{liq}(\ph_{GCP}) + \ph_{GCP} \frac{d  y^{HS}_{liq}}{d\ph}(\ph_{GCP}) \right)
 G_0^{HS}(\a^*_{HS}(\ph_{GCP})) \right] \\
&= \frac{3.03430 \, \ph_{GCP}}{\ph_{GCP} - \ph}.
\end{split}\eeq
The prefactor $3.03430$ is extremely close to the space dimension $d=3$, as predicted by free volume
theory; the value $d$ is obtained in the small $A$ expansion~\cite{PZ10}.

\subsection{Soft spheres at zero temperature}

Close to $\ph_{GCP}$ we have $\Sigma_0^{HS}(\ph) = 62.9577 (\ph_{GCP} - \ph)$ 
and $\SS_1(\ph_{GCP}) = 3767.51$, hence we immediately deduce from Eqs.~(\ref{eq:tausmall}) that:
\beq\begin{split}
&\t^*(\ph) = - \Sigma_0^{HS}(\ph) / (2 \SS_1(\ph))
=  0.00835535   (\ph - \ph_{GCP}) \ , \\
&e_{GS}(\ph)  = ( \Sigma_0^{HS}(\ph) )^2/ (4 \SS_1(\ph)) = 0.263017 (\ph - \ph_{GCP})^2 \ .
\end{split}\eeq
The pressure is obtained from Eq.~(\ref{def_p_glass}):
\beq
P_{glass}(T=0,\ph) = \lim_{T\to 0} [ \r \, T \, p_{glass}(T,\ph) ] = \frac{6 \ph}{\pi } \left[ p^{HS}_{liq}(\ph) - 4 \ph  
\left( y^{HS}_{liq}(\ph) + \ph \frac{d  y^{HS}_{liq}}{d\ph}(\ph) \right)
 G_0(\a^*(\ph),\t^*(\ph)) \right]  \t^*(\ph),
\eeq
and expanding around $\ph_{GCP}$ we get
\beq\begin{split}
P_{glass}(T=0,\ph) & =\frac{6 \ph_{GCP}}{\pi }  \left[ p^{HS}_{liq}(\ph_{GCP}) - 4 \ph_{GCP}  
\left( y^{HS}_{liq}(\ph_{GCP}) + \ph_{GCP} \frac{d  y^{HS}_{liq}}{d\ph}(\ph_{GCP}) \right)
 G_0^{HS}(\a^*_{HS}(\ph_{GCP}))  \right]   \t^*(\ph) \\
& = 0.403001 (\ph-\ph_{GCP}).
\end{split}\eeq
Clearly the same scaling could have been obtained from the exact relation $P_{glass}(T=0,\ph) = \frac{6 \ph^2}\pi \frac{de_{GS}}{d\ph}$.

\subsection{Correlation functions}

From these values we can deduce the scaling of the peak of $g_{glass}(r)$
for $\ph \to \ph_{GCP}^+$:
\beq
g_{max}(\ph \to \ph_{GCP}^+) = y^{HS}_{liq}(\ph) \left( 1 + \frac{4 \a^*(\ph)}{\tau^*(\ph)}\right) = \frac{1.09949}{\ph - \ph_{GCP}},
\eeq
and for $\ph \to \ph_{GCP}^-$:
\beq
g_{max}(\ph \to \ph_{GCP}^-) = y^{HS}_{liq}(\ph) \frac{1}{m^*_{HS}(\ph)} \D_{\a^*}(0) = \frac{1.09922}{ \ph_{GCP} - \ph}.
\eeq
Note that even though the prefactors are almost identical numerically, we were not able to prove that they actually coincide.

\end{widetext}

\bibliography{long_biblio}

\begin{thebibliography}{86}
\expandafter\ifx\csname natexlab\endcsname\relax\def\natexlab#1{#1}\fi
\expandafter\ifx\csname bibnamefont\endcsname\relax
  \def\bibnamefont#1{#1}\fi
\expandafter\ifx\csname bibfnamefont\endcsname\relax
  \def\bibfnamefont#1{#1}\fi
\expandafter\ifx\csname citenamefont\endcsname\relax
  \def\citenamefont#1{#1}\fi
\expandafter\ifx\csname url\endcsname\relax
  \def\url#1{\texttt{#1}}\fi
\expandafter\ifx\csname urlprefix\endcsname\relax\def\urlprefix{URL }\fi
\providecommand{\bibinfo}[2]{#2}
\providecommand{\eprint}[2][]{\url{#2}}

\bibitem[{\citenamefont{Bernal}(1959)}]{Be59}
\bibinfo{author}{\bibfnamefont{J.~D.} \bibnamefont{Bernal}},
  \bibinfo{journal}{Nature} \textbf{\bibinfo{volume}{183}},
  \bibinfo{pages}{141} (\bibinfo{year}{1959}).

\bibitem[{\citenamefont{Bernal and Mason}(1960)}]{BM60}
\bibinfo{author}{\bibfnamefont{J.~D.} \bibnamefont{Bernal}} \bibnamefont{and}
  \bibinfo{author}{\bibfnamefont{J.}~\bibnamefont{Mason}},
  \bibinfo{journal}{Nature} \textbf{\bibinfo{volume}{188}},
  \bibinfo{pages}{910} (\bibinfo{year}{1960}).

\bibitem[{\citenamefont{Jaeger et~al.}(1996)\citenamefont{Jaeger, Nagel, and
  Behringer}}]{rmpgrains}
\bibinfo{author}{\bibfnamefont{H.~M.} \bibnamefont{Jaeger}},
  \bibinfo{author}{\bibfnamefont{S.~R.} \bibnamefont{Nagel}}, \bibnamefont{and}
  \bibinfo{author}{\bibfnamefont{R.~P.} \bibnamefont{Behringer}},
  \bibinfo{journal}{Rev. Mod. Phys.} \textbf{\bibinfo{volume}{68}},
  \bibinfo{pages}{1259} (\bibinfo{year}{1996}).

\bibitem[{\citenamefont{Larson}(1999)}]{booklarson}
\bibinfo{author}{\bibfnamefont{R.~G.} \bibnamefont{Larson}},
  \emph{\bibinfo{title}{The Structure and Rheology of Complex Fluids}}
  (\bibinfo{publisher}{Oxford University Press}, \bibinfo{year}{1999}).

\bibitem[{\citenamefont{Hansen and McDonald}(1986)}]{hansen}
\bibinfo{author}{\bibfnamefont{J.~P.} \bibnamefont{Hansen}} \bibnamefont{and}
  \bibinfo{author}{\bibfnamefont{I.~R.} \bibnamefont{McDonald}},
  \emph{\bibinfo{title}{Theory of simple liquids}}
  (\bibinfo{publisher}{Academic Press}, \bibinfo{address}{London},
  \bibinfo{year}{1986}).

\bibitem[{\citenamefont{Aste and Weaire}(2000)}]{bookpursuit}
\bibinfo{author}{\bibfnamefont{T.}~\bibnamefont{Aste}} \bibnamefont{and}
  \bibinfo{author}{\bibfnamefont{D.}~\bibnamefont{Weaire}},
  \emph{\bibinfo{title}{The pursuit of perfect packing}}
  (\bibinfo{publisher}{Taylor and Francis}, \bibinfo{year}{2000}).

\bibitem[{\citenamefont{Torquato and Stillinger}(2010)}]{TS10}
\bibinfo{author}{\bibfnamefont{S.}~\bibnamefont{Torquato}} \bibnamefont{and}
  \bibinfo{author}{\bibfnamefont{F.~H.} \bibnamefont{Stillinger}},
  \bibinfo{journal}{Rev. Mod. Phys.} \textbf{\bibinfo{volume}{82}},
  \bibinfo{pages}{2633} (\bibinfo{year}{2010}).

\bibitem[{\citenamefont{Edwards and Oakeshott}(1989)}]{EO89}
\bibinfo{author}{\bibfnamefont{S.~F.} \bibnamefont{Edwards}} \bibnamefont{and}
  \bibinfo{author}{\bibfnamefont{R.~B.~S.} \bibnamefont{Oakeshott}},
  \bibinfo{journal}{Physica A} \textbf{\bibinfo{volume}{157}},
  \bibinfo{pages}{1080} (\bibinfo{year}{1989}).

\bibitem[{\citenamefont{Edwards}(1998)}]{Ed98}
\bibinfo{author}{\bibfnamefont{S.~F.} \bibnamefont{Edwards}},
  \bibinfo{journal}{Physica A} \textbf{\bibinfo{volume}{249}},
  \bibinfo{pages}{226} (\bibinfo{year}{1998}).

\bibitem[{\citenamefont{Song et~al.}(2008)\citenamefont{Song, Wang, and
  Makse}}]{SWM08}
\bibinfo{author}{\bibfnamefont{C.}~\bibnamefont{Song}},
  \bibinfo{author}{\bibfnamefont{P.}~\bibnamefont{Wang}}, \bibnamefont{and}
  \bibinfo{author}{\bibfnamefont{H.~A.} \bibnamefont{Makse}},
  \bibinfo{journal}{Nature} \textbf{\bibinfo{volume}{453}},
  \bibinfo{pages}{629} (\bibinfo{year}{2008}).

\bibitem[{\citenamefont{Henkes and Chakraborty}(2009)}]{HC09}
\bibinfo{author}{\bibfnamefont{S.}~\bibnamefont{Henkes}} \bibnamefont{and}
  \bibinfo{author}{\bibfnamefont{B.}~\bibnamefont{Chakraborty}},
  \bibinfo{journal}{Phys. Rev. E} \textbf{\bibinfo{volume}{79}},
  \bibinfo{pages}{061301} (\bibinfo{year}{2009}).

\bibitem[{\citenamefont{Berthier and Biroli}(2011)}]{BB10}
\bibinfo{author}{\bibfnamefont{L.}~\bibnamefont{Berthier}} \bibnamefont{and}
  \bibinfo{author}{\bibfnamefont{G.}~\bibnamefont{Biroli}},
  \bibinfo{journal}{Rev. Mod. Phys.} \textbf{\bibinfo{volume}{83}},
  \bibinfo{pages}{587} (\bibinfo{year}{2011}).

\bibitem[{\citenamefont{Liu and Nagel}(1998)}]{LN98}
\bibinfo{author}{\bibfnamefont{A.~J.} \bibnamefont{Liu}} \bibnamefont{and}
  \bibinfo{author}{\bibfnamefont{S.~R.} \bibnamefont{Nagel}},
  \bibinfo{journal}{Nature} \textbf{\bibinfo{volume}{396}}, \bibinfo{pages}{21}
  (\bibinfo{year}{1998}).

\bibitem[{\citenamefont{Liu et~al.}(2011)\citenamefont{Liu, Nagel, van
  Saarloos, and Wyart}}]{LNSWW10}
\bibinfo{author}{\bibfnamefont{A.~J.} \bibnamefont{Liu}},
  \bibinfo{author}{\bibfnamefont{S.~R.} \bibnamefont{Nagel}},
  \bibinfo{author}{\bibfnamefont{W.}~\bibnamefont{van Saarloos}},
  \bibnamefont{and} \bibinfo{author}{\bibfnamefont{M.}~\bibnamefont{Wyart}}, in
  \emph{\bibinfo{booktitle}{Dynamical Heterogeneities in Glasses, Colloids, and
  Granular Media}}, edited by
  \bibinfo{editor}{\bibfnamefont{L.}~\bibnamefont{Berthier}},
  \bibinfo{editor}{\bibfnamefont{G.}~\bibnamefont{Biroli}},
  \bibinfo{editor}{\bibfnamefont{J.-P.} \bibnamefont{Bouchaud}},
  \bibinfo{editor}{\bibfnamefont{L.}~\bibnamefont{Cipeletti}},
  \bibnamefont{and} \bibinfo{editor}{\bibfnamefont{W.}~\bibnamefont{van
  Saarloos}} (\bibinfo{publisher}{Oxford University Press},
  \bibinfo{year}{2011}).

\bibitem[{\citenamefont{Jacquin and Berthier}(2010)}]{JB10}
\bibinfo{author}{\bibfnamefont{H.}~\bibnamefont{Jacquin}} \bibnamefont{and}
  \bibinfo{author}{\bibfnamefont{L.}~\bibnamefont{Berthier}},
  \bibinfo{journal}{Soft Matter} \textbf{\bibinfo{volume}{6}},
  \bibinfo{pages}{2970} (\bibinfo{year}{2010}).

\bibitem[{\citenamefont{Berthier
  et~al.}(2010{\natexlab{a}})\citenamefont{Berthier, Jacquin, and
  Zamponi}}]{BJZ10}
\bibinfo{author}{\bibfnamefont{L.}~\bibnamefont{Berthier}},
  \bibinfo{author}{\bibfnamefont{H.}~\bibnamefont{Jacquin}}, \bibnamefont{and}
  \bibinfo{author}{\bibfnamefont{F.}~\bibnamefont{Zamponi}},
  \bibinfo{journal}{J. Stat. Mech.} p. \bibinfo{pages}{P01004}
  (\bibinfo{year}{2010}{\natexlab{a}}).

\bibitem[{\citenamefont{Stoessel and Wolynes}(1984)}]{SW84}
\bibinfo{author}{\bibfnamefont{J.~P.} \bibnamefont{Stoessel}} \bibnamefont{and}
  \bibinfo{author}{\bibfnamefont{P.~G.} \bibnamefont{Wolynes}},
  \bibinfo{journal}{J. Chem. Phys.} \textbf{\bibinfo{volume}{80}},
  \bibinfo{pages}{4502} (\bibinfo{year}{1984}).

\bibitem[{\citenamefont{Singh et~al.}(1985)\citenamefont{Singh, Stoessel, and
  Wolynes}}]{SSW85}
\bibinfo{author}{\bibfnamefont{Y.}~\bibnamefont{Singh}},
  \bibinfo{author}{\bibfnamefont{J.~P.} \bibnamefont{Stoessel}},
  \bibnamefont{and} \bibinfo{author}{\bibfnamefont{P.~G.}
  \bibnamefont{Wolynes}}, \bibinfo{journal}{Phys. Rev. Lett.}
  \textbf{\bibinfo{volume}{54}}, \bibinfo{pages}{1059} (\bibinfo{year}{1985}).

\bibitem[{\citenamefont{Kirkpatrick and Wolynes}(1987)}]{KW87a}
\bibinfo{author}{\bibfnamefont{T.~R.} \bibnamefont{Kirkpatrick}}
  \bibnamefont{and} \bibinfo{author}{\bibfnamefont{P.~G.}
  \bibnamefont{Wolynes}}, \bibinfo{journal}{Phys. Rev. A}
  \textbf{\bibinfo{volume}{35}}, \bibinfo{pages}{3072} (\bibinfo{year}{1987}).

\bibitem[{\citenamefont{Kirkpatrick and Thirumalai}(1987)}]{KT87a}
\bibinfo{author}{\bibfnamefont{T.~R.} \bibnamefont{Kirkpatrick}}
  \bibnamefont{and}
  \bibinfo{author}{\bibfnamefont{D.}~\bibnamefont{Thirumalai}},
  \bibinfo{journal}{Phys. Rev. Lett.} \textbf{\bibinfo{volume}{58}},
  \bibinfo{pages}{2091} (\bibinfo{year}{1987}).

\bibitem[{\citenamefont{Kirkpatrick et~al.}(1989)\citenamefont{Kirkpatrick,
  Thirumalai, and Wolynes}}]{KTW89}
\bibinfo{author}{\bibfnamefont{T.~R.} \bibnamefont{Kirkpatrick}},
  \bibinfo{author}{\bibfnamefont{D.}~\bibnamefont{Thirumalai}},
  \bibnamefont{and} \bibinfo{author}{\bibfnamefont{P.~G.}
  \bibnamefont{Wolynes}}, \bibinfo{journal}{Phys. Rev. A}
  \textbf{\bibinfo{volume}{40}}, \bibinfo{pages}{1045} (\bibinfo{year}{1989}).

\bibitem[{\citenamefont{M\'ezard et~al.}(1987)\citenamefont{M\'ezard, Parisi,
  and Virasoro}}]{MPV87}
\bibinfo{author}{\bibfnamefont{M.}~\bibnamefont{M\'ezard}},
  \bibinfo{author}{\bibfnamefont{G.}~\bibnamefont{Parisi}}, \bibnamefont{and}
  \bibinfo{author}{\bibfnamefont{M.~A.} \bibnamefont{Virasoro}},
  \emph{\bibinfo{title}{Spin glass theory and beyond}}
  (\bibinfo{publisher}{World Scientific}, \bibinfo{address}{Singapore},
  \bibinfo{year}{1987}).

\bibitem[{\citenamefont{Castellani and Cavagna}(2005)}]{CC05}
\bibinfo{author}{\bibfnamefont{T.}~\bibnamefont{Castellani}} \bibnamefont{and}
  \bibinfo{author}{\bibfnamefont{A.}~\bibnamefont{Cavagna}},
  \bibinfo{journal}{J. Stat. Mech.} p. \bibinfo{pages}{P05012}
  (\bibinfo{year}{2005}).

\bibitem[{\citenamefont{Lubchenko and Wolynes}(2007)}]{LW07}
\bibinfo{author}{\bibfnamefont{V.}~\bibnamefont{Lubchenko}} \bibnamefont{and}
  \bibinfo{author}{\bibfnamefont{P.~G.} \bibnamefont{Wolynes}},
  \bibinfo{journal}{Ann. Rev. Phys. Chem.} \textbf{\bibinfo{volume}{58}},
  \bibinfo{pages}{235} (\bibinfo{year}{2007}).

\bibitem[{\citenamefont{Cavagna}(2009)}]{Ca09}
\bibinfo{author}{\bibfnamefont{A.}~\bibnamefont{Cavagna}},
  \bibinfo{journal}{Phys. Rep.} \textbf{\bibinfo{volume}{476}},
  \bibinfo{pages}{51} (\bibinfo{year}{2009}).

\bibitem[{\citenamefont{Biroli and Bouchaud}(2009)}]{BB09}
\bibinfo{author}{\bibfnamefont{G.}~\bibnamefont{Biroli}} \bibnamefont{and}
  \bibinfo{author}{\bibfnamefont{J.-P.} \bibnamefont{Bouchaud}},
  \bibinfo{journal}{{\tt arXiv:0912.2542}}  (\bibinfo{year}{2009}).

\bibitem[{\citenamefont{Monasson}(1995)}]{Mo95}
\bibinfo{author}{\bibfnamefont{R.}~\bibnamefont{Monasson}},
  \bibinfo{journal}{Phys. Rev. Lett.} \textbf{\bibinfo{volume}{75}},
  \bibinfo{pages}{2847} (\bibinfo{year}{1995}).

\bibitem[{\citenamefont{M\'ezard and Parisi}(1996)}]{MP96}
\bibinfo{author}{\bibfnamefont{M.}~\bibnamefont{M\'ezard}} \bibnamefont{and}
  \bibinfo{author}{\bibfnamefont{G.}~\bibnamefont{Parisi}},
  \bibinfo{journal}{J. Phys. A} \textbf{\bibinfo{volume}{29}},
  \bibinfo{pages}{6515} (\bibinfo{year}{1996}).

\bibitem[{\citenamefont{Cardenas et~al.}(1998)\citenamefont{Cardenas, Franz,
  and Parisi}}]{CFP98}
\bibinfo{author}{\bibfnamefont{M.}~\bibnamefont{Cardenas}},
  \bibinfo{author}{\bibfnamefont{S.}~\bibnamefont{Franz}}, \bibnamefont{and}
  \bibinfo{author}{\bibfnamefont{G.}~\bibnamefont{Parisi}},
  \bibinfo{journal}{J. Phys. A} \textbf{\bibinfo{volume}{31}},
  \bibinfo{pages}{163} (\bibinfo{year}{1998}).

\bibitem[{\citenamefont{M\'ezard and Parisi}(1999{\natexlab{a}})}]{MP99a}
\bibinfo{author}{\bibfnamefont{M.}~\bibnamefont{M\'ezard}} \bibnamefont{and}
  \bibinfo{author}{\bibfnamefont{G.}~\bibnamefont{Parisi}},
  \bibinfo{journal}{J. Chem. Phys.} \textbf{\bibinfo{volume}{111}},
  \bibinfo{pages}{1076} (\bibinfo{year}{1999}{\natexlab{a}}).

\bibitem[{\citenamefont{M\'ezard and Parisi}(1999{\natexlab{b}})}]{MP99b}
\bibinfo{author}{\bibfnamefont{M.}~\bibnamefont{M\'ezard}} \bibnamefont{and}
  \bibinfo{author}{\bibfnamefont{G.}~\bibnamefont{Parisi}},
  \bibinfo{journal}{Phys. Rev. Lett.} \textbf{\bibinfo{volume}{82}},
  \bibinfo{pages}{747} (\bibinfo{year}{1999}{\natexlab{b}}).

\bibitem[{\citenamefont{van Hecke}(2010)}]{Vh09}
\bibinfo{author}{\bibfnamefont{M.}~\bibnamefont{van Hecke}},
  \bibinfo{journal}{J. Phys.: Condens. Matter} \textbf{\bibinfo{volume}{22}},
  \bibinfo{pages}{033101} (\bibinfo{year}{2010}).

\bibitem[{\citenamefont{Parisi and Zamponi}(2010)}]{PZ10}
\bibinfo{author}{\bibfnamefont{G.}~\bibnamefont{Parisi}} \bibnamefont{and}
  \bibinfo{author}{\bibfnamefont{F.}~\bibnamefont{Zamponi}},
  \bibinfo{journal}{Rev. Mod. Phys.} \textbf{\bibinfo{volume}{82}},
  \bibinfo{pages}{789} (\bibinfo{year}{2010}).

\bibitem[{\citenamefont{Mari and Kurchan}(2011)}]{MK11}
\bibinfo{author}{\bibfnamefont{R.}~\bibnamefont{Mari}} \bibnamefont{and}
  \bibinfo{author}{\bibfnamefont{J.}~\bibnamefont{Kurchan}},
  \bibinfo{journal}{{\tt arXiv:1104.3420}}  (\bibinfo{year}{2011}).

\bibitem[{\citenamefont{Mari et~al.}(2009)\citenamefont{Mari, Krzakala, and
  Kurchan}}]{MKK09}
\bibinfo{author}{\bibfnamefont{R.}~\bibnamefont{Mari}},
  \bibinfo{author}{\bibfnamefont{F.}~\bibnamefont{Krzakala}}, \bibnamefont{and}
  \bibinfo{author}{\bibfnamefont{J.}~\bibnamefont{Kurchan}},
  \bibinfo{journal}{Phys. Rev. Lett.} \textbf{\bibinfo{volume}{103}},
  \bibinfo{pages}{025701} (\bibinfo{year}{2009}).

\bibitem[{\citenamefont{M\'ezard et~al.}(2011)\citenamefont{M\'ezard, Parisi,
  Tarzia, and Zamponi}}]{solvable}
\bibinfo{author}{\bibfnamefont{M.}~\bibnamefont{M\'ezard}},
  \bibinfo{author}{\bibfnamefont{G.}~\bibnamefont{Parisi}},
  \bibinfo{author}{\bibfnamefont{M.}~\bibnamefont{Tarzia}}, \bibnamefont{and}
  \bibinfo{author}{\bibfnamefont{F.}~\bibnamefont{Zamponi}},
  \bibinfo{journal}{J. Stat. Mech.} p. \bibinfo{pages}{P03002}
  (\bibinfo{year}{2011}).

\bibitem[{\citenamefont{Durian}(1995)}]{Du95}
\bibinfo{author}{\bibfnamefont{D.~J.} \bibnamefont{Durian}},
  \bibinfo{journal}{Phys. Rev. Lett.} \textbf{\bibinfo{volume}{75}},
  \bibinfo{pages}{4780} (\bibinfo{year}{1995}).

\bibitem[{\citenamefont{O'Hern et~al.}(2002)\citenamefont{O'Hern, Langer, Liu,
  and Nagel}}]{OLLN02}
\bibinfo{author}{\bibfnamefont{C.~S.} \bibnamefont{O'Hern}},
  \bibinfo{author}{\bibfnamefont{S.~A.} \bibnamefont{Langer}},
  \bibinfo{author}{\bibfnamefont{A.~J.} \bibnamefont{Liu}}, \bibnamefont{and}
  \bibinfo{author}{\bibfnamefont{S.~R.} \bibnamefont{Nagel}},
  \bibinfo{journal}{Phys. Rev. Lett.} \textbf{\bibinfo{volume}{88}},
  \bibinfo{pages}{075507} (\bibinfo{year}{2002}).

\bibitem[{\citenamefont{Berthier and Witten}(2009{\natexlab{a}})}]{BW09b}
\bibinfo{author}{\bibfnamefont{L.}~\bibnamefont{Berthier}} \bibnamefont{and}
  \bibinfo{author}{\bibfnamefont{T.~A.} \bibnamefont{Witten}},
  \bibinfo{journal}{Europhys. Lett.} \textbf{\bibinfo{volume}{86}},
  \bibinfo{pages}{10001} (\bibinfo{year}{2009}{\natexlab{a}}).

\bibitem[{\citenamefont{Berthier and Witten}(2009{\natexlab{b}})}]{BW09a}
\bibinfo{author}{\bibfnamefont{L.}~\bibnamefont{Berthier}} \bibnamefont{and}
  \bibinfo{author}{\bibfnamefont{T.~A.} \bibnamefont{Witten}},
  \bibinfo{journal}{Phys. Rev. E} \textbf{\bibinfo{volume}{80}},
  \bibinfo{pages}{021502} (\bibinfo{year}{2009}{\natexlab{b}}).

\bibitem[{\citenamefont{Zhang et~al.}(2009)\citenamefont{Zhang, Xu, Chen,
  Yunker, Alsayed, Aptowicz, Habdas, Liu, Nagel, and Yodh}}]{ZXCYAAHLNY09}
\bibinfo{author}{\bibfnamefont{Z.}~\bibnamefont{Zhang}},
  \bibinfo{author}{\bibfnamefont{N.}~\bibnamefont{Xu}},
  \bibinfo{author}{\bibfnamefont{D.~T.~N.} \bibnamefont{Chen}},
  \bibinfo{author}{\bibfnamefont{P.}~\bibnamefont{Yunker}},
  \bibinfo{author}{\bibfnamefont{A.~M.} \bibnamefont{Alsayed}},
  \bibinfo{author}{\bibfnamefont{K.~B.} \bibnamefont{Aptowicz}},
  \bibinfo{author}{\bibfnamefont{P.}~\bibnamefont{Habdas}},
  \bibinfo{author}{\bibfnamefont{A.~J.} \bibnamefont{Liu}},
  \bibinfo{author}{\bibfnamefont{S.~R.} \bibnamefont{Nagel}}, \bibnamefont{and}
  \bibinfo{author}{\bibfnamefont{A.~G.} \bibnamefont{Yodh}},
  \bibinfo{journal}{Nature} \textbf{\bibinfo{volume}{459}},
  \bibinfo{pages}{230} (\bibinfo{year}{2009}).

\bibitem[{\citenamefont{Fernandez-Nieves
  et~al.}(2011)\citenamefont{Fernandez-Nieves, Wyss, Mattsson, and
  Weitz}}]{bookmicrogel}
\bibinfo{editor}{\bibfnamefont{A.}~\bibnamefont{Fernandez-Nieves}},
  \bibinfo{editor}{\bibfnamefont{H.}~\bibnamefont{Wyss}},
  \bibinfo{editor}{\bibfnamefont{J.}~\bibnamefont{Mattsson}}, \bibnamefont{and}
  \bibinfo{editor}{\bibfnamefont{D.~A.} \bibnamefont{Weitz}}, eds.,
  \emph{\bibinfo{title}{Microgel suspensions}} (\bibinfo{publisher}{Wiley-VCH},
  \bibinfo{year}{2011}).

\bibitem[{\citenamefont{Donev et~al.}(2005{\natexlab{a}})\citenamefont{Donev,
  Torquato, and Stillinger}}]{DTS05}
\bibinfo{author}{\bibfnamefont{A.}~\bibnamefont{Donev}},
  \bibinfo{author}{\bibfnamefont{S.}~\bibnamefont{Torquato}}, \bibnamefont{and}
  \bibinfo{author}{\bibfnamefont{F.~H.} \bibnamefont{Stillinger}},
  \bibinfo{journal}{Phys. Rev. E} \textbf{\bibinfo{volume}{71}},
  \bibinfo{pages}{011105} (\bibinfo{year}{2005}{\natexlab{a}}).

\bibitem[{\citenamefont{Silbert et~al.}(2006)\citenamefont{Silbert, Liu, and
  Nagel}}]{SLN06}
\bibinfo{author}{\bibfnamefont{L.~E.} \bibnamefont{Silbert}},
  \bibinfo{author}{\bibfnamefont{A.~J.} \bibnamefont{Liu}}, \bibnamefont{and}
  \bibinfo{author}{\bibfnamefont{S.~R.} \bibnamefont{Nagel}},
  \bibinfo{journal}{Phys. Rev. E} \textbf{\bibinfo{volume}{73}},
  \bibinfo{pages}{041304} (\bibinfo{year}{2006}).

\bibitem[{\citenamefont{Heussinger and Barrat}(2009)}]{HB09}
\bibinfo{author}{\bibfnamefont{C.}~\bibnamefont{Heussinger}} \bibnamefont{and}
  \bibinfo{author}{\bibfnamefont{J.-L.} \bibnamefont{Barrat}},
  \bibinfo{journal}{Phys. Rev. Lett.} \textbf{\bibinfo{volume}{102}},
  \bibinfo{pages}{218303} (\bibinfo{year}{2009}).

\bibitem[{\citenamefont{Gao et~al.}(2006)\citenamefont{Gao, Blawzdziewicz, and
  O'Hern}}]{GBO06}
\bibinfo{author}{\bibfnamefont{G.-J.} \bibnamefont{Gao}},
  \bibinfo{author}{\bibfnamefont{J.}~\bibnamefont{Blawzdziewicz}},
  \bibnamefont{and} \bibinfo{author}{\bibfnamefont{C.~S.}
  \bibnamefont{O'Hern}}, \bibinfo{journal}{Phys. Rev. E}
  \textbf{\bibinfo{volume}{74}}, \bibinfo{pages}{061304}
  (\bibinfo{year}{2006}).

\bibitem[{\citenamefont{Krzakala and Kurchan}(2007)}]{KK07a}
\bibinfo{author}{\bibfnamefont{F.}~\bibnamefont{Krzakala}} \bibnamefont{and}
  \bibinfo{author}{\bibfnamefont{J.}~\bibnamefont{Kurchan}},
  \bibinfo{journal}{Phys. Rev. E} \textbf{\bibinfo{volume}{76}},
  \bibinfo{pages}{021122} (\bibinfo{year}{2007}).

\bibitem[{\citenamefont{Monasson}(2007)}]{Mo07}
\bibinfo{author}{\bibfnamefont{R.}~\bibnamefont{Monasson}}, in
  \emph{\bibinfo{booktitle}{Complex Systems}}, edited by
  \bibinfo{editor}{\bibfnamefont{J.-P.} \bibnamefont{Bouchaud}},
  \bibinfo{editor}{\bibfnamefont{M.}~\bibnamefont{M\'ezard}}, \bibnamefont{and}
  \bibinfo{editor}{\bibfnamefont{J.}~\bibnamefont{Dalibard}}
  (\bibinfo{publisher}{Elsevier}, \bibinfo{year}{2007}).

\bibitem[{\citenamefont{M\'ezard and Montanari}(2009)}]{MM09}
\bibinfo{author}{\bibfnamefont{M.}~\bibnamefont{M\'ezard}} \bibnamefont{and}
  \bibinfo{author}{\bibfnamefont{A.}~\bibnamefont{Montanari}},
  \emph{\bibinfo{title}{Information, Physics, and Computation}}
  (\bibinfo{publisher}{Oxford University Press}, \bibinfo{year}{2009}).

\bibitem[{\citenamefont{M\'ezard and Parisi}(2003)}]{MP03}
\bibinfo{author}{\bibfnamefont{M.}~\bibnamefont{M\'ezard}} \bibnamefont{and}
  \bibinfo{author}{\bibfnamefont{G.}~\bibnamefont{Parisi}},
  \bibinfo{journal}{J. Stat. Phys.} \textbf{\bibinfo{volume}{111}},
  \bibinfo{pages}{1} (\bibinfo{year}{2003}).

\bibitem[{\citenamefont{Krzakala et~al.}(2007)\citenamefont{Krzakala,
  Montanari, Ricci-Tersenghi, Semerjian, and Zdeborov\'a}}]{KMRSZ07}
\bibinfo{author}{\bibfnamefont{F.}~\bibnamefont{Krzakala}},
  \bibinfo{author}{\bibfnamefont{A.}~\bibnamefont{Montanari}},
  \bibinfo{author}{\bibfnamefont{F.}~\bibnamefont{Ricci-Tersenghi}},
  \bibinfo{author}{\bibfnamefont{G.}~\bibnamefont{Semerjian}},
  \bibnamefont{and}
  \bibinfo{author}{\bibfnamefont{L.}~\bibnamefont{Zdeborov\'a}},
  \bibinfo{journal}{Proc. Nat. Acad. Sci.} \textbf{\bibinfo{volume}{104}},
  \bibinfo{pages}{10318} (\bibinfo{year}{2007}).

\bibitem[{\citenamefont{Krzakala and Zdeborov\'a}(2008)}]{KZ08}
\bibinfo{author}{\bibfnamefont{F.}~\bibnamefont{Krzakala}} \bibnamefont{and}
  \bibinfo{author}{\bibfnamefont{L.}~\bibnamefont{Zdeborov\'a}},
  \bibinfo{journal}{EPL} \textbf{\bibinfo{volume}{81}}, \bibinfo{pages}{57005}
  (\bibinfo{year}{2008}).

\bibitem[{\citenamefont{Ricci-Tersenghi}(2010)}]{ricci10}
\bibinfo{author}{\bibfnamefont{F.}~\bibnamefont{Ricci-Tersenghi}},
  \bibinfo{journal}{Science} \textbf{\bibinfo{volume}{330}},
  \bibinfo{pages}{1639} (\bibinfo{year}{2010}).

\bibitem[{\citenamefont{Jacquin et~al.}(2011)\citenamefont{Jacquin, Berthier,
  and Zamponi}}]{JBZ11}
\bibinfo{author}{\bibfnamefont{H.}~\bibnamefont{Jacquin}},
  \bibinfo{author}{\bibfnamefont{L.}~\bibnamefont{Berthier}}, \bibnamefont{and}
  \bibinfo{author}{\bibfnamefont{F.}~\bibnamefont{Zamponi}},
  \bibinfo{journal}{Phys. Rev. Lett.} \textbf{\bibinfo{volume}{106}},
  \bibinfo{pages}{135702} (\bibinfo{year}{2011}).

\bibitem[{\citenamefont{Clusel et~al.}(2009)\citenamefont{Clusel, Corwin,
  Siemens, and Brujic}}]{CCSB09}
\bibinfo{author}{\bibfnamefont{M.}~\bibnamefont{Clusel}},
  \bibinfo{author}{\bibfnamefont{E.~I.} \bibnamefont{Corwin}},
  \bibinfo{author}{\bibfnamefont{A.~O.~N.} \bibnamefont{Siemens}},
  \bibnamefont{and} \bibinfo{author}{\bibfnamefont{J.}~\bibnamefont{Brujic}},
  \bibinfo{journal}{Nature} \textbf{\bibinfo{volume}{460}},
  \bibinfo{pages}{611} (\bibinfo{year}{2009}).

\bibitem[{\citenamefont{Wyart et~al.}(2005{\natexlab{a}})\citenamefont{Wyart,
  Nagel, and Witten}}]{WNW05}
\bibinfo{author}{\bibfnamefont{M.}~\bibnamefont{Wyart}},
  \bibinfo{author}{\bibfnamefont{S.}~\bibnamefont{Nagel}}, \bibnamefont{and}
  \bibinfo{author}{\bibfnamefont{T.}~\bibnamefont{Witten}},
  \bibinfo{journal}{Europhysics Letters} \textbf{\bibinfo{volume}{72}},
  \bibinfo{pages}{486} (\bibinfo{year}{2005}{\natexlab{a}}).

\bibitem[{\citenamefont{Wyart et~al.}(2005{\natexlab{b}})\citenamefont{Wyart,
  Silbert, Nagel, and Witten}}]{WSNW05}
\bibinfo{author}{\bibfnamefont{M.}~\bibnamefont{Wyart}},
  \bibinfo{author}{\bibfnamefont{L.~E.} \bibnamefont{Silbert}},
  \bibinfo{author}{\bibfnamefont{S.~R.} \bibnamefont{Nagel}}, \bibnamefont{and}
  \bibinfo{author}{\bibfnamefont{T.~A.} \bibnamefont{Witten}},
  \bibinfo{journal}{Phys. Rev. E} \textbf{\bibinfo{volume}{72}},
  \bibinfo{pages}{051306} (\bibinfo{year}{2005}{\natexlab{b}}).

\bibitem[{\citenamefont{Brito and Wyart}(2006)}]{BW06}
\bibinfo{author}{\bibfnamefont{C.}~\bibnamefont{Brito}} \bibnamefont{and}
  \bibinfo{author}{\bibfnamefont{M.}~\bibnamefont{Wyart}},
  \bibinfo{journal}{Europhysics Letters (EPL)} \textbf{\bibinfo{volume}{76}},
  \bibinfo{pages}{149} (\bibinfo{year}{2006}).

\bibitem[{\citenamefont{Wyart}(2005)}]{Wyart}
\bibinfo{author}{\bibfnamefont{M.}~\bibnamefont{Wyart}},
  \bibinfo{journal}{Annales de Physique} \textbf{\bibinfo{volume}{30}},
  \bibinfo{pages}{1} (\bibinfo{year}{2005}), \eprint{{\tt
  ar{X}iv:cond-mat/0512155}}.

\bibitem[{\citenamefont{O'Hern et~al.}(2003)\citenamefont{O'Hern, Silbert, Liu,
  and Nagel}}]{OSLN03}
\bibinfo{author}{\bibfnamefont{C.~S.} \bibnamefont{O'Hern}},
  \bibinfo{author}{\bibfnamefont{L.~E.} \bibnamefont{Silbert}},
  \bibinfo{author}{\bibfnamefont{A.~J.} \bibnamefont{Liu}}, \bibnamefont{and}
  \bibinfo{author}{\bibfnamefont{S.~R.} \bibnamefont{Nagel}},
  \bibinfo{journal}{Phys. Rev. E} \textbf{\bibinfo{volume}{68}},
  \bibinfo{pages}{011306} (\bibinfo{year}{2003}).

\bibitem[{\citenamefont{Vitelli et~al.}(2010)\citenamefont{Vitelli, Xu, Wyart,
  Liu, and Nagel}}]{VXWLN10}
\bibinfo{author}{\bibfnamefont{V.}~\bibnamefont{Vitelli}},
  \bibinfo{author}{\bibfnamefont{N.}~\bibnamefont{Xu}},
  \bibinfo{author}{\bibfnamefont{M.}~\bibnamefont{Wyart}},
  \bibinfo{author}{\bibfnamefont{A.~J.} \bibnamefont{Liu}}, \bibnamefont{and}
  \bibinfo{author}{\bibfnamefont{S.~R.} \bibnamefont{Nagel}},
  \bibinfo{journal}{Phys. Rev. E} \textbf{\bibinfo{volume}{81}},
  \bibinfo{pages}{021301} (\bibinfo{year}{2010}).

\bibitem[{\citenamefont{Gao et~al.}(2009)\citenamefont{Gao, Blawzdziewicz,
  O'Hern, and Shattuck}}]{GBOS09}
\bibinfo{author}{\bibfnamefont{G.-J.} \bibnamefont{Gao}},
  \bibinfo{author}{\bibfnamefont{J.}~\bibnamefont{Blawzdziewicz}},
  \bibinfo{author}{\bibfnamefont{C.~S.} \bibnamefont{O'Hern}},
  \bibnamefont{and} \bibinfo{author}{\bibfnamefont{M.}~\bibnamefont{Shattuck}},
  \bibinfo{journal}{Phys. Rev. E} \textbf{\bibinfo{volume}{80}},
  \bibinfo{pages}{061304} (\bibinfo{year}{2009}).

\bibitem[{\citenamefont{Jin and Makse}(2010)}]{JM10}
\bibinfo{author}{\bibfnamefont{Y.}~\bibnamefont{Jin}} \bibnamefont{and}
  \bibinfo{author}{\bibfnamefont{H.}~\bibnamefont{Makse}},
  \bibinfo{journal}{Physica A: Statistical Mechanics and its Applications}
  \textbf{\bibinfo{volume}{389}} (\bibinfo{year}{2010}).

\bibitem[{\citenamefont{Schreck
  et~al.}(2011{\natexlab{a}})\citenamefont{Schreck, O'Hern, and
  Silbert}}]{SOS11}
\bibinfo{author}{\bibfnamefont{C.~F.} \bibnamefont{Schreck}},
  \bibinfo{author}{\bibfnamefont{C.~S.} \bibnamefont{O'Hern}},
  \bibnamefont{and} \bibinfo{author}{\bibfnamefont{L.~E.}
  \bibnamefont{Silbert}}, \bibinfo{journal}{Phys. Rev. E}
  \textbf{\bibinfo{volume}{84}}, \bibinfo{pages}{011305}
  (\bibinfo{year}{2011}{\natexlab{a}}).

\bibitem[{\citenamefont{Jiao et~al.}(2011)\citenamefont{Jiao, Stillinger, and
  Torquato}}]{JST11}
\bibinfo{author}{\bibfnamefont{Y.}~\bibnamefont{Jiao}},
  \bibinfo{author}{\bibfnamefont{F.}~\bibnamefont{Stillinger}},
  \bibnamefont{and} \bibinfo{author}{\bibfnamefont{S.}~\bibnamefont{Torquato}},
  \bibinfo{journal}{Journal of Applied Physics} \textbf{\bibinfo{volume}{109}},
  \bibinfo{pages}{013508} (\bibinfo{year}{2011}).

\bibitem[{\citenamefont{Kauzmann}(1948)}]{Ka48}
\bibinfo{author}{\bibfnamefont{W.}~\bibnamefont{Kauzmann}},
  \bibinfo{journal}{Chem. Rev.} \textbf{\bibinfo{volume}{43}},
  \bibinfo{pages}{219} (\bibinfo{year}{1948}).

\bibitem[{\citenamefont{Castellana et~al.}(2010)\citenamefont{Castellana,
  Decelle, Franz, M\'ezard, and Parisi}}]{CDFMP10}
\bibinfo{author}{\bibfnamefont{M.}~\bibnamefont{Castellana}},
  \bibinfo{author}{\bibfnamefont{A.}~\bibnamefont{Decelle}},
  \bibinfo{author}{\bibfnamefont{S.}~\bibnamefont{Franz}},
  \bibinfo{author}{\bibfnamefont{M.}~\bibnamefont{M\'ezard}}, \bibnamefont{and}
  \bibinfo{author}{\bibfnamefont{G.}~\bibnamefont{Parisi}},
  \bibinfo{journal}{Phys. Rev. Lett.} \textbf{\bibinfo{volume}{104}},
  \bibinfo{pages}{127206} (\bibinfo{year}{2010}).

\bibitem[{\citenamefont{Cammarota et~al.}(2011)\citenamefont{Cammarota, Biroli,
  Tarzia, and Tarjus}}]{CBTT11}
\bibinfo{author}{\bibfnamefont{C.}~\bibnamefont{Cammarota}},
  \bibinfo{author}{\bibfnamefont{G.}~\bibnamefont{Biroli}},
  \bibinfo{author}{\bibfnamefont{M.}~\bibnamefont{Tarzia}}, \bibnamefont{and}
  \bibinfo{author}{\bibfnamefont{G.}~\bibnamefont{Tarjus}},
  \bibinfo{journal}{Phys. Rev. Lett.} \textbf{\bibinfo{volume}{106}},
  \bibinfo{pages}{115705} (\bibinfo{year}{2011}).

\bibitem[{\citenamefont{Xu et~al.}(2011)\citenamefont{Xu, Frenkel, and
  Liu}}]{XFL11}
\bibinfo{author}{\bibfnamefont{N.}~\bibnamefont{Xu}},
  \bibinfo{author}{\bibfnamefont{D.}~\bibnamefont{Frenkel}}, \bibnamefont{and}
  \bibinfo{author}{\bibfnamefont{A.~J.} \bibnamefont{Liu}},
  \bibinfo{journal}{Phys. Rev. Lett.} \textbf{\bibinfo{volume}{106}},
  \bibinfo{pages}{245502} (\bibinfo{year}{2011}).

\bibitem[{\citenamefont{Foini et~al.}(2011)\citenamefont{Foini, Semerjian, and
  Zamponi}}]{FSZ11}
\bibinfo{author}{\bibfnamefont{L.}~\bibnamefont{Foini}},
  \bibinfo{author}{\bibfnamefont{G.}~\bibnamefont{Semerjian}},
  \bibnamefont{and} \bibinfo{author}{\bibfnamefont{F.}~\bibnamefont{Zamponi}},
  \bibinfo{journal}{Phys. Rev. B} \textbf{\bibinfo{volume}{83}},
  \bibinfo{pages}{094513} (\bibinfo{year}{2011}).

\bibitem[{\citenamefont{Berthier
  et~al.}(2010{\natexlab{b}})\citenamefont{Berthier, Flenner, Jacquin, and
  Szamel}}]{BFJS10}
\bibinfo{author}{\bibfnamefont{L.}~\bibnamefont{Berthier}},
  \bibinfo{author}{\bibfnamefont{E.}~\bibnamefont{Flenner}},
  \bibinfo{author}{\bibfnamefont{H.}~\bibnamefont{Jacquin}}, \bibnamefont{and}
  \bibinfo{author}{\bibfnamefont{G.}~\bibnamefont{Szamel}},
  \bibinfo{journal}{Phys. Rev. E} \textbf{\bibinfo{volume}{81}},
  \bibinfo{pages}{031505} (\bibinfo{year}{2010}{\natexlab{b}}).

\bibitem[{\citenamefont{Gibbs}(1956)}]{Gi56}
\bibinfo{author}{\bibfnamefont{J.~H.} \bibnamefont{Gibbs}},
  \bibinfo{journal}{J. Chem. Phys.} \textbf{\bibinfo{volume}{25}},
  \bibinfo{pages}{185} (\bibinfo{year}{1956}).

\bibitem[{\citenamefont{Gibbs and Di~Marzio}(1958)}]{GD58}
\bibinfo{author}{\bibfnamefont{J.~H.} \bibnamefont{Gibbs}} \bibnamefont{and}
  \bibinfo{author}{\bibfnamefont{E.~A.} \bibnamefont{Di~Marzio}},
  \bibinfo{journal}{J. Chem. Phys.} \textbf{\bibinfo{volume}{28}},
  \bibinfo{pages}{373} (\bibinfo{year}{1958}).

\bibitem[{\citenamefont{Adam and Gibbs}(1965)}]{AG65}
\bibinfo{author}{\bibfnamefont{G.}~\bibnamefont{Adam}} \bibnamefont{and}
  \bibinfo{author}{\bibfnamefont{J.~H.} \bibnamefont{Gibbs}},
  \bibinfo{journal}{J. Chem. Phys.} \textbf{\bibinfo{volume}{43}},
  \bibinfo{pages}{139} (\bibinfo{year}{1965}).

\bibitem[{\citenamefont{Martinez and Angell}(2001)}]{MA01}
\bibinfo{author}{\bibfnamefont{L.-M.} \bibnamefont{Martinez}} \bibnamefont{and}
  \bibinfo{author}{\bibfnamefont{C.~A.} \bibnamefont{Angell}},
  \bibinfo{journal}{Nature} \textbf{\bibinfo{volume}{410}},
  \bibinfo{pages}{663} (\bibinfo{year}{2001}).

\bibitem[{\citenamefont{Salsburg and Wood}(1962)}]{SW62}
\bibinfo{author}{\bibfnamefont{Z.~W.} \bibnamefont{Salsburg}} \bibnamefont{and}
  \bibinfo{author}{\bibfnamefont{W.~W.} \bibnamefont{Wood}},
  \bibinfo{journal}{J. Chem. Phys.} \textbf{\bibinfo{volume}{37}},
  \bibinfo{pages}{798} (\bibinfo{year}{1962}).

\bibitem[{\citenamefont{Olsson and Teitel}(2007)}]{OT07}
\bibinfo{author}{\bibfnamefont{P.}~\bibnamefont{Olsson}} \bibnamefont{and}
  \bibinfo{author}{\bibfnamefont{S.}~\bibnamefont{Teitel}},
  \bibinfo{journal}{Phys. Rev. Lett.} \textbf{\bibinfo{volume}{99}},
  \bibinfo{pages}{178001} (\bibinfo{year}{2007}).

\bibitem[{\citenamefont{Cheng}(2010)}]{Ch10a}
\bibinfo{author}{\bibfnamefont{X.}~\bibnamefont{Cheng}},
  \bibinfo{journal}{Phys. Rev. E} \textbf{\bibinfo{volume}{81}},
  \bibinfo{pages}{031301} (\bibinfo{year}{2010}).

\bibitem[{\citenamefont{Chaudhuri et~al.}(2010)\citenamefont{Chaudhuri,
  Berthier, and Sastry}}]{CBS09}
\bibinfo{author}{\bibfnamefont{P.}~\bibnamefont{Chaudhuri}},
  \bibinfo{author}{\bibfnamefont{L.}~\bibnamefont{Berthier}}, \bibnamefont{and}
  \bibinfo{author}{\bibfnamefont{S.}~\bibnamefont{Sastry}},
  \bibinfo{journal}{Phys. Rev. Lett.} \textbf{\bibinfo{volume}{104}},
  \bibinfo{pages}{165701} (\bibinfo{year}{2010}).

\bibitem[{\citenamefont{Xu et~al.}(2010)\citenamefont{Xu, Vitelli, Liu, and
  Nagel}}]{XVLN10}
\bibinfo{author}{\bibfnamefont{N.}~\bibnamefont{Xu}},
  \bibinfo{author}{\bibfnamefont{V.}~\bibnamefont{Vitelli}},
  \bibinfo{author}{\bibfnamefont{A.~J.} \bibnamefont{Liu}}, \bibnamefont{and}
  \bibinfo{author}{\bibfnamefont{S.~R.} \bibnamefont{Nagel}},
  \bibinfo{journal}{Europhys. Lett.} \textbf{\bibinfo{volume}{90}},
  \bibinfo{pages}{56001} (\bibinfo{year}{2010}).

\bibitem[{\citenamefont{Schreck
  et~al.}(2011{\natexlab{b}})\citenamefont{Schreck, Bertrand, O'Hern, and
  Shattuck}}]{SBOS11}
\bibinfo{author}{\bibfnamefont{C.~F.} \bibnamefont{Schreck}},
  \bibinfo{author}{\bibfnamefont{T.}~\bibnamefont{Bertrand}},
  \bibinfo{author}{\bibfnamefont{C.~S.} \bibnamefont{O'Hern}},
  \bibnamefont{and} \bibinfo{author}{\bibfnamefont{M.~D.}
  \bibnamefont{Shattuck}}, \bibinfo{journal}{Phys. Rev. Lett.}
  \textbf{\bibinfo{volume}{107}}, \bibinfo{pages}{078301}
  (\bibinfo{year}{2011}{\natexlab{b}}).

\bibitem[{\citenamefont{Yoshino and M\'ezard}(2010)}]{MY10}
\bibinfo{author}{\bibfnamefont{H.}~\bibnamefont{Yoshino}} \bibnamefont{and}
  \bibinfo{author}{\bibfnamefont{M.}~\bibnamefont{M\'ezard}},
  \bibinfo{journal}{Phys. Rev. Lett.} \textbf{\bibinfo{volume}{105}},
  \bibinfo{pages}{015504} (\bibinfo{year}{2010}).

\bibitem[{\citenamefont{Szamel and Flenner}(2011)}]{SF11}
\bibinfo{author}{\bibfnamefont{G.}~\bibnamefont{Szamel}} \bibnamefont{and}
  \bibinfo{author}{\bibfnamefont{E.}~\bibnamefont{Flenner}},
  \bibinfo{journal}{{\tt arXiv:1105.3578}}  (\bibinfo{year}{2011}).

\bibitem[{\citenamefont{Donev et~al.}(2005{\natexlab{b}})\citenamefont{Donev,
  Torquato, and Stillinger}}]{DTS05b}
\bibinfo{author}{\bibfnamefont{A.}~\bibnamefont{Donev}},
  \bibinfo{author}{\bibfnamefont{S.}~\bibnamefont{Torquato}}, \bibnamefont{and}
  \bibinfo{author}{\bibfnamefont{F.~H.} \bibnamefont{Stillinger}},
  \bibinfo{journal}{Phys. Rev. Lett.} \textbf{\bibinfo{volume}{95}},
  \bibinfo{pages}{090604} (\bibinfo{year}{2005}{\natexlab{b}}).

\bibitem[{\citenamefont{Berthier et~al.}(2011)\citenamefont{Berthier,
  Chaudhuri, Coulais, Dauchot, and Sollich}}]{BCCDS11}
\bibinfo{author}{\bibfnamefont{L.}~\bibnamefont{Berthier}},
  \bibinfo{author}{\bibfnamefont{P.}~\bibnamefont{Chaudhuri}},
  \bibinfo{author}{\bibfnamefont{C.}~\bibnamefont{Coulais}},
  \bibinfo{author}{\bibfnamefont{O.}~\bibnamefont{Dauchot}}, \bibnamefont{and}
  \bibinfo{author}{\bibfnamefont{P.}~\bibnamefont{Sollich}},
  \bibinfo{journal}{Phys. Rev. Lett.} \textbf{\bibinfo{volume}{106}},
  \bibinfo{pages}{120601} (\bibinfo{year}{2011}).

\bibitem[{\citenamefont{Stillinger and Weber}(1985)}]{SW85}
\bibinfo{author}{\bibfnamefont{F.~H.} \bibnamefont{Stillinger}}
  \bibnamefont{and} \bibinfo{author}{\bibfnamefont{T.~A.} \bibnamefont{Weber}},
  \bibinfo{journal}{Phys. Rev. B} \textbf{\bibinfo{volume}{31}},
  \bibinfo{pages}{5262} (\bibinfo{year}{1985}).

\end{thebibliography}

\end{document}